\documentclass[twocolumn]{aastex631}
\usepackage{booktabs}
\usepackage{enumitem}
\usepackage[para]{threeparttable}
\let\tablenum\relax
\usepackage{siunitx}

\usepackage{amsmath}

\shorttitle{Investigating the morphology of kilonova host galaxies}
\shortauthors{Skobe et al.}
\graphicspath{{./}{figures/}}

\newcommand{\manchester}{
    Jodrell Bank Centre for Astrophysics, Alan Turing Building, University of Manchester, Oxford Road, Manchester M13 9PL, UK
}

\begin{document}

\title{The host galaxies and merger environments of short and long gamma-ray bursts producing kilonovae}

\correspondingauthor{Hannah Skobe}
\email{hskobe@andrew.cmu.edu}

\author[0000-0003-0516-3485]{Hannah Skobe}
    \affiliation{McWilliams Center for Cosmology and Astrophysics, Department of Physics, Carnegie Mellon University, Pittsburgh, PA 15213, USA}

\author[0000-0002-9700-0036]{Brendan O'Connor}
    \altaffiliation{McWilliams Fellow}
    \affiliation{McWilliams Center for Cosmology and Astrophysics, Department of Physics, Carnegie Mellon University, Pittsburgh, PA 15213, USA}

\author[0000-0002-6011-0530]{Antonella Palmese}
    \affiliation{McWilliams Center for Cosmology and Astrophysics, Department of Physics, Carnegie Mellon University, Pittsburgh, PA 15213, USA}

\author[0009-0008-8642-5275]{Lewi Westcott}
    \affiliation{\manchester}

\author[0000-0003-1949-7638]{Christopher J. Conselice}
    \affiliation{\manchester}

\author[0000-0001-5228-6598]{Katelyn Breivik}
    \affiliation{McWilliams Center for Cosmology and Astrophysics, Department of Physics, Carnegie Mellon University, Pittsburgh, PA 15213, USA}

\received{2026 April 8}
\revised{2026 August 12}
\accepted{2026 August 17}






\begin{abstract}

Gamma-ray bursts (GRBs) have traditionally been classified by their prompt emission duration and spectral hardness, with short GRBs (sGRB; $\lesssim2 \ \rm{s}$) originating from compact object mergers and long GRBs (LGRB; $> 2 \ \rm{s}$) from massive star core-collapse. Recent kilonova (KN) associations with long-duration GRBs have challenged this standard picture. We analyze the host galaxies of nine GRBs with associated kilonova candidates at $z<0.6$, including five sGRB-KNe and four LGRB-KNe. Using both parametric and non-parametric modeling of the host light distributions, we investigate the progenitor environments of these events and test whether their hosts show evidence for recent galaxy interactions that could favor dynamical formation channels or isolated pathways following merger-driven star formation episodes for neutron star binaries. We find that five of the nine hosts display tidal features that show they have likely undergone recent mergers, suggesting that merger-driven, dynamical formation pathways may contribute in some systems. We find no clear morphological distinction between sGRB-KN and LGRB-KN hosts as both populations span a wide range of morphologies, including ellipticals, spirals, and interacting systems with tidal features. Multi-Sérsic modeling of the host light profiles further shows that host-normalized offsets inferred from single-Sérsic fits can be either under- or overestimated when the transient is associated with a specific subcomponent of a complex host light profile. These results highlight the importance of decomposing host morphology into physically relevant components when interpreting GRB environments and galactocentric offsets.


\end{abstract}




\section{Introduction} 
\label{sec:intro} 

Short gamma-ray bursts (sGRB) are usually thought to be the explosive counterparts of compact object mergers \citep{Blinnikov1984, Paczynski1986, Eichler1989, Narayan1992}, involving either binary neutron star (BNS, \citealt{Ruffert1999, Rosswog2003}) or neutron star-black hole (NS-BH, \citealt{Faber2006, Shibata2011}) binaries. A relativistic jet is produced as a result of the merger and is initially observed as a prompt flash of gamma-rays \citep{Rezzolla2011, Paschalidis2015, Ruiz2016}. The discovery of gravitational waves (GW) from a BNS merger, GW170817, combined with the discovery of its electromagnetic counterpart, sGRB 170817A, provided conclusive evidence of a direct link between BNS mergers and sGRBs \citep{Abbott2017, Goldstein2017,Savchenko2017}. 

Compact object mergers are not solely detectable with GWs; they are also identifiable via the characteristic optical and near-infrared signatures of kilonovae \citep{Li1998, Metzger2010, Barnes2013, Tanaka2013, Grossman2014, Kasen2017}, which are fueled by the radioactive decay of heavy elements produced in the cataclysmic merger. GW170817 was followed by the kilonova AT2017gfo \citep{Andreoni2017, Arcavi2017, Chornock2017, Coulter2017, Covino2017, Cowperthwaite2017, Drout2017, Evans2017, Kasliwal2017, Lipunov2017, Nicholl2017, Pian2017, Shappee2017, Smartt2017, Soares-Santos2017, Tanvir2017, Troja17, Utsumi2017, Valenti2017}, providing robust evidence of a connection between sGRB and kilonova (hereafter referred to as sGRB-KN), though this connection was already strongly suggested \citep[e.g.,][]{Tanvir2013,Berger2013kilonova}. 

sGRBs are traditionally classified based on their prompt gamma-ray durations of $\lesssim 2 \ \rm{s}$ \citep{Kouveliotou1993}. Bursts of longer durations ($> 2$ s) are classified as long gamma-ray bursts (LGRB) and are associated with Type Ic-BL supernovae \citep{Galama1998,Hjorth2003,Stanek2003,Woosley2006}. Thus LGRBs are thought to originate from the death of massive stars due to core-collapse, where a bipolar relativistic jet is formed and emerges through the stellar layers detected as gamma-ray emission \citep{Woosley1993, MacFadyen2001}. These two observational classes have been the prevailing system for categorizing GRBs for the past few decades \citep{Kouveliotou1993}. 

In recent years, however, a new class of GRBs has been identified: merger-driven LGRBs \citep[hereafter, LGRB-KNe; GRBs 211211A and 230307A;][]{Rastinejad2022,Troja2022,Gillanders2023,Levan2023,Yang2024}. While these events have been suggested for two decades \citep{GalYam2006}, these two new events provide incontrovertible evidence of their kilonova emission. These BNS or BH-NS merger-driven transients have long lasting gamma-ray emission ($\sim$ 10s of seconds) as seen in LGRBs, but have a number of attributes that are standard for sGRBs, including their spectral lag, hardness, and minimum variability timescale \citep[e.g.,][]{Troja2022,Yang2024}. What is striking is the association of kilonovae with a handful of these hybrid long-duration gamma-ray transients \citep{Rastinejad2022,Troja2022,Levan2023,Gillanders2023,Yang2024}, linking these LGRB-KNe to compact object mergers (hence the name merger-driven). These LGRB-KNe challenge the long-held assumptions regarding the gamma-ray duration of compact object mergers and raise questions regarding the diversity of GRB progenitors. These events have motivated ongoing studies focused on classifying GRBs based off of their prompt emission characteristics using machine learning (see \citealt{Horvath_classification_2009, Jespersen_unambiguous_2020, Dimple2023, Garcia-cifuentes_identification_2023, Luo_identifying_2023, Steinhardt_classification_2023, Nuessle2024, Li_advancing_2025, Negro_prompt_2025, Zhu_unsupervised_2025}).

The environments of these GRB-KNe play a significant role in our understanding of their progenitors \citep[e.g.,][]{Prochaska2006,DAvanzo2009,Berger2010a,Fong2013,FongBerger2013,Tunnicliffe2014,OConnor2022,Fong2022,Nugent2022,Jeong2024}. Studies have shown that sGRBs are found in a diverse set of host galaxies \citep[e.g.,][]{Prochaska2006,Leibler2010,Fong2013}, with $\sim 84\%$ in star-forming systems, $\sim 6\%$ in transitioning, and $\sim 10 \%$ in quiescent, reflecting a range from younger, star-forming galaxies to older, quiescent environments \citep{Fong2013,Fong2022, Nugent2022}. This diversity demonstrates the range of possible progenitor formation channels that create sGRBs and/or the subsequent broad delay-time distribution \citep{Nakar2006, HaoYuan2013, Wanderman2015, Beniamini2016, BeniaminiPiran2019, Zevin2022, Beniamini2024, Pracchia2026}. In contrast, LGRBs have been observed in young, predominately star-forming galaxies \citep[e.g.,][]{Christensen2004, Fruchter2006, Levesque2014, Lyman2017}, with low stellar mass and metallicity \citep[e.g.,][]{Le_Floc_h_2003, Perley2016, Savaglio2009, Levesque2010}, solidifying the theory that massive stars are their progenitors \citep{Woosley1993, MacFadyen2001}. Such environments reflect the short lifespan of the massive stars that produce collapsars, which in turn generate LGRBs. To corroborate the progenitors of LGRB-KNe, it is equally important to study their host galaxies and environments. By analyzing the diversity of the environments in which LGRB-KNe are found, we can deduce which properties are shared between LGRB-KNe and LGRBs or sGRBs, if any. Previous studies have explored the relationship between the morphology and environments of host galaxies and the progenitors of GRBs (see \citealt{Wainwright2007, Japelj_2018, Schneider2022}).

In this paper, we study the morphology and properties of nine host galaxies of sGRB-KNe and LGRB-KNe. We employ both parametric and non-parametric methods to evaluate the morphology of these hosts and classify them within the Hubble sequence. We are interested in exploring whether the morphology of our host sample shows evidence of a history of galaxy mergers, as seen in GRB 170817A with its concentric shell structure \citep{Palmese2017,Kilpatrick2022}. The turbulent nature of these environments could suggest additional formation channels such as a burst of stellar formation triggered by the dynamical interactions or capture from the infalling host. We explore further into the stellar population properties of the hosts such as stellar mass and star formation rate (SFR). The rate of stellar production of a galaxy can inform us on the delay times of the mergers. Moreover, we examine the offsets of the sGRB-KNe and LGRB-KNe. The spatial relation of the GRB to the nucleus or star-forming regions of the host provide an insight to the formation channel of the progenitors and their delay times. The offsets are also suggestive to the natal kick strength of their progenitors. We use the combine analyses of these features to reveal distinctions and similarities between GRB-KNe and the classical GRB host populations at large. In particular, we aim to elucidate whether LGRB-KNe are distinct from the LGRB host population and originate from compact object mergers similar to sGRBs.

The paper is laid out as follows: in Section \ref{sec:obs}, we discuss our sample selection and give a brief overview of the GRB-KN events. In Section \ref{sec:methods}, we explain the data reduction (\S \ref{subsec:data}) and the morphology classification (\S \ref{subsec:modelmethods}) for the parametric (\S \ref{subsubsec:par}) modeling of the host galaxies and non-parametric (\S \ref{subsubsec:nonpar}) statistics. Section \ref{sec:host_morph} focuses on the host morphology. We report the results of the morphological statistics (\S \ref{subsec:nonparmorfclass}), the limitations (\S \ref{subsec:limits}), and implications of our morphology classification (\S \ref{subsec:morf}). Section \ref{sec:host_properties} contains our analysis of the host and GRB properties. We evaluate the half-light radii (\S \ref{subsec:r50}), the host galaxy properties (\S \ref{subsec:properties}), the galactocentric offsets (\S \ref{subsec:offsets}), the formation channels (\S \ref{subsec:interpoff}), the high-energy properties (\S \ref{subsec:hep}), and the kilonova properties (\S \ref{subsec:KN}). Lastly, our conclusion and comparison across the GRB populations are given in Section \ref{sec:conclusions}.

\section{Sample Selection} 
\label{sec:obs} 

There are a number of kilonova candidates presented in the literature with varying levels of certainty based on the security of their distance scale (from host association) or the quality of the dataset (i.e., multi-epoch multi-color information), see \citet{Troja2023} for a recent review. A secure distance scale and multi-epoch color information are required to distinguish between other possible explanations than a kilonova, including intrinsic dust in the host galaxy or a high redshift. To perform these checks and confirm kilonova associations with GRBs, deep, space-based imaging is required. \citet{Rastinejad2025} provides a classification based on the number of required data points (2 or more optical or near-infrared detections at $>$\,$2$ d). For our sample, we follow the dataset of \cite{Rastinejad2025} and supplement with the kilonova candidate GRB 150101B given the classification from \cite{Troja2023}. The remaining GRBs in \cite{Troja2023} are not kilonova candidates and therefore overlooked. Each event in our sample has either \textit{HST} or \textit{JWST} imaging in order to be classified as a kilonova candidate. For our morphological analysis, we require the resolution of deep, space-based imaging to evaluate the structures of the galaxies. In the following subsections, we provide a brief overview of the events in our sample including a summary of the past studies of the host galaxies, the available space-based data from \textit{HST} and \textit{JWST}, and a summary of the kilonova properties.

We include in our sample: GRB 050709 \citep{Jin2016}, GRB 060614 \citep{Jin2015,Yang2015}, GRB 130603B \citep{Tanvir2013,Berger2013kilonova}, GRB 150101B \citep{Troja2018b}, GRB 160821B \citep{Kasliwal2017a,Lamb2019,Troja2019b}, GRB 200522A \citep{OConnor2021kn,Fong2020}, and the recent long duration GRBs 211211A \citep{Troja2022,Rastinejad2022} and 230307A \citep{Levan2023,Gillanders2023,Yang2024}. We also include the neutron star merger event GW170817 \citep{Coulter2017,Drout2017,Arcavi2017,Evans2017,Soares-Santos2017,Troja2017,Pian2017}. Our sample contains two spectroscopically confirmed, high-confidence kilonova events (GRBs 170817A and 230307A), with the remaining kilonova photometrically confirmed. 

Despite other claimed kilonova candidates in the literature, e.g., GRBs  070809 \citep{Jin2020} and 080503 \citep{Perley2009,Gao2015}, we exclude these events due to their uncertain distance scale and the multiple (equally likely) possible host associations \citep[see, e.g.,][]{Berger2010a,FongBerger2013}. For a further discussion of the varying levels of quality among kilonova candidates, see \citet{Troja2023,Rastinejad2025}. 

We break our sample into subpopulations based on their prompt gamma-ray emission properties: sGRB-KN ($\lesssim 2$ s) and LGRB-KN ($> 2$ s). Five events are classified as sGRB-KN (GRBs 130603B, 150101B, 160821B, 170817A, and 200522A), and four events are classified as LGRB-KN (GRBs 050709, 060614, 211211A, and 230307A). The existence of peculiar LGRBs lacking supernovae extends back to the early days of the \textit{Swift} mission \citep[e.g., GRB 060614;][]{Gehrels2006,GalYam2006}. These sources are now understood to likely be merger-driven events, similar to GRBs 211211A and 230307A. In particular, the class of sGRBs with a long duration tail of emission, or extended emission, following the initial short burst (hereafter sGRBEE; \citealt{Norris2006,Norris2010}) have long been treated as belonging to the class of merger-driven events. As such, we likewise consider sGRBEEs 050709 and 060614 \citep{Gehrels_2006,Norris2010,Berger2014,Gompertz2020,Gompertz2023} to belong to the class of LGRB-KNe for the purpose of the subsample comparisons presented in this work.

\subsection{GRB 050709}

GRB 050709 triggered the \textit{High Energy Transient Explorer II} (\textit{HETE-II}) at 22:36:37 UT on 2005 July 9 with a $T_{90} = 0.07 \ \rm{s}$ \citep{Hjorth2005, Villasenor2005}. The transient was localized at RA, Dec (J2000) $= 23^h01^m26.96^s, \ang{-38;58;39.5}$, a distance of $1.35\pm0.02\arcsec$ ($3.76\pm.056 \ \rm{kpc}$) from the host galaxy \citep{Fox2005,Hjorth2012}. The host galaxy has been identified as a blue dwarf (log($M_*/M_\odot$) $= 8.55_{-0.01}^{+0.01}$) at $z = 0.161$ with a SFR of $0.024_{-0.001}^{+0.001} \ M_{\odot} \rm{yr}^{-1}$ \citep{Leibler2010, Nugent2022}. The \textit{HST} Advanced Camera for Surveys (ACS) observed the transient location July and August 2005 in the $F814W$ band (see Table \ref{tab:obslog} in Appendix \ref{app:obs}). 

GRB 050709 exhibited an initially short-hard burst followed by a long-soft bump $T_{90} = 130\pm7 \ \rm{s}$ \citep{Villasenor2005}. \cite{Villasenor2005} argued that the delayed gamma-rays were the onset of the afterglow while \cite{Norris2006} interpreted this as extended emission. The following literature tends to consider GRB 050709 one of the first sGRBEEs \citep[e.g.,][]{Troja2008,Berger2014,Matsumoto2020,Gompertz2020,Gompertz2023}. We adopt the sGRBEE classification in this work. Initially GRB 050709 did not have an obviously identified kilonova. However, a later re-analysis identified a candidate kilonova due to its red excess at $\sim 7-10 \ \rm{days}$ \citep{Jin2016}. The associated kilonova had a peak luminosity of $\sim 10^{41} \ \rm{erg \ s^{-1}}$ in the $F814W$ (approximately $I$) band at $t > 2.5 \ \rm{days}$ with an estimated ejecta mass of $M_{\rm{ej}} \sim 0.05 \ M_\odot$ \citep{Jin2016}.

\subsection{GRB 060614}

GRB 060614 triggered the \textit{Swift} Burst Alert Telescope (BAT) at 12:43:48.5 UT on 2006 June 14 with a $T_{90} = 102\pm3 \ \rm{s}$ \citep{Barthelmy2006,Gehrels2006,GalYam2006}. The transient was localized at RA, Dec (J2000) $=21^h23^m31.8^s, \ang{-53;02;04.4}$ \citep{Mangano2007}, a distance of $0.31\pm0.35\arcsec$ ($0.70 \pm 0.79 \ \rm{kpc}$) from the host galaxy \citep{Gehrels_2006, GalYam2006, Fong2022}. The host is a faint dwarf galaxy (log($M_*/M_\odot$) $= 7.85_{-0.04}^{+0.03}$) at $z=0.125$ with a SFR of $ 0.005_{-0.001}^{+0.001} \ M_{\odot} \rm{yr}^{-1}$ \citep{GalYam2006, Nugent2022}. \textit{HST}/ACS observed the transient location September through November 2006 in the $F606W$ and $F814W$ bands, and the \textit{HST} Wide Field Camera 3 (WFC3) observed October 2010 in the $F160W$ band (see Table \ref{tab:obslog} in Appendix \ref{app:obs}).

GRB 060614, despite a long duration of $102 \ \rm{s}$ \citep{Gehrels_2006}, was not accompanied by a supernova \citep{Fynbo2006,GalYam2006,DellaValle2006}, a counterpart expected of collapsars. Additionally, with its spectral hardness and lag, GRB 060614 was immediately treated as separate from the class of LGRBs and more characteristic of sGRBs \citep{Gehrels2006}. Thus, GRB 060614 is classified as a sGRBEE. Follow-up revealed reddening of the optical light curve at $\sim 13 \ \rm{days}$, indicating a possible kilonova that peaked at $t \lesssim 4 \ \rm{days}$ after the GRB \citep{Jin2015, Yang2015}. The kilonova had an inferred ejecta mass of $M_{\rm{ej}} \sim 0.1 \ M_\odot$ \citep{Gehrels_2006, Jin2015, Yang2015}.

\subsection{GRB 130603B}

GRB 130603B triggered \textit{Swift}/BAT at 15:49:14 UT on 2013 June 3 with a $T_{90} = 0.18\pm0.02 \ \rm{s}$ \citep{Tanaka2013, Berger2013kilonova}. The transient was localized at RA, Dec (J2000) $=11^h28^m48.16^s, \ang{+17;04;18.2}$ \citep{Tanaka2013}, a distance of $1.07\pm0.04\arcsec$ ($5.40\pm0.20 \ \rm{kpc}$) from the host galaxy \citep{Cucchiara_2013}. The host, at $z=0.356$ \citep{deUgartePostigo2014}, has a stellar mass of log($M_* / M_\odot$) $= 9.82_{-0.04}^{+0.05}$ and a SFR of $0.44_{-0.09}^{+0.22} \ M_{\odot} \rm{yr}^{-1}$  \citep{Cucchiara_2013}. \textit{HST}/WFC3 observed the transient location in June and July 2013 in the $F606W$ and $F160W$ bands (see Table \ref{tab:obslog} in Appendix \ref{app:obs}).

The optical afterglow of GRB 130603B reddened at later times and, along with an excess of near-IR emission seen $t \sim 7 \ \rm{days}$ after the GRB, suggested a kilonova brighter than AT2017gfo by a factor of $\sim 2$ \citep{Berger2013kilonova, Tanvir2013}. The kilonova suggests an ejecta mass of $M_{\rm{ej}} \sim 0.03-0.08 \ M_\odot$ \citep{Berger2013kilonova, Tanvir2013, Hotokezaka2013kilonova, Jin2016}

\subsection{GRB 150101B}

GRB 150101B triggered \textit{Swift}/BAT at 15:23 UT on 2015 January 1 with a $T_{90} = 0.012\pm0.009 \ \rm{s}$  \citep{Fong2016kilonova}. The transient was localized at RA, Dec (J2000) $=12^h32^m10.4^s, \ang{-10;58;48}$ \citep{Lien2016}, a distance of $3.07\pm0.03\arcsec$ ($7.35\pm0.072 \ \rm{kpc}$) southeast from the host galaxy \citep{Fong2016kilonova, Troja150101B}. The host is categorized as an early-type galaxy (log($M_*/M_\odot$) $= 11.13_{-0.02}^{+0.02}$) at $z=0.1341$  with a SFR of $0.22_{-0.02}^{+0.02} \ M_{\odot} \rm{yr}^{-1}$ \citep{Fong2016kilonova}. \textit{HST}/WFC3 observed the transient location December 2015 in the $F606W$ and $F160W$ bands (see Table \ref{tab:obslog} in Appendix \ref{app:obs}).

Multi-band follow-up observations of GRB 150101B identified an optical excess, brighter by a factor of $\sim 2$ than AT2017gfo, the byproduct of a kilonova \citep{Troja150101B, Troja2023}. The inferred kilonova luminosity suggests an ejecta mass of $M_{\rm{ej}} \gtrsim 0.02\ M_\odot$ \citep{Troja2023}.

\subsection{GRB 160821B}

GRB 160821B triggered \textit{Swift}/BAT at 22:29:13 UT on 2016 August 21 with a $T_{90} = 0.48\pm0.07 \ \rm{s}$ \citep{Lien2016, Troja2019b}. The transient was localized at RA, Dec (J2000) $=18^h39^m53.994^s, \ang{+62;23;34.427}$, a distance of $5.61\pm0.01\arcsec$ ($15.74\pm0.03 \ \rm{kpc}$) from the host galaxy \citep{Troja2019b}. The host galaxy is classified as a star-forming spiral (log($M_*/M_\odot$) $= 9.245_{-0.004}^{+0.003}$) at $z=0.1613$ with a SFR of $0.24_{-0.01}^{+0.01} \ M_{\odot} \rm{yr}^{-1}$ \citep{Nugent2022}. \textit{HST}/WFC3 observed the transient location August through December 2016 and August 2018 in the $F110W$, $F160W$, $F606W$ bands (see Table \ref{tab:obslog} in Appendix \ref{app:obs}).

Studies of the optical and near-IR afterglow of GRB 160821B identified an associated kilonova \citep{Kasliwal2017a,Lamb2019,Troja2019b} with a predicted ejecta mass of $M_{\rm{ej}} < 0.03 \ M_\odot$ \citep{Troja2019b}. The kilonova is fainter by a factor of $\sim 2$ than AT2017gfo, but matches AT2017gfo with respect to color and timescales \citep{Troja2019b, Troja2023}.

\subsection{GRB 170817A/GW170817}

GW170817 triggered LIGO at 12:41:04 UT on 2017 August 17 \citep{Abbott2017} and observations with DECam, and other telescopes, began at 23:13 UT \citep{Soares-Santos2017}. GRB 170817A was promptly detected by the \textit{Fermi}-GBM and INTErnational Gamma-ray Astrophysics Laboratory (\textit{INTEGRAL}) on 2017 August 17 at 12:41:06 UT \citep{Abbott2017, Goldstein2017, Savchenko2017} with a $T_{90} = 2.0\pm0.5 \ \rm{s}$ \citep{Margutti2021}. The transient was localized at RA, Dec (J2000) $= 13^h09^m48.09^s, \ang{-23;22;53.38}$, a distance of $10.317\pm0.005\arcsec$ ($2.125\pm0.0001 \ \rm{kpc}$) from the host galaxy \citep{Soares-Santos2017}. The host is an elliptical galaxy (log($M_*/M_\odot$) $= 10.61_{-0.02}^{+0.01}$) at $z=0.0098$ with a SFR of $0.019_{-0.005}^{+0.004} \ M_{\odot} \rm{yr}^{-1}$ \citep{Palmese2017,Kilpatrick2022,Nugent2022}. \textit{HST}/ACS and \textit{HST}/WFC3 observed the transient location from 2017 to 2021 in the $F110W$, $F160W$, $F606W$ bands (see Table \ref{tab:obslog} in Appendix \ref{app:obs}).

The associated kilonova AT2017gfo has been extensively studied for this transient \citep{Andreoni2017, Arcavi2017, Chornock2017, Coulter2017, Covino2017, Cowperthwaite2017, Drout2017, Evans2017, Kasliwal2017, Lipunov2017, Nicholl2017, Pian2017, Shappee2017, Smartt2017, Soares-Santos2017, Tanvir2017, Troja17, Utsumi2017, Valenti2017}, with a blue component of $M_{\rm{ej}}^{\rm{blue}} \sim 0.01 \ M_\odot$ and $v_{\rm{ej}}^{\rm{blue}} \sim 0.3c$ and red component of $M_{\rm{ej}}^{\rm{blue}} \sim 0.04 \ M_\odot$ and $v_{\rm{ej}}^{\rm{blue}} \sim 0.1c$ \citep{Cowperthwaite2017}. It remains to be the gold standard of kilonovae in terms of its solid association to a binary neutron star merger, early identification and long-term multi-band follow-up, and spectroscopic sequence of optical and near-infrared spectra. 

\subsection{GRB 200522A}

GRB 200522A triggered \textit{Swift}/Bat at 11:41:34 UT on 2020 May 22 with a $T_{90} = 0.62\pm0.08 \ \rm{s}$  \citep{OConnor2021kn, Fong2020}. The transient was localized at RA, Dec (J2000) $= 00^h22^m40.3^s, \ang{-00;15;49.9}$, a distance of $0.143\pm0.029\arcsec$ ($0.93\pm0.19 \ \rm{kpc}$) from the host galaxy \citep{Fong2020}. The host is a young, star-forming galaxy (log($M_*/M_\odot$) $= 9.66_{-0.01}^{+0.01}$) at $z=0.554$ with a SFR of $2.23_{-0.05}^{+0.06} \ M_{\odot} \rm{yr}^{-1}$ \citep{OConnor2021kn, Fong2020}. \textit{HST}/WFC3 observed the transient location May through July 2020 in the $F125W$ and $F160W$ bands (see Table \ref{tab:obslog} in Appendix \ref{app:obs}).

The luminosity and reddened counterpart of GRB 200522A, possibly attributed to dust extinction \citep{OConnor2021kn} or perhaps a magnetar-boosted kilonova \citep{Fong2020}, were plausibly related to a kilonova with peak luminosity of $\sim (1.3-1.7) \ \times \ 10^{42} \ \rm{erg \ s^{-1}}$ \citep{Fong2020}. Simulations suggest an ejecta mass of $M_{\rm{ej}} \lesssim 0.1 \ M_\odot$ \citep{OConnor2021kn, Bruni2021}.

\subsection{GRB 211211A}

GRB 211211A triggered \textit{Swift}/BAT at 13:09 UT on 2021 December 11 with a $T_{90} =51.37\pm0.80 \ \rm{s}$ \citep{Rastinejad2022,Troja2022}. The transient was localized at RA, Dec (J2000) $= 14^h09^m10.467^s, \ang{+27;53;21.050}$, a distance of $5.44 \pm 0.02\arcsec$ ($7.91\pm0.03 \ \rm{kpc}$) from the host galaxy \citep{Rastinejad2022,Troja2022}. The host is a dwarf galaxy (log($M_*/M_\odot$) $= 8.84_{-0.05}^{+0.10}$) at $z=0.0762$ with a SFR of $0.07_{-0.01}^{+0.01} \ M_{\odot} \rm{yr}^{-1}$ \citep{Rastinejad2022,Troja2022}. \textit{HST}/ACS and \textit{HST}/WFC3 observed the transient location April 2022 in the $F814W$, $F606W$, $F160W$, and $F140W$ bands (see Table \ref{tab:obslog} in Appendix \ref{app:obs}).

GRB 211211A immediately caught the attention of the GRB community due to its extreme brightness, which was, at the time, the second brightest \textit{Swift}-detected GRB behind GRB 130427A. The afterglow of GRB 211211A was rapidly localized to a nearby galaxy ($z=0.0762$), which led to extensive follow-up by the community. Similar to GRB 060614, the GRB had a significantly longer duration of $\sim 51 \ \rm{s}$ with the majority of its emission dominated by later pulses, and it was not immediately clear that the event was not a collapsar.  Late-time near-infrared emission in $K$ band showed an extreme red color and was the key signpost of its kilonova counterpart \citep{Rastinejad2022}. Additionally, no supernova emission was uncovered and deep late-time \textit{HST} imaging acquired by two different teams ruled out a faint background host galaxy \citep{Troja2022, Rastinejad2022}. The lack of background host galaxy was key to solidifying its distance scale (i.e., ruling out higher redshifts) and solidifying the interpretation of this red excess as a kilonova. While long-duration GRBs had previously failed to display supernova emission \citep[e.g., GRB 060614;][]{Fynbo2006,GalYam2006,DellaValle2006}, the evidence that GRB 211211A provided for an extreme red excess that could only be associated with kilonova emission was significantly stronger than the limited datasets that had previously existed. As such, GRB 211211A marked a defining moment for the field.

In the end, GRB 211211A serves as a robust kilonova with high quality multi-band imaging allowing the identification of kilonova excess in data acquired as early as 5 hours after discovery \citep{Troja2022}. The kilonova had an estimated ejecta mass of $M_{\rm{ej}} \sim 0.047 \ M_\odot$ \citep{Troja2022, Rastinejad2022}.

\subsection{GRB 230307A}

GRB 230307A triggered the Fermi Gamma-ray Burst Monitor (GBM) at 15:44:06.67 UT on 2023 March 7 with a $T_{90} = 35 \ \rm{s}$ \citep{Levan2023}. The transient was localized at RA, Dec (J2000) $=04^h03^m26.02^s, \ang{-75;22;42.76}$, a distance of $30.20\pm0.01\arcsec$ ($38.90\pm0.01 \ \rm{kpc}$) from the host galaxy \citep{Yang2024}. The host is classified as a low-mass spiral galaxy (log($M_*/M_\odot$) $= 9.38\pm0.16$) at $z=0.065$ dominated by an old stellar population with a SFR of $0.20\pm0.03 \ M_{\odot} \rm{yr}^{-1}$ \citep{Levan2023, Yang2024}. \textit{HST}/WFC3 observed the transient location April and May 2023 in the $F105W$ and $F140W$ bands. The \textit{JWST} Near Infrared Camera (NIRcam) observed April and May 2023 in the $F070W$, $F115W$, $F150W$, $F277W$, $F356W$, and $F444W$ bands (see Table \ref{tab:obslog} in Appendix \ref{app:obs}).

Similar to GRB 211211A, the extreme brightness of GRB 230307A, which placed it in the top few \textit{Fermi}-detected bursts, immediately caught the attention of the community. The afterglow localization initially posed some uncertainty regarding its classification, as while the extreme brightness implied a nearby redshift (and was further constrained by Gemini spectroscopy; \citealt{Gillanders2023,Yang2024}) the only obvious low redshift galaxy existed at a very large $30\arcsec$ offset \citep{Levan2023,Yang2024}. Since a LGRB-KN was known following GRB 211211A at that point, this hypothesis immediately manifested. \textit{JWST} imaging and spectroscopy later confirmed the kilonova association and that the low redshift galaxy was indeed the correct host \citep{Levan2023,Gillanders2023,Yang2024}. GRB 230307A is the second spectroscopically confirmed kilonova (following AT2017gfo) and the first observed and detected by \textit{JWST} \citep{Levan2023}, with an estimated ejecta mass of $M_{\rm{ej, tot}} \sim 0.08 \ M_\odot$ and ejecta velocity of $0.2-0.3c$ \citep{Yang2024}. Together with GRBs 060614 and 211211A, it solidified even further the class of LGRB-KNe. 

\begin{figure*}
    \centering
    \includegraphics[width=\textwidth]{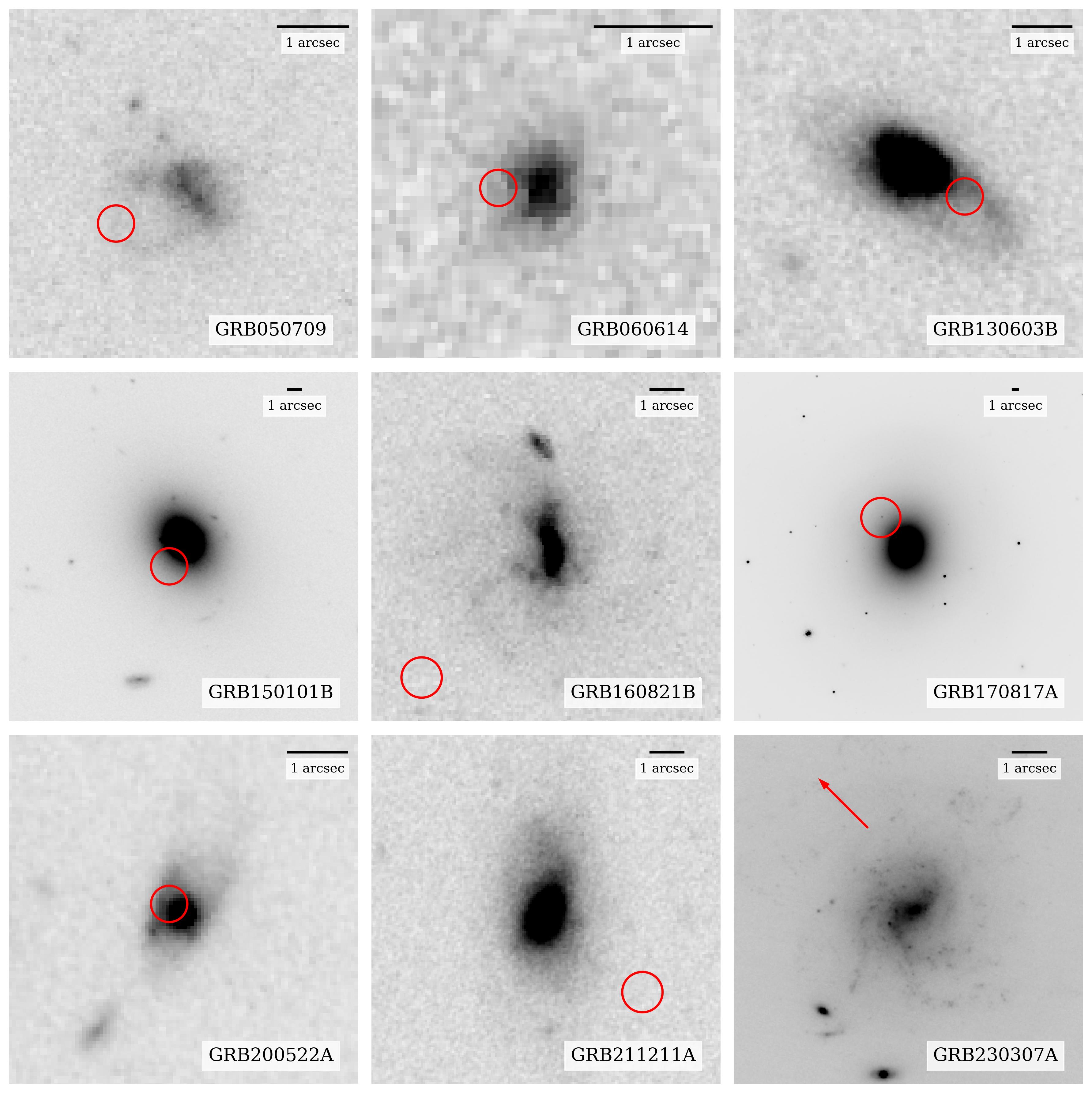}
    \caption{Mosaic of \textit{HST}/$F160W$ image cutouts (\textit{HST}/$F814W$ for GRB 050709; \textit{JWST}/$F150W$ for GRB 230307A) of each host galaxy in our sample. The red circles indicate the approximate location of the transients, with the exception of GRB 230307A being at $\sim30\arcsec$ offset outside the scope of the cutout. The red circles are the same pixel size relative to the pixel scale of each image. The localization uncertainties are not used for the circle radii as they are too small and would obscure the local environment.}
    \label{fig:grid}
\end{figure*}

\section{Methods} 
\label{sec:methods}

\subsection{Data Analysis}
\label{subsec:data}

We focus on a sample of these nine kilonova candidates (\S \ref{sec:obs}) with deep, space-based optical and near-infrared (NIR) imaging from \textit{HST} and \textit{JWST}. All events in our sample have publicly available archival \textit{HST} imaging, whereas only GRB 230307A has archival \textit{JWST} imaging \citep{Levan2023,Yang2024}. In Table \ref{tab:obslog} in Appendix \ref{app:obs}, we display the log of all observations used in our analysis. For transients with a small offset from their host galaxy, we focus on late-time imaging epochs where the transient has faded so that it does not impact the results of our morphological analysis. A finding chart for each of these events is displayed in Figure \ref{fig:grid}. 

For all events, we obtained publicly available \textit{HST}/ACS and \textit{HST}/WFC3 imaging from the Mikulski Archive for Space Telescopes (MAST)\footnote{\url{https://archive.stsci.edu/index.html}}. We process and stack the calibrated \textsc{\_flt.fits} (or \textsc{\_flc.fits}) files following standard procedures using the \texttt{DrizzlePac} software package\footnote{\url{https://www.stsci.edu/scientific-community/software/drizzlepac.html}} \citep{Gonzaga2012}, which includes \texttt{AstroDrizzle} and \texttt{TweakReg}. The individual exposures within each epoch are first aligned to a common world-coordinate system with \texttt{TweakReg}. We next employ \texttt{AstroDrizzle} to produce the final drizzled image by combining all \textsc{\_flt.fits} (or \textsc{\_flc.fits}) from that epoch in the same filter. The final pixel scale is $0.03\arcsec$/pix for optical imaging with WFC3 and $0.06\arcsec$/pix for NIR imaging, both using \texttt{pixfrac}\,$=$\,$0.8$. For optical imaging with ACS, the final pixel scale is $0.05\arcsec$/pix using \texttt{pixfrac}\,$=$\,$1.0$. For GW170817, we combine all exposures across all epochs (see Table \ref{tab:obslog} in Appendix \ref{app:obs}) in the $F606W$ and $F160W$ filters following \citet{Kilpatrick2022}, whereas for all other events we combine only single epoch of late-time template images (due to the higher potential for transient contamination of the host galaxy's morphology when including earlier epochs). In the case of GW170817, the transient emission is extremely sub-dominant ($\sim$\,$26-27$ AB mag; \citealt{Fong2019,Kilpatrick2022}) compared to the brightness of the host galaxy ($\sim$\,$11-12$ AB mag; \citealt{Blanchard2017,Levan2017}; see Figure \ref{fig:grid}) and does not impact our inferences.

For GRB 230307A, we obtain the publicly available pipeline processed JWST Level-3 mosaics from the MAST Archive\footnote{\url{https://mast.stsci.edu/search/ui/\#/jwst}}. NIRCam imaging is available for GRB 230307A \citep{Levan2023,Yang2024} in the $F070W$, $F115W$, $F150W$, $F277W$, $F356W$ and $F444W$ filters. The short wavelength channel ($F070W$, $F115W$, and $F150W$) has a pixel size of $0.031\arcsec$/pix, and the long wavelength channel ($F277W$, $F356W$ and $F444W$) has pixel size $0.063\arcsec$/pix. As the transient is located at a large offset ($\sim$\,$30\arcsec$) from the host galaxy, it does not impact the results of our morphological analysis of the host. 

Across all available filters, seven hosts of our sample have \textit{HST}/WFC3 imaging in the NIR band $F160W$, with the exception of GRBs 050709 and 230307A. We use the only available filter for GRB 050709 in the optical band $F814W$. For GRB 230307A, there is \textit{JWST}/NIRCam imaging in band $F150W$, which is the most comparable filter with $F160W$. The NIR filters correspond to a similar rest-frame wavelengths, with $F160W$ ranging $\sim989-1522 \ \rm{nm}$ and $F150W$ for GRB 230307A at $\sim1409 \ \rm{nm}$. The NIR band $F160W$ is solely chosen on the basis that the majority of our sample has data in this filter, and we aim to be consistent in our analysis of the host sample.

The point spread functions (PSF) of the images are generated with \texttt{spike} \citep{Polzin2025}. We use \texttt{spike} to create a unique PSF at the source position of each event. The PSFs are co-added and processed with the same STDPSF pipeline used for \textit{HST} and \textit{JWST} images \citep{Anderson2016, Libralato2023, Libralato2024}. \texttt{spike} requires the original calibrated files (\textsc{\_flt.fits} or \textsc{\_flc.fits}) of the images downloaded from MAST, the source's astronomical coordinates (e.g., to determine placement on the detector), and drizzle parameters. The drizzle parameters for the \textit{HST} PSFs are matched to our custom parameters used for processing the images from WFC3 and ACS (see above). We use the default drizzle parameters set in the pipeline for the \textit{JWST}/NIRCam images.

\subsection{Galaxy Morphology Modeling}
\label{subsec:modelmethods}

We apply both parametric (\S \ref{subsubsec:par}) and non-parametric models (\S \ref{subsubsec:nonpar})  to classify the morphology of the host galaxies. The parametric method models the light profile of the galaxy using predefined analytic functions (e.g., a Sérsic profile, \citealt{Sersic1963, Sersic1968}). These models are characterized by a set of parameters that define physical quantities of the galaxy. Parametric models rely heavily on the light profile chosen to reconstruct the galaxy, so a non-parametric method is also employed to counter any assumptions made. Non-parametric models calculate the morphological parameters of the galaxy through statistical measurements based on the galaxy light distribution; there are no predefined assumptions of the galaxy structure. The parametric model is additionally used to visualize the components of the galaxy and mark by eye any noticeable residual features, such as dust lanes or shells that could suggest a history of recent galaxy mergers. In turn, the non-parametric method provides a statistical classification of the hosts into the categories of early (elliptical), intermediate/late (spiral), and merging galaxies.

For our non-parametric modeling, we use the NIR filters as discussed in \S \ref{subsec:data}. As structural components of the galaxies can become more visible in different wavelengths, selecting a band common to all events mitigates this effect. However, we conduct parametric modeling across all available filters in our sample (see Table \ref{tab:obslog} in Appendix \ref{app:obs}). 

\subsubsection{Parametric Methods} 
\label{subsubsec:par}

For our parametric modeling, we use the software \texttt{GALFIT} \citep{Peng2002}. \texttt{GALFIT} is a fitting algorithm that breaks down the galaxy into its structural components using two-dimensional profiles. The result is a list of fitted parameters, a model image of the galaxy components, and a residual image.

\texttt{GALFIT} takes in a number of inputs such as the data image, PSF, and a bad pixel segmentation map, along with initial fitting parameters. In order for \texttt{GALFIT} to create a sigma image of the data, an array of the standard deviation of the flux of each pixel, we convert the native pixel units to the raw photon counts using the exposure time and gain for \textit{HST} images. The sigma image is a necessity in minimizing the $\chi^2$ value (see \texttt{GALFIT} user guide\footnote{\url{https://users.obs.carnegiescience.edu/peng/work/galfit/README.pdf}}). The PSF is similarly converted to keep units consistent. 
The data images passed in are cropped to a given box size that encapsulates the entirety of the galaxy light (Figure \ref{fig:grid}). The PSF files are of the same cutout size. \texttt{Photutils} \citep{Bradley_2024} is used in the creation of image segmentation, creating a mask to highlight any extraneous sources to be ignored by \texttt{GALFIT}.

The components for each galaxy are cumulatively stacked upon one another. Initial parameters for a Sérsic component consist of the $x$ and $y$ center of the source in detector pixels, the integrated magnitude, half-light radius ($\rm{r}_{50}$), Sérsic index, axis ratio, and the position angle. A sky component is included in every \texttt{GALFIT} model to account for the background. We begin with a single Sérsic profile to fit the overall light of the host. If the structure proves to be more complex, a secondary Sérsic component is added. This process can continue until the reduced-$\chi^2$ is minimized sufficiently close to 1.0 and the residual image looks uniform. Once the general light profile is captured by \texttt{GALFIT}, the residual is evaluated for structural components not visible in the data image (see Figure \ref{fig:galfit} for residuals in Appendix \ref{app:galfit}). Dust lanes, spiral arms, and shells are examples of the structures that are seen in our sample. 

\texttt{GALFIT} has the ability to apply azimuthal profile functions such as bending and fourier modes to the light profiles to fit more complex structures. In our sample, these modes were used for galaxies with visible spiral arms, such as in GRB 160821B and GRB 230307A (see Figure \ref{fig:grid}). However, for the extent of this morphological study, a simpler model composed of Sérsic profiles is sufficient to obtain the $\rm{r}_{50}$ of the galaxies and to detect any underlying features (see Table \ref{tab:stats}; Figure \ref{fig:galfit} in Appendix \ref{app:galfit}).

To compare with the $\rm{r}_{50}$ results from previous studies of GRB host galaxies \citep[e.g.,][]{Fong2022}, we ran additional \texttt{GALFIT} models with only a single Sérsic profile. The single Sérsic is able to capture the extent of the host light, but concentrates on the central, brightest region of the light profile. While a multi-component \texttt{GALFIT} model is useful to reveal the more complex morphology of the hosts, it can impact the $\rm{r}_{50}$ results. A multi-component model of our sample typically comprises of a Sérsic profile to capture the central light of the densely populated core and a secondary Sérsic to model the diffuse light of a disk, if present. This provides an $\rm{r}_{50}$ for each profile. The choice of which radii to use impacts the measurements of the host-normalized offsets and our interpretation of the offset distributions for the various GRBs (see \S \ref{subsec:offsets} for further discussion).

\subsubsection{Non-parametric Methods}
\label{subsubsec:nonpar}

The non-parametric morphological statistics are calculated using two software codes: \texttt{statmorph} \citep{Rodriguez_Gomez_2018} and \texttt{Morfometryka} \citep{Ferrari2015}. The two statistics of interest in this study are Gini-$\rm{M}_{20}$ and CAS, calculated by \texttt{statmorph} and \texttt{Morfometryka} respectively (see Table \ref{tab:stats}). The Gini-$\rm{M}_{20}$ statistics evaluate both the shape and distribution of light \citep{Lotz2004}. The Gini coefficient measures the distribution of the galaxy light, where highly concentrated light in uneven distributions have high Gini values. $\rm{M}_{20}$ is a measurement of the concentration of the brightest 20\% of the galaxy light. Smooth galaxies produce lower $\rm{M}_{20}$ values, while clumpier, more disturbed galaxies have higher $\rm{M}_{20}$ values. The Gini-$\rm{M}_{20}$ statistic classifies galaxies into three major groups: elliptical, spiral, and major or minor mergers.

Similarly, the CAS statistics are a method for classifying galaxies \citep{Kent1985, Conselice2000, Conselice2003}. The concentration (C) parameter measures the concentration of light in the galaxy by taking the ratio of the $\rm{r}_{20}$ and $\rm{r}_{80}$ radii. Higher concentration values indicate light centered at the core, while lower concentration values indicate light spread out through the galaxy. The asymmetry (A) parameter uses the residual image between the original galaxy and the galaxy rotated by $180^\circ$ to evaluate the symmetry. Asymmetry increases due to mergers and local dynamics \citep{De_Propris_2007,Conselice_2009}. The clumpiness (S) parameter measures the patchiness of the galaxy light by comparing the original galaxy image against the same image with image convolution applied. A galaxy with a high clumpiness value indicates the presence of multiple star-forming regions. The CAS statistics group galaxies into early-type (elliptical), late-type (spiral), or major mergers.

An important choice to make in using these statistics is the fraction of Petrosian radius, $\rm{R_p}$ to  use when measuring the parameters. The Petrosian radius is the radius where the ratio of local surface brightness to mean surface brightness equals a fixed value. This creates an objective measure of the galaxy size insensitive to distance, diminishing the impact of observational limitations. This allows for comparison of galaxies regardless of redshift or brightness. Depending on the factor of $\rm{R_p}$, this can impact how the non-parametric statistics are measured. For both \texttt{statmorph} and \texttt{Morfometryka}, the background is calculated within some factor of $\rm{R_p}$ of the galaxy center. This background is then accounted for in the measurements of these statistics. Thus, we chose a factor of $1.5\rm{R_p}$, matching the fixed $\rm{R_p}$ in \texttt{statmorph}.

For \texttt{statmorph}, we pass in the data image, a segmentation map, PSF, and the factor of $1.5\rm{R_p}$. To be consistent with the data used for \texttt{GALFIT}, the data image and PSF are converted to units of counts and cropped to a given cutout size that capture the entirety of the host galaxy. The segmentation map is created with \texttt{Photutils}, highlighting the galaxy of interest. Since \texttt{statmorph} does not calculate the errors for the Gini-$\rm{M}_{20}$ statistic, we follow the method outlined in \cite{Lyman2017}. For each host, we reduce the \textsc{\_flt.fits} or \textsc{\_flc.fits} files using \texttt{lacosmic} to identify and remove cosmic rays \citep{vanDokkum2001}, \texttt{TweakReg} to align the multiple exposures, and drizzle with \texttt{AstroDrizzle} using our custom parameters described in \S\ref{subsec:data}. The stacked image are resampled 200 times using the ERR header and randomly selecting a value in the variance of each pixel. Each of the resampled images are passed through \texttt{statmorph}, and individual statistics – $\rm{r}_{50}$, Gini, $\rm{M}_{20}$ – are saved. The final values of the statistics we report are of the mean and $1\sigma$ of the 200 resampled images (see Table \ref{tab:stats}). 

Tied to the selection of $\rm{R_p}$, since the A and S parameters are sensitive to how the sky background is derived, it is just as important to choose a software that handles the background properly. We opt for \texttt{Morfometryka} to calculate the CAS statistics as the background is sampled from multiple random sections across the image \citep{Sazonova2024}. \texttt{Morfometryka} takes in the data image and the PSF. The software segments the images, removing any outliers, and uses the properties of the segment to define an ellipse with a major axis of $1.5\rm{R_p}$ that matches the shape of the galaxy. \texttt{Morfometryka} also does not provide errors on the given statistics, so to calculate these we take a Monte Carlo approach in order to quantify our uncertainties. For each source, and iteration, we inject random Gaussian noise, which is scaled to the measured sky background, $\sigma_{\rm sky}$. From each new image the full analysis pipeline is executed again, including segmentation as well as determining $\rm{R_p}$ and the centroid of the source, giving us a new measure of the parameters of interest. This is then resampled 1000 times. We then take our $1\sigma$ errors from the distribution of our measured parameters (see Table \ref{tab:stats}).

\section{Host Morphology} 
\label{sec:host_morph}

\begin{table*} \vspace{25mm}
    \begin{rotatetable}
        \centering
        \caption{Compilation of the CAS and Gini-$M_{20}$ statistics of the GRB-KN host galaxy sample from \texttt{statmorph} and \texttt{Morfometryka}, and their half-light radii and Sérsic indices from \texttt{GALFIT}.}
        \label{tab:stats}
        \begin{tabular}{lllllllll}
            \hline
            \hline 
            \\[-2.5mm]
            \multicolumn{2}{c}{} & \multicolumn{3}{c}{\texttt{Morfometryka}} & \multicolumn{2}{c}{\texttt{statmorph}} & \multicolumn{2}{c}{\texttt{GALFIT}} \\
            \cmidrule(lr){3-5}\cmidrule(lr){6-7}\cmidrule(lr){8-9}
            \multicolumn{1}{c}{\textbf{GRB}} & \multicolumn{1}{c}{\textbf{Filter}} & \multicolumn{1}{c}{\textbf{\textit{C}}} & \multicolumn{1}{c}{\textbf{\textit{A}}} & \multicolumn{1}{c}{\textbf{\textit{S}}} & \multicolumn{1}{c}{\textbf{\textit{G}}} & \multicolumn{1}{c}{\textbf{$M_{20}$}} & \multicolumn{1}{c}{\textbf{$\rm{r}_{50}$ \ ($\arcsec$)}} & \multicolumn{1}{c}{\textbf{Sérsic index}} \\[1.5mm]
            \hline
            050709 & $F814W$ & $2.68_{-0.04}^{+0.02}$ & $0.29_{-0.02}^{+0.02}$ & $0.029_{-0.001}^{+0.002}$ & $0.54\pm0.01$ & $-1.50\pm0.03$ & $0.830\pm0.008$ & $n_1:0.330\pm0.009$ \\ 
            & & & & & & & $0.421\pm0.004$ & $n_2:0.46\pm0.01$ \\[3.5mm] 
            
            060614 & $F160W$ & $2.72_{-0.06}^{+0.07}$ & $0.11_{-0.02}^{+0.02}$ & $0.009_{-0.001}^{+0.001}$ & $0.52\pm0.02$ & $-1.74\pm0.04$ & $0.332\pm0.008$ & $n_1:1.06\pm0.05$\\[1.5mm] 
            & $F606W$ & $2.69_{-0.04}^{+0.04}$ & $0.17_{-0.02}^{+0.02}$ & $0.0128_{-0.0008}^{0.0013}$ & $0.55\pm0.01$ & $-1.83\pm0.03$ & $0.425\pm0.008$ & $n_1:1.16\pm0.02$\\[3.5mm] 
            
            130603B & $F160W$ & $2.73_{-0.02}^{+0.02}$ & $0.20_{-0.01}^{+0.01}$ & $0.0214_{-0.0004}^{+0.0004}$ & $0.55\pm0.01$ & $-1.80\pm0.03$ & $0.460\pm0.003$ &  $n_1:1.218\pm0.009$\\ 
            & & & & & & & $1.63\pm0.01$ & $n_2:0.088\pm0.007$ \\[1.5mm] 
            & $F606W$ & $2.72_{-0.02}^{+0.02}$ & $0.31_{-0.01}^{+0.01}$ & $0.030_{-0.008}^{+0.0003}$ & $0.55\pm0.01$ & $-1.79\pm0.02$ & $0.367\pm0.006$ & $n_1:0.64\pm0.02$\\ 
            & & & & & & & $1.20\pm0.04$ & $n_2:1.45\pm0.03$ \\[3.5mm] 
            
            150101B & $F160W$ & $4.214_{-0.004}^{+0.001}$ & $0.0251_{-0.0003}^{+0.0003}$ & $0.0856_{-0.0096}^{+0.0001}$ & $0.589\pm0.001$ & $-2.459\pm0.002$ & $4.98\pm0.08$ & $n_1:7.71\pm0.02$ \\ 
            & & & & & & & $2.75\pm0.01$ & $n_2:1.96\pm0.01$ \\[1.5mm] 
            & $F606W$ & $3.345_{-0.003}^{+0.003}$ & $0.041_{-0.002}^{+0.002}$ & $0.1165_{-0.0002}^{+0.0003}$ & $0.623\pm0.002$ & $-2.637\pm0.002$ & $6.17\pm0.07$ & $n_1:5.46\pm0.02$ \\[3.5mm] 
            
            160821B & $F160W$ & $2.99_{-0.01}^{+0.01}$ & $0.33_{-0.01}^{+0.01}$ & $0.058_{-0.001}^{+0.001}$ & $0.557\pm0.004$ & $-1.70\pm0.04$ & $2.52\pm0.04$ & $n_1:0.75\pm0.01$\\ 
            & & & & & & & $0.964\pm0.007$ & $n_2:1.19\pm0.01$ \\[1.5mm] 
            & $F606W$ & $3.35_{-0.04}^{+0.02}$ & $0.34_{-0.01}^{+0.01}$ & $0.13_{-0.02}^{+0.01}$ & $0.550\pm0.001$ & $-1.33\pm0.01$ & $2.69\pm0.03$ & $n_1:0.77\pm0.01$ \\ 
            & & & & & & & $1.10\pm0.01$ & $n_2:1.40\pm0.01$ \\[3.5mm] 
            
            170817A & $F160W$ & $4.2843_{-0.0010}^{+0.0003}$ & $0.0454_{-0.0003}^{+0.0003}$ & $0.095_{-0.002}^{+0.000}$ & $0.616\pm0.002$ & $-1.07\pm0.01$ & $0.79\pm0.06$ & $n_1:4.78\pm0.10$ \\ 
            & & & & & & & $16.33\pm0.17 $& $n_2:4.47\pm0.03$ \\[1.5mm] 
            & $F606W$ & $3.3189_{-0.0012}^{+0.0006}$ & $0.0787_{-0.0008}^{+0.0005}$ & $0.06738_{-0.00003}^{+0.00002}$ & $0.5969\pm 0.0002$ & $-1.615\pm0.002$ & $13.11\pm0.02$ & $n_1:3.322\pm0.002$ \\[3.5mm] 
            
            200522A & $F160W$ & $2.77_{-0.01}^{+0.01}$ & $0.24_{-0.01}^{+0.01}$ & $0.0142_{-0.0003}^{+0.0004}$ & $0.51\pm0.01$ & $-1.84\pm0.01$ & $0.666\pm0.008$ & $n_1:0.08\pm0.02$ \\ 
            & & & & & & & $1.04\pm0.01$ & $n_2:0.28\pm0.01$ \\ 
            & & & & & & & $0.252\pm0.004$ & $n_3:1.67\pm0.03$ \\[3.5mm] 
            
            211211A & $F160W$ & $3.034_{-0.004}^{+0.004}$ & $0.163_{-0.005}^{+0.005}$ & $0.0345_{-0.0005}^{+0.0005}$ & $0.538\pm0.005$ & $-1.81\pm0.01$ & $0.642\pm0.003$ & $n_1:0.972\pm0.004$ \\ 
            & & & & & & & $2.174\pm0.005$ & $n_2:0.564\pm0.003$ \\[1.5mm] 
            & $F606W$ & $3.15_{-0.01}^{+0.01}$ & $0.17_{-0.01}^{+0.01}$ & $0.06_{-0.00}^{+0.01}$ & $0.55\pm0.01$ & $-1.81\pm0.01$ & $2.34\pm0.01$ & $n_1:0.556\pm0.005$ \\[1.5mm] 
            & & & & & & & $0.732\pm0.008$ & $n_2:1.13\pm0.01$ \\[3.5mm] 
            
            230307A & $F150W$ & $2.644_{-0.002}^{+0.002}$ & $0.188_{-0.004}^{+0.004}$ & $0.0517_{-0.0004}^{+0.0004}$ & $0.5382\pm0.0002$ & $-1.549\pm0.002$ & $2.382\pm0.007$ & $n_1:0.974\pm0.003$ \\ 
            & & & & & & & $1.25\pm0.02$ & $n_2:1.26\pm0.01$ \\[1.5mm] 
            & $F070W$ & $2.463_{-0.004}^{+0.004}$ & $0.18_{-0.01}^{+0.01}$ & $0.052_{-0.001}^{+0.003}$ & $0.5232\pm0.0002$ & $-1.706\pm0.003$ & $3.54\pm0.07$ & $n_1:1.35\pm0.01$\\ 
            & & & & & & & $4.23\pm0.10$& $n_2:1.77\pm0.02$ \\ 
            \hline
        \end{tabular}
    \end{rotatetable}
\end{table*} 

\subsection{Non-parametric Morphological Classification} \label{subsec:nonparmorfclass}

The classification of the host galaxies for the Gini-$\rm{M}_{20}$ and CAS statistics are determined by the boundaries defined in \cite{Lotz2008} (Eq. \ref{eqn:GiniM20}) and \cite{Bershady_2000} (Eq. \ref{eqn:CA}) respectively:

\begin{equation}\label{eqn:GiniM20}
\begin{aligned}
    \rm{Mergers:} & \ \rm{G > -0.14 \ M_{20}   + 0.33} && \\
    \rm{E/S0/Sa:} & \ \rm{G \leq -0.14 \ M_{20} + 0.33 \ \rm{and}} && \\
    & \ \rm{G > \ \ 0.14 \ M_{20} \ + 0.80} \\
    \rm{Sb/Sc/Sd/Irr:} & \ \rm{G \leq -0.14 \ M_{20} + 0.33\ \rm{and}} && \\
    & \ \rm{G \leq 0.14 \ M_{20} \ + 0.80}
\end{aligned}
\end{equation}

\begin{equation} \label{eqn:CA}
\begin{aligned}
    \rm{Intermediate/Late:} & \ \rm{C = 2.44 \log A + 5.49} && \\
    \rm{Early:} & \ \rm{A < 0.07} && \\
    \rm{Intermediate:} & \ \rm{ 0.07 < A < 0.35} && \\
    \rm{Major \ Mergers:} & \ \rm{A > 0.35} &&
\end{aligned}
\end{equation}

\begin{figure*}
    \centering
    \includegraphics[width=0.8\textwidth]{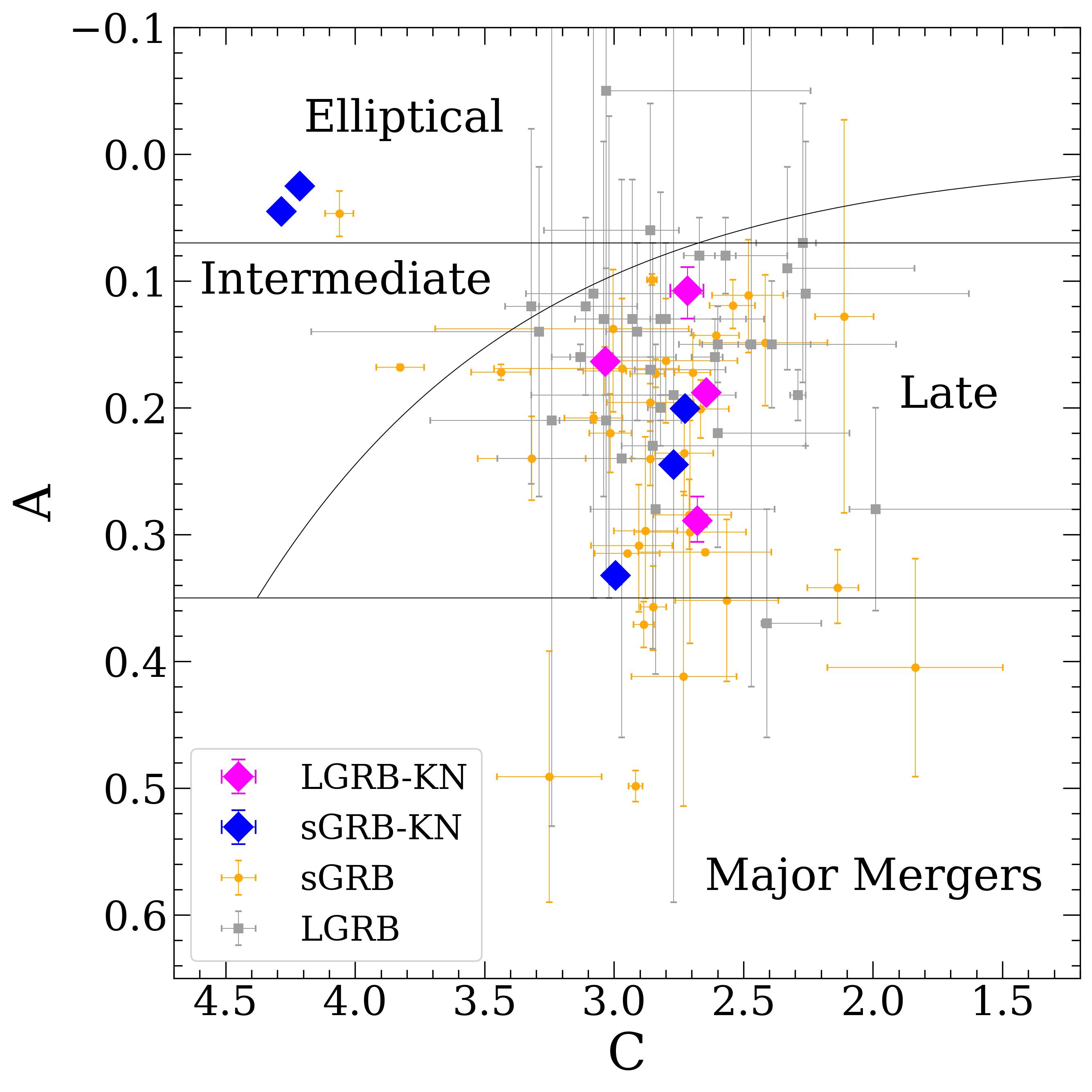}
    \caption{Morphology classification based on concentration versus asymmetry of our host sample – sGRB-KNe (blue) and LGRB-KNe (magenta) – compared to a sGRB (orange; \citealt{Palmese2017}, O'Connor et al. in prep) and LGRB (grey; \citealt{Lyman2017}) host population. The boundaries are defined by Eq. \ref{eqn:CA} of \cite{Bershady_2000}. The bulk of the host sample ($\sim77\%$) is classified as late-type galaxies, with the exception of GRB 150101B and GRB 170817A as both early-type.}
    \label{fig:CAS}
\end{figure*}

In both regimes, the boundaries were determined by eye based on the distribution of Hubble types in the two-dimensional parameter space \citep{Bershady_2000, Lotz2008}. Both statistic methods divide the parameter space into three regions for early-type galaxies (elliptical, lenticular), intermediate-type (spiral), and mergers (tidal features, irregular shapes, multiple nuclei). In the CA parameter space, there is an additional division within the intermediate region. The curved boundary in Figure \ref{fig:CAS} defined in Eq. \ref{eqn:CA} separates intermediate-type above the line and late-type below.

The concentration and asymmetry statistics for our hosts are given in Table \ref{tab:stats} and plotted in Figure \ref{fig:CAS} against sGRB (\citealt{Palmese2017}, O'Connor et al. in prep) and LGRB host populations \citep{Lyman2017}. Both populations are observed with space-based \textit{HST} imaging, with 37 sGRBs in the bands $F160W$ and $F814W$ and 39 LGRBs in band $F160W$. These populations capture fainter, higher redshift galaxies, which result in larger CAS uncertainties than our GRB-KN host sample due to their lower signal-to-noise. We note that the parameter errors for our sample of GRB-KN events (reported in Table \ref{tab:stats}) are too small to be visible in the figure for most objects. The morphological boundaries from \cite{Bershady_2000} are included (Eq. \ref{eqn:CA}). The majority of our sample ($\sim77\%$) falls within the late-type (spiral) region. The spread in concentration for our GRB-KN host sample falls between the sGRB and LGRB host populations, whereas the asymmetry range is narrower than both. Two hosts stand out, GRB 150101B and GRB 170817A, classified in the early-type (elliptical) region of the CA parameter space. These classifications are consistent with the literature \citep{Carter1988,Troja2019b}. Based on the CA statistic, the LGRB host population is classified as $\sim88\%$ late-type (spiral), $\sim9\%$ early-type (elliptical), and $\sim3\%$ mergers as found by \cite{Lyman2017}. Similarly, $\sim78\%$ of the sGRB host population is in the late-type region, with $\sim19\%$ of the sGRB host galaxies identified as major mergers and $\sim3\%$ early-type (elliptical). The sGRB and LGRB hosts have a similar spread in the CA parameter space with a concentration in the late-type region.

\begin{figure*}
    \centering
    \includegraphics[width=\textwidth]{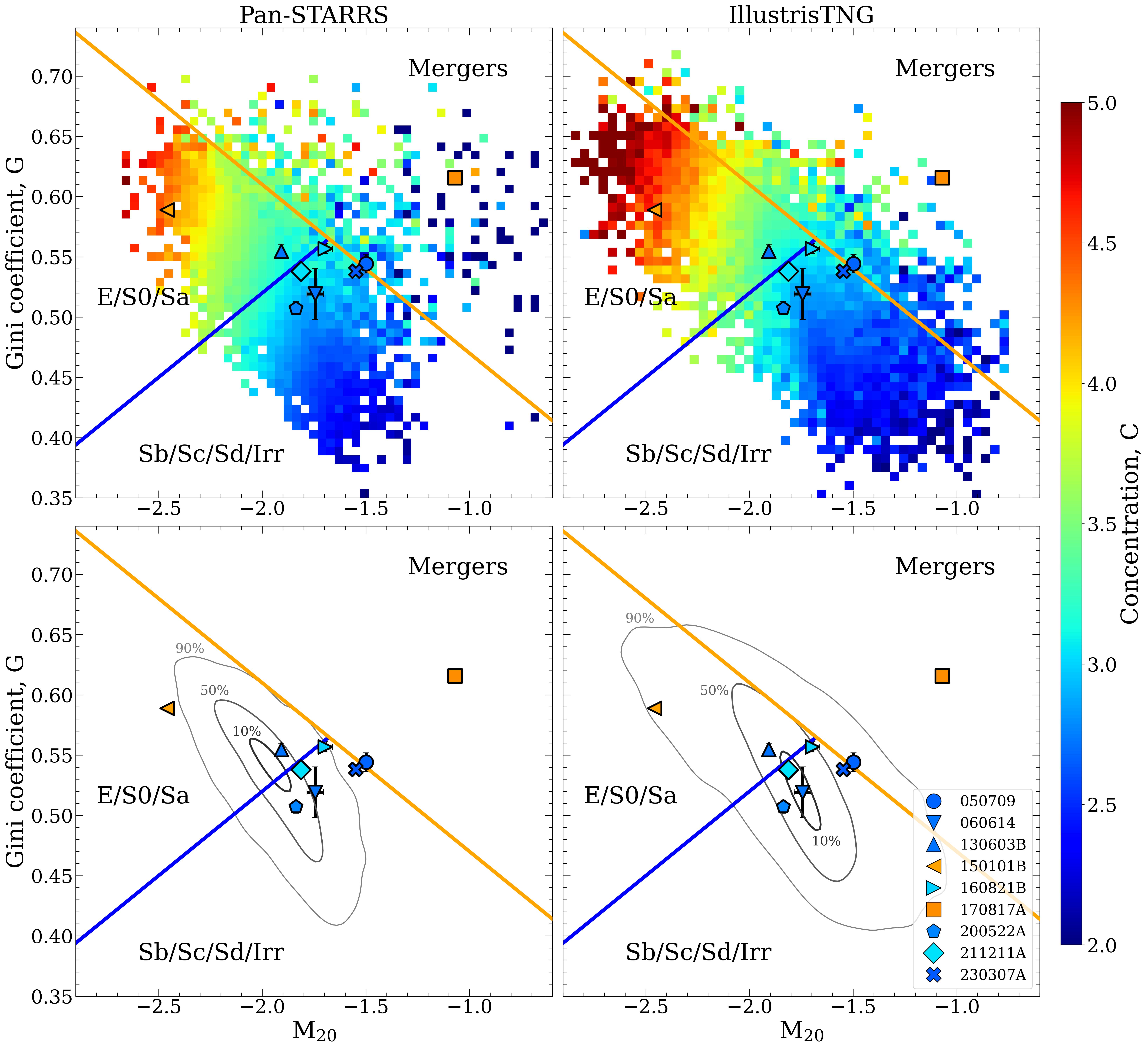}
    \caption{
    A Gini-$\rm{M}_{20}$ comparison between our sample with the observed galaxies in Pan-STARRS (left panels) and the simulated galaxies from IllustrisTNG (right panels) from \cite{Rodriguez_Gomez_2018}. Each host in our sample has a unique marker. The  color for our sample indicates the concentration value from \texttt{MORFOMETRYKA} while the color for the populations is the median concentration values for 2D bins calculated with \texttt{statmorph}. The bottom panels shows the density of the Pan-STARRS and IllustrisTNG populations with the contours depicting 10/50/90\% of the galaxies.}
    \label{fig:GiniM20Pop}
\end{figure*}

The Gini-$\rm{M_{20}}$ statistics are given in Table \ref{tab:stats}. We plot our host sample against the Gini-$\rm{M_{20}}$ statistics of the observed galaxies in Pan-STARRS and the simulated galaxies in IllustrisTNG found in \cite{Rodriguez_Gomez_2018} (see Figure \ref{fig:GiniM20Pop}) with the morphological borders from \cite{Lotz2008}. The galaxies are colored by the concentration value as used in Figure \ref{fig:CAS} and identified with unique markers. Error bars are included, but too small to be visible for some hosts. There are several galaxies that are near the borders of the classification regions (GRBs 050709, 160821B, 211211A, 230307A). GRB 150101B and GRB 130603B are distinctly in the early-type (elliptical) region. While this is expected for GRB 150101B \citep{Fong2016kilonova,Troja2018b}, GRB 130603B instead is usually classified as disturbed or irregular \citep{Cucchiara_2013,Tanvir2013,deUgartePostigo2014}. In addition, we find that GRB 170817A clearly lies in the major or minor merger parameter space \citep[e.g.,][]{Palmese2017, Ebrova2020}. GRB 050709, though along the border, is also classified as having undergone a galaxy merger \citep[see also][]{NicuesaGuelbenzu2021}, whereas it is generally considered a late-type (spiral) galaxy \citep{Hjorth2005,Villasenor2005}. However, the possibility of the host of GRB 050709 having undergone a relatively recent galaxy merger is plausible given its clearly disturbed morphology \citep[see also][]{NicuesaGuelbenzu2021}. The remainder of our sample is concentrated in the late-type region. The Gini-$\rm{M_{20}}$ classification reports $\sim55\%$ of our host sample as late-types (spiral) while the CA classification reports $\sim77\%$. It is important to note that the CAS system classifies only major mergers and the Gini-$\rm{M_{20}}$ system identifies both major and minor mergers. Thus, this difference in classification is due to the two galaxies – GRBs 050709 and 170817A – labeled as minor mergers in the Gini-$\rm{M_{20}}$ parameter space.

The Pan-STARRS population, comprised of 10,829 galaxies, has stellar mass range of log$(M_{\star}/M_{\odot}) \sim 9.8-11.3$ at $z \sim 0.045-0.05$; IllustrisTNG is comprised of 12,470 simulated galaxies with $M_\star > 10^{9.5} \ M_\odot$ at $z=0.0485$ \citep{Rodriguez_Gomez_2018}. The spread of mergers in the Gini-$\rm{M_{20}}$ parameter space is greater in Pan-STARRS than in IllustrisTNG. The reduced number of mergers in IllustrisTNG is an artifact of the simulated image procedure that does not generate early-stage mergers as discussed in \cite{Rodriguez_Gomez_2018}. IllustrisTNG extends to lower $\rm{M_{20}}$ values compared to Pan-STARRS, primarily affecting more massive galaxies.  \cite{Rodriguez_Gomez_2018} suggest this could be due to insufficient quenching in the model, and residual star formation is left at the center of these galaxies. In the top panels of Figure \ref{fig:GiniM20Pop}, the background populations are colored with the median concentration values for each 2D bin. The populations show that there is a correlation between concentration and the Gini-$\rm{M_{20}}$ parameter space. For both populations, our host sample is consistent in concentration except for GRB 130603B that has a lower concentration than the surrounding galaxies in Figure \ref{fig:GiniM20Pop}. This could be attributed to the different software used to calculate the concentration value as discussed in \S \ref{subsubsec:nonpar}. 

The bottom panels of Figure \ref{fig:GiniM20Pop} illustrate the density of the galaxy populations with our host sample mapped over. The contours indicate where 10/50/90\% of the galaxies lie. The IllustrisTNG population has a large spread with 90\% of the population distributed broadly in the Gini-$\rm{M_{20}}$ parameter space. Pan-STARRS is more concentrated than IllustrisTNG, where 90\% of the galaxies are found in a tighter range of Gini and $\rm{M_{20}}$ values. For both populations, the majority of galaxies are classified as early- or late-type with a handful classified as galaxy mergers. Looking at the 10\% contour, the Pan-STARRS distribution is skewed towards early-type (elliptical) galaxies, while IllustrisTNG finds more late-type (spiral) galaxies with lower $\rm{M_{20}}$ values. Our host sample falls outside the densest region of the Pan-STARRS population due to lower values of $\rm{M_{20}}$ values, but it appears to better trace the distribution of IllustrisTNG objects. Our host sample may not align precisely with the observed population since we are considering the $F160W$ band at a range of redshifts from $z \sim 0.0098-0.554$ while \cite{Rodriguez_Gomez_2018} selected the \textit{i}-band from Pan-STARRS at $z \sim 0.045-0.05$. Additionally, the observed population does not extend to lower mass galaxies and our sample has a number of hosts that are $< 10^9 \ M_\odot$.

\subsection{Limitations of the Non-Parametric Galaxy Classification}
\label{subsec:limits}

We want to address caveats that comes with the use of the non-parametric morphological statistics. As mentioned in \S \ref{subsec:nonparmorfclass}, the boundaries that have been defined in the CA and Gini-$\rm{M_{20}}$ space by \cite{Bershady_2000} and \cite{Lotz2008}, respectively, are not strict. The boundaries were established by evaluating the Hubble types in each parameter space by eye. For the CAS statistics, \cite{Bershady_2000} studied a sample of low-redshift galaxies. \cite{Lotz2008} classified Extended Growth Strip galaxies in the redshift range $0.2 < z < 1.2$. Classifying galaxies outside these selections can compromise the accuracy of the results. In addition to these selections, depending on the mass ratio of the two merging galaxies and the observed time frame in the relaxation time of the merger, the asymmetry of the system can be too low to be classified as a major merger in the CAS system \citep{Conselice2006}. This likewise can lead to misalignment of classifications between non-parametric statistic systems (as discussed in \S \ref{subsec:nonparmorfclass}), and to an oversight of minor galaxy mergers. Lastly, these 2D spaces are a slice of the multi-dimensional parameter space used to describe galaxy morphology and galaxy type. Therefore, these statistics only capture a fraction of this complex picture and are insufficient to classify the morphology of our sample alone \citep[see also][]{Lyman2017}. For this paper, we take these classifications at face value and recognize that a galaxy could have a more nuanced classification, and the boundaries between different galaxy types are not held firm. 

\subsection{Morphological Classifications of Our Sample} 
\label{subsec:morf}
  
\subsubsection{CAS}
\label{subsubsec:cas}

Here, we discuss the morphology classification based on the concentration and asymmetry statistics (see \S \ref{subsubsec:nonpar}, \citealt{Bershady_2000}). Upon first glance, the three host populations – sGRBs, LGRBs, and GRB-KNe – show a broad distribution across both concentration and asymmetry (Figure \ref{fig:CAS}). This is the case for the sGRB host population, which have the largest range of concentration and asymmetry compared to the LGRBs and GRB-KNe, spanning the classification boundaries. The majority of sGRBs ($\sim78\%$) can be found in the late-type (spiral) region. This is consistent with the mix of environments found for sGRBs \citep{Fong2013,OConnor2022,Fong2022,Nugent2022} due to their wide rage of progenitor delay times and formation pathways. Moreover, while a large fraction of sGRB host galaxies are classified as star-forming \citep[e.g,][]{Leibler2010, Nugent2022, Jeong2024}, a characteristic of late-type (spiral) galaxies, a closer investigation shows that LGRBs are slightly more concentrated in the late-type (spiral) region with an overall lower range of concentration and asymmetry than the sGRB host population. This is in line with LGRBs being predominately found in star-forming galaxies \citep{Perley2013,Lyman2017}. 

While galaxy mergers can trigger star formation episodes that produce sGRB and LGRB progenitors, our sample notably does not fall in the major merger region, but primarily spans the late-type and early-type regions. The host of GRB 160821B is the one object which is consistent with the CA major merger definition within uncertainties in $F606W$. This is not conclusive to suggest that GRB-KNe are not typically found in hosts that have undergone recent galaxy mergers \citep[e.g., GW170817;][see \S \ref{subsec:interpoff}]{Palmese2017}. Rather, we have a small number of events and the non-parametric morphology statistics are not mutually exclusive as discussed further in \S \ref{subsubsec:Gini-M20}. Along these lines, it is worth noting that, beyond GRB 160821B, another three out of the nine galaxies (namely, the hosts of GRB 050709, 130603B, 200522A) in our sample show a relatively large asymmetry of $\sim 0.25-0.3$ in at least one band, which is close to the limit for considering a galaxy a major merger according to the CA statistics. That said, galaxies with tidal features do not necessarily mean they are products of mergers, but could simply be irregularly shaped. It is also worthy to note that \cite{Conselice2003} found the optical wavelength was sufficient to resolve the structure of nearby galaxies for classification with CAS statistics.   This remains true for light in the optical with small changes from red to blue, but can change dramatically when ultraviolet light is probed \citep{2018ApJ...864..123M}. 

Due to our choice of filter (see \S \ref{subsec:data} for detailed discussion) where the probed rest-frame wavelength range is $\sim693 - 1522 \ \rm{nm}$, we expect the classification boundaries to be slightly different from what is defined in \cite{Bershady_2000} and our values of C and A to be systematically shifted.

\subsubsection{Gini-$M_{20}$} 
\label{subsubsec:Gini-M20}

Here, we discuss the Gini-$\rm{M_{20}}$ statistics and its morphological classification (\S \ref{subsubsec:nonpar}, \citealt{Lotz2008}). Analyzing our host sample alone in Figure \ref{fig:GiniM20Pop}, the Gini-$\rm{M_{20}}$ classification is relatively consistent with the CA statistics. Most hosts lie in the late-type (spiral) region, with a couple of systems identified as early-type (elliptical) or major/minor mergers. GRB 170817A stands as an outlier, classified as a galaxy merger rather than an early-type galaxy, as it is in the CA parameter space. Returning to our earlier point (\S \ref{subsec:limits}), there is a limitation to the classification of these non-parametric morphological statistics. NGC 4993, the host galaxy of GRB 170817A, is classified as a early-type  galaxy \citep{Carter1988} and simultaneously identified as a galaxy that has undergone a recent merger indicated by the shell-like structures around the nucleus \citep[see, e.g.,][]{Quinn1984, Pop2018}. This determination was made by previous studies of NGC 4993 \citep[e.g.,][]{Palmese2017,Levan2017,Blanchard2017,Ebrova2020,Kilpatrick2022} that occurred after the discovery of GW170817.

Though the Gini-$\rm{M_{20}}$ space classifies the host of GRB 170817A as a galaxy merger and the CA statistics classifies the host as an early type, neither classification is incorrect. This dichotomy showcases the necessary caution needed when utilizing these non-parametric statistics. With the GRB 170817A host as an example, we can easily extend this dual-classification to several of the hosts in our sample. Many of the host galaxies in our sample (GRBs 050709, 130603B, 160821B, 200522A) have visible evidence of tidal features in both the observed images and the residuals from \texttt{GALFIT} (see Figure \ref{fig:galfit} in Appendix \ref{app:galfit}). Though classified as early- or late-type galaxies, these hosts could also have undergone a recent major or minor merger. GRB 050709, while not as clear as GRB 170817A, is similarly classified as a merger by Gini-$\rm{M_{20}}$ \citep[consistent with][]{NicuesaGuelbenzu2021}, but as a late-type by CA \citep[consistent with][]{Hjorth2005,Villasenor2005}. Depending on the classification of host galaxies, this can impact our understanding of the likely delay times and progenitors of  GRBs and their overall population. Using GRB 170817A as an example, if the host was classified solely as a early-type (quiescent) galaxy this would be suggestive that the merger has a higher probability of having a long delay time between formation and merger \citep[e.g.,][]{Levan2017,Blanchard2017}. However, the evidence of NGC 4993 having undergone a recent galaxy merger \citep[e.g.,][]{ Palmese2017,Levan2017,Blanchard2017,Ebrova2020} allows for an alternative progenitor formation channel from dynamical interactions and a shorter delay time of $\lesssim 200 \ \rm{Myr}$ \citep[e.g.][]{Palmese2017} or longer delay time of $\gtrsim 200 \ \rm{Myr}$ \citep[e.g.][]{Ebrova2020, Kilpatrick2022}.

We next compare our sample with the observed and simulated galaxy populations Pan-STARRS and IllustrisTNG, respectively. Of our sample, five hosts fall within 90\% of the Pan-STARRS population, with only two hosts within 50\% and none in 10\%, while four hosts are in the outskirts and beyond. We perform a bootstrap resampling of the Pan-STARRS population and our host sample with 10000 draws of 9 galaxies, calculating the mean Gini and $\rm{M_{20}}$ statistic for each draw. An AD test comparing these mean distributions of the two statistics confirms that our host sample does not come from the same distribution as the Pan-STARRS population ($P_{AD, Gini} \sim 0.001, P_{AD, \rm{M_{20}}} \sim 0.001$).

In other words, it appears that the GRB-KN hosts are skewed towards more disturbed/merging morphology compared to  the generic galaxy population. Our hosts explore a range of redshifts whereas the Pan-STARRS population in Figure \ref{fig:GiniM20Pop} is at $z\sim0.05$, somewhat limiting the extent of this comparison. However, galaxy evolution is not expected to dramatically alter our conclusions given that (i) 7/9 of the hosts are at $z<0.17$, and (ii) the two hosts at higher redshift are actually the two within the 50\% contours of the PanSTARRS populations, hence they are not the objects skewing the distribution towards mergers. Perhaps more importantly, it is the galaxy mass range that limits the comparison since our sample includes lower mass galaxies than the typical PanSTARRS stellar mass range.

For IllustrisTNG, we find the opposite, where all but one host are encompassed by 90\% of the galaxy population. The locus of the IllustrisTNG population models the Pan-STARRS morphologies reasonably well, but there is a clear larger spread towards lower and higher $\rm{M_{20}}$ values (and consequently lower and higher Gini values). As previously discussed (\S \ref{subsec:nonparmorfclass}), these discrepancies are thought to be artifacts from feedback models implemented in the simulation. The specifics of these model mechanisms are beyond the scope of this paper but are worth bringing attention to as they impact our analysis. Thus, we keep to PanSTARRS as our fiducial population as IllustrisTNG returns more massive galaxies than what is observed.

\section{Host and GRB Properties} 
\label{sec:host_properties}

As morphology alone is insufficient to classify our sample, we examine other host properties to illuminate distinctions between the GRB classes. In the following sections, we investigate the size, mass, stellar formation rate, and redshift of the the host galaxies. We explore further into the high-energy and kilonova properties of the GRB events themselves. With these analyses, we can examine each factor that make up a GRB-KN event individually.

\subsection{Half-light Radii}
\label{subsec:r50}

The $\rm{r_{50}}$ of our sample are determined using \texttt{GALFIT}. For each component that comprises a \texttt{GALFIT} model of a given host galaxy, the Sérsic profile has a calculated $\rm{r_{50}}$ (see Table \ref{tab:stats} for multiple Sérsic profiles, Table \ref{tab:offsets} for single Sérsic). Our sample contains a range of varying galaxy sizes, with the largest $\rm{r_{50}}$ at $13.11\pm0.02\arcsec$ for GRB 170817A and the smallest at $0.332\pm0.008\arcsec$ for GRB 060614, which each respective Sérisc profile captures the entirety of the galaxy light (see Table \ref{tab:stats}). Considering the $\rm{r_{50}}$ for the single Sérsic \texttt{GALFIT} models, there is a mixture of galaxy sizes for both LGRB-KN and sGRB-KN hosts.

We can compare the angular, physical, and host-normalized\footnote{The host-normalized offset is defined as $\rm{R/r}_{50}$.} offsets of the transients with respect to the host galaxies (see Table \ref{tab:offsets}). The angular and physical offsets are taken from Table 4 in \cite{Fong2022}. Using the host redshifts and assuming a standard WMAP9 cosmology of $H_0 = 69.6 \ \rm{km \ s^{-1} \ Mpc^{-1}}$ and $\Omega_{\rm{M}} = 0.286$ \citep{Hinshaw2013, Bennett2014}, we calculate the physical size of the $\rm{r}_{50}$ from our single Sérsic \texttt{GALFIT} models to use in the computation of the host-normalized offsets (see Table \ref{tab:offsets}). For star-forming hosts in our sample (see Table \ref{tab:properties}), we find a median angular offset of $1.07\arcsec$, a median physical offset of $3.76 \ \rm{kpc}$, and a median host-normalized offset of $1.62 \ \rm{R/r}_{50}$. The quiescent hosts (see Table \ref{tab:properties}) have a median angular offset of $6.69\arcsec$, a median physical offset of $4.74 \ \rm{kpc}$, and a median host-normalized offset of $0.44 \ \rm{R/r}_{50}$. This supports previous work that finds quiescent hosts to have overall larger observed physical offsets compared to star-forming hosts \citep[e.g.,][]{OConnor2022,Nugent2022}.

\begin{figure*}
    \centering
    \includegraphics[width=\textwidth]{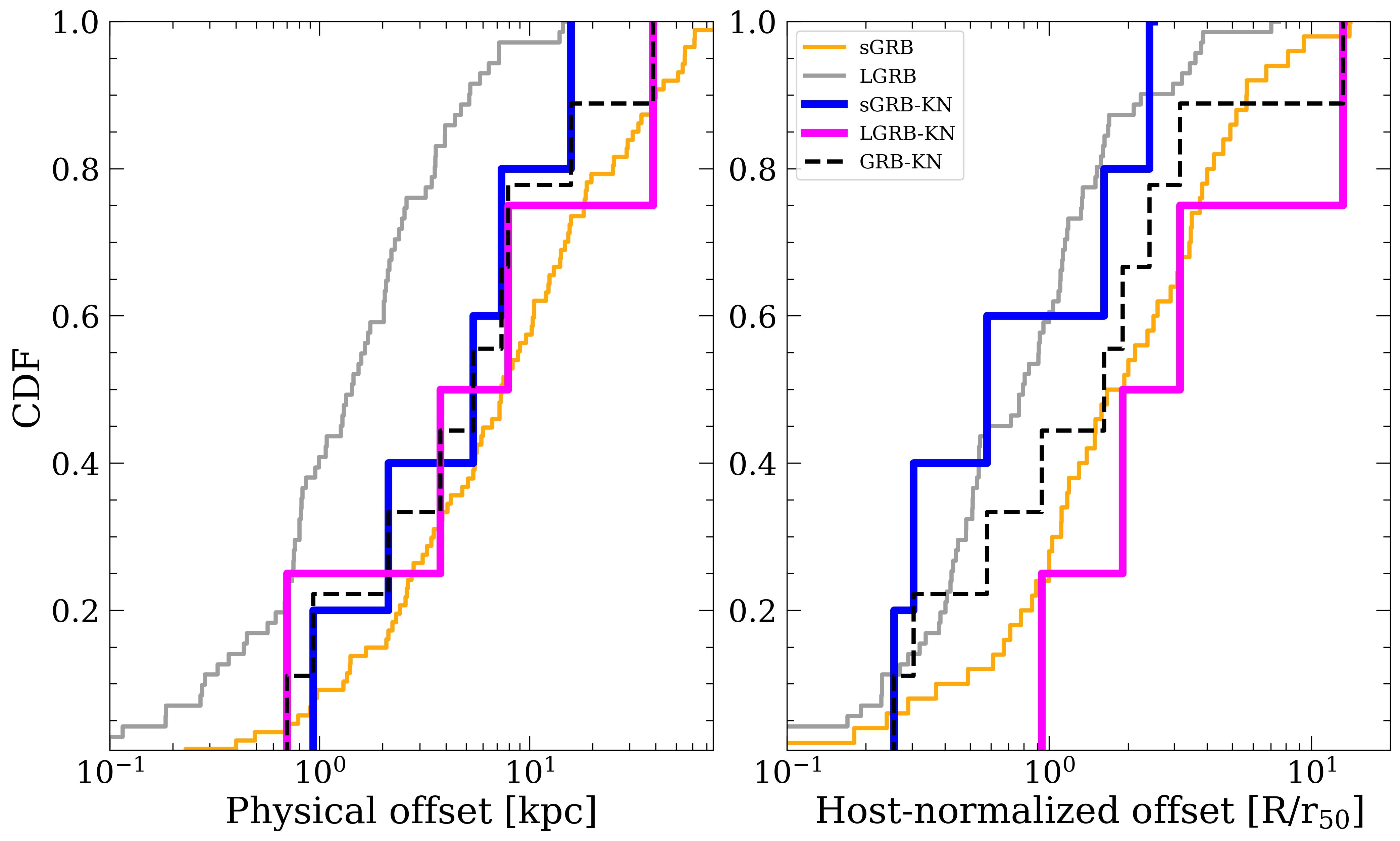}
    \caption{The CDF for the observed physical (left panel) and host-normalized (right panel) offsets of our sample. We calculate the host-normalized offset by taking the ratio of physical offset, R, to the $\rm{r_{50}}$ of the galaxy. The CDF shows sGRBs (orange; \citealt{Gompertz2020}, \citealt{Fong2022}, \citealt{Nugent2022}, \citealt{OConnor2022}), LGRBs (grey; \citealt{Blanchard2016}), sGRB-KNe (blue), and LGRB-KNe (magenta). We use filter $F160W$ ($F814W$ for GRB 050709 and $F150W$ for GRB 230307A) for our sample. The dashed-line shows the GRB-KN host sample as a whole.}
    \label{fig:CDF_offsets}
\end{figure*}

In Figure \ref{fig:CDF_offsets}, we show the cumulative distribution function (CDF) for the physical and host-normalized offsets of our sample given a single Sérsic profile. In both panels, we compare our sample against 87 sGRBs \citep{Gompertz2020, Fong2022, Nugent2022, OConnor2022} and 105 LGRBs \citep{Blanchard2016}. The host-normalized offsets of these populations are determined from a single Sérsic model of the galaxies. For our CDFs, we perform Anderson-Darling (AD) tests to determine if the distributions differ significantly, with the null hypothesis being that they originate from the same underlying distribution. We reject the null hypothesis when the p-value is $<0.05$. We opt for an AD test as it is more sensitive than the Kolmogorov-Smirnov test to differences in the tails of distributions. With large offsets from GRBs 211211A and 230307A, we want to fully explore the extent of the distributions. In these AD tests, we compare our sample against the classical sGRB and LGRB host populations, with an emphasis on LGRB-KNe to determine from which population distribution these events originate.

Figure \ref{fig:CDF_offsets} shows for LGRBs there is a much tighter distribution around small offsets \citep[i.e., tracing instead their formation radii][]{Fruchter2006,Blanchard2016} than for the sGRB host population \citep[e.g.,][]{FongBerger2013}. For our host sample, the sGRB-KNe have a larger spread in offsets that matches the both the distribution of classical sGRBs ($P_{\rm{AD, host}} \sim 0.11$) and classical LGRBs ($P_{\rm{AD, host}} \sim 0.25$) in the host-noramlized CDF (Figure \ref{fig:CDF_offsets}, right panel). However, sGRB-KNe only differ from LGRBs in the physical space ($P_{\rm{AD, phys}} \sim 0.02$). It can be seen that there are LGRB-KNe found at significantly larger distances from their hosts \citep{Rastinejad2022,Troja2022,Levan2023,Yang2024} than what is typical of LGRBs \citep[e.g.,][]{Fruchter2006,Blanchard2016,Lyman2017}. The AD tests for both the physical and host-normalized offsets show that LGRB-KNe and LGRBs have distinct distributions ($P_{\rm{AD, phys}} \sim 0.02$, $P_{\rm{AD, host}} \sim 0.01$) while LGRB-KNe are from the same underlying population as sGRBs ($P_{\rm{AD, phys}} \sim 0.25$, $P_{\rm{AD, host}} \sim 0.25$).

\begin{table*}
    \centering
    \caption{Angular, physical, and host-normalized offsets of the GRB-KN host galaxy sample.}
    \label{tab:offsets}
    \begin{tabular}{llcccll}
        \hline
        \hline 
        \\[-2.5mm]
        \multicolumn{1}{c}{\textbf{GRB}} & \multicolumn{1}{c}{\textbf{Filter}} & \multicolumn{1}{c}{\textbf{\textit{z}}} & \multicolumn{1}{c}{\textbf{Offset}} & \multicolumn{1}{c}{\textbf{Offset}} & \multicolumn{1}{c}{\textbf{$\rm{r}_{50}$}} & \multicolumn{1}{c}{\textbf{Offset}}\\
        & & & \multicolumn{1}{c}{($\arcsec$)} & \multicolumn{1}{c}{(kpc)} & \multicolumn{1}{c}{(kpc)} & \multicolumn{1}{c}{(R/$\rm{r}_{50}$)}\\[1.5mm]
        \hline
        050709 & $F814W$ & 0.161 & $1.35\pm0.02$ & $3.76\pm0.056$ & $1.97\pm0.03$ & $1.91\pm0.06$\\[3.5mm] 
        
        060614 & $F160W$ & 0.125 & $0.31\pm0.35$ & $0.70\pm0.79$ & $0.75\pm0.02$ & $0.94\pm0.79$\\ 
        & $F606W$ & & & & $0.96\pm0.02$ & $0.73\pm0.79$\\ 
        & $F814W$ & & & & $0.78\pm0.03$ & $0.90\pm0.79$\\[3.5mm] 
        
        130603B & $F160W$ & 0.356 & $1.07\pm0.04$ & $5.40\pm0.20$ & $3.34\pm0.02$ & $1.62\pm0.20$\\ 
        & $F606W$ & & & & $3.34\pm0.04$ & $1.62\pm0.20$\\[3.5mm] 
        
        150101B & $F160W$ & 0.1341 & $3.07\pm0.03$ & $7.35\pm0.072$ & $12.69\pm0.03$ & $0.58\pm0.08$\\ 
        & $F606W$ & & & &$14.78\pm0.16$ & $0.50\pm0.18$\\[3.5mm] 
        
        160821B & $F110W$ & 0.1613 & $5.61\pm0.01$ & $15.74\pm0.03$ & $7.36\pm0.12$ & $2.14\pm0.12$\\ 
        & $F160W$ & & & & $6.53\pm0.10$ & $2.41\pm0.11$\\ 
        & $F606W$ & & & & $8.90\pm0.14$ & $1.77\pm0.14$\\[3.5mm] 
        
        170817A & $F110W$ & 0.0098 & $10.317\pm0.005$ & $2.125\pm0.001$ & $5.679\pm0.006$ & $0.374\pm0.006$\\
        & $F160W$ & & & & $6.99\pm0.04$ & $0.30\pm0.04$\\ 
        & $F606W$ & & & & $2.652\pm0.004$ & $0.801\pm0.004$\\[3.5mm] 
        
        200522A & $F125W$ & 0.554 & $0.143\pm0.029$ & $0.93\pm0.19$ & $3.77\pm0.04$ & $0.25\pm0.19$\\ 
        & $F160W$ & & & & $3.64\pm0.04$ & $0.26\pm0.20$\\[3.5mm] 
        
        211211A & $F140W$ & 0.0762 & $5.44\pm0.02$ & $7.92\pm0.029$ & $2.47\pm0.01$ & $3.21\pm0.03$\\ 
        & $F160W$ & & & & $2.51\pm0.01$ & $3.15\pm0.03$\\ 
        & $F606W$ & & & & $3.21\pm0.03$ & $2.47\pm0.04$\\ 
        & $F814W$ & & & & $3.11\pm0.08$ & $2.55\pm0.09$\\[3.5mm] 
        
        230307A & $F105W$ & 0.065 & $30.20\pm0.01$ & $38.90\pm0.01$ & $3.21\pm0.02$ & $12.12\pm0.03$\\ 
        & $F140W$ & & & & $3.92\pm0.06$ & $9.93\pm0.06$\\ 
        & $F070W$ & & & & $3.63\pm0.03$ & $10.72\pm0.03$\\ 
        & $F115W$ & & & & $3.07\pm0.02$ & $12.68\pm0.02$\\ 
        & $F150W$ & & & & $2.95\pm0.02$ & $13.20\pm0.02$\\ 
        & $F277W$ & & & & $3.38\pm0.02$ & $11.52\pm0.03$\\ 
        & $F356W$ & & & & $3.48\pm0.03$ & $11.19\pm0.03$ \\ 
        & $F444W$ & & & & $3.70\pm0.04$ & $10.50\pm0.04$\\[3.5mm] 
        \hline
    \end{tabular}
\end{table*}

\subsection{Comparison of Host Mass, Star Formation Rate, and Redshift} 
\label{subsec:properties}

\begin{figure*}
    \centering
    \includegraphics[width=\linewidth]{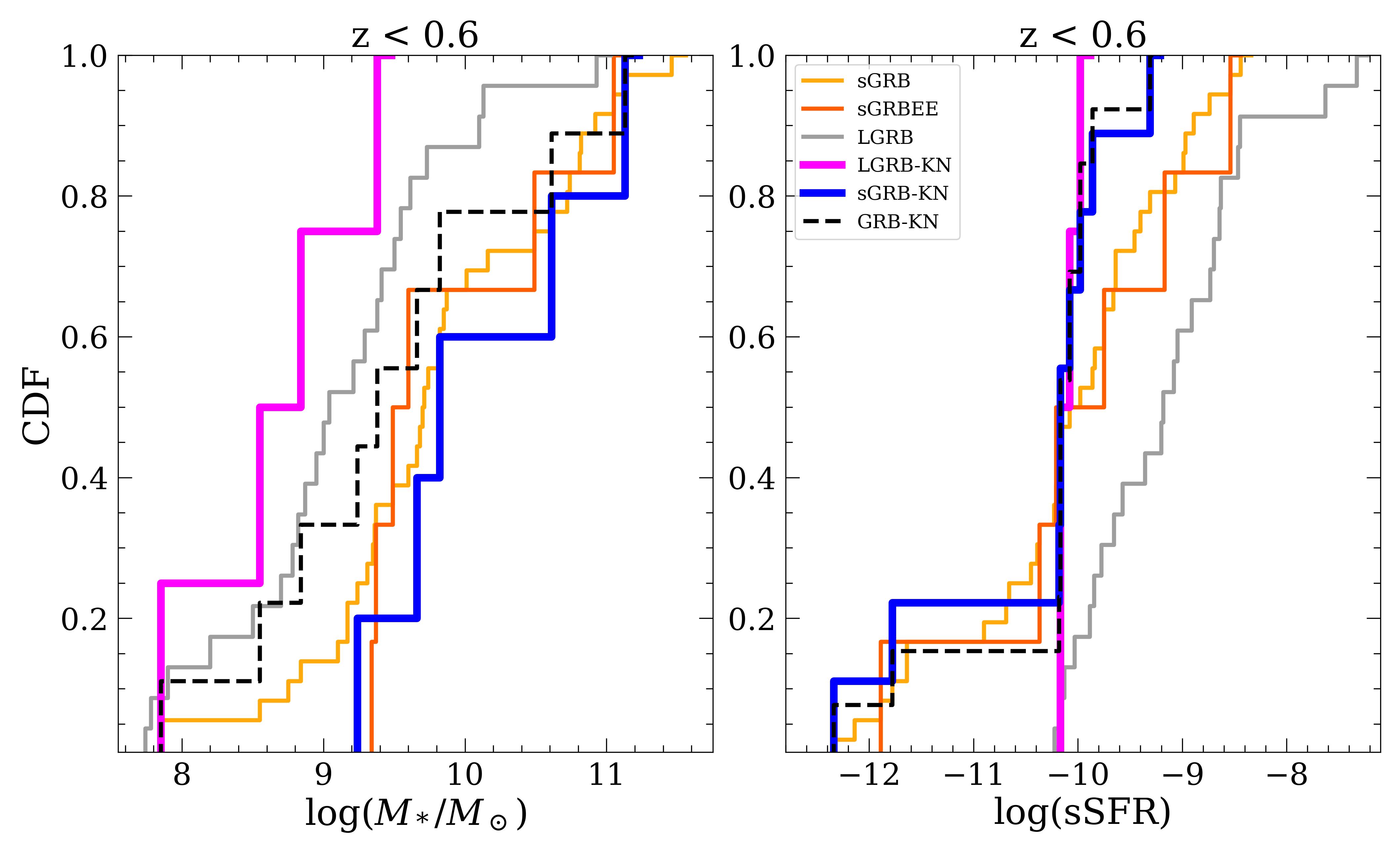}
    \caption{The CDF of stellar mass (left panel) and sSFR (right panel) for sGRBs (orange; \citealt{OConnor2020}, \citealt{Fong2022}, \citealt{Nugent2022}), LGRBs (grey; \citealt{Svensson2010}, \citealt{Perley2013}, \citealt{Vergani2015}, \citealt{Wang2014}, \citealt{Niino2017}), sGRBEEs (dark orange; \citealt{OConnor2020}, \citealt{Fong2022}, \citealt{Nugent2022}), sGRB-KNe (blue), and LGRB-KNe (magenta) with a redshift cut at $z < 0.6$. The dashed-line shows the GRB-KN host sample as a whole.
    }
    \label{fig:CDF_mass_sSFR}
\end{figure*}

\begin{figure*}
    \centering
    \includegraphics[width=\linewidth]{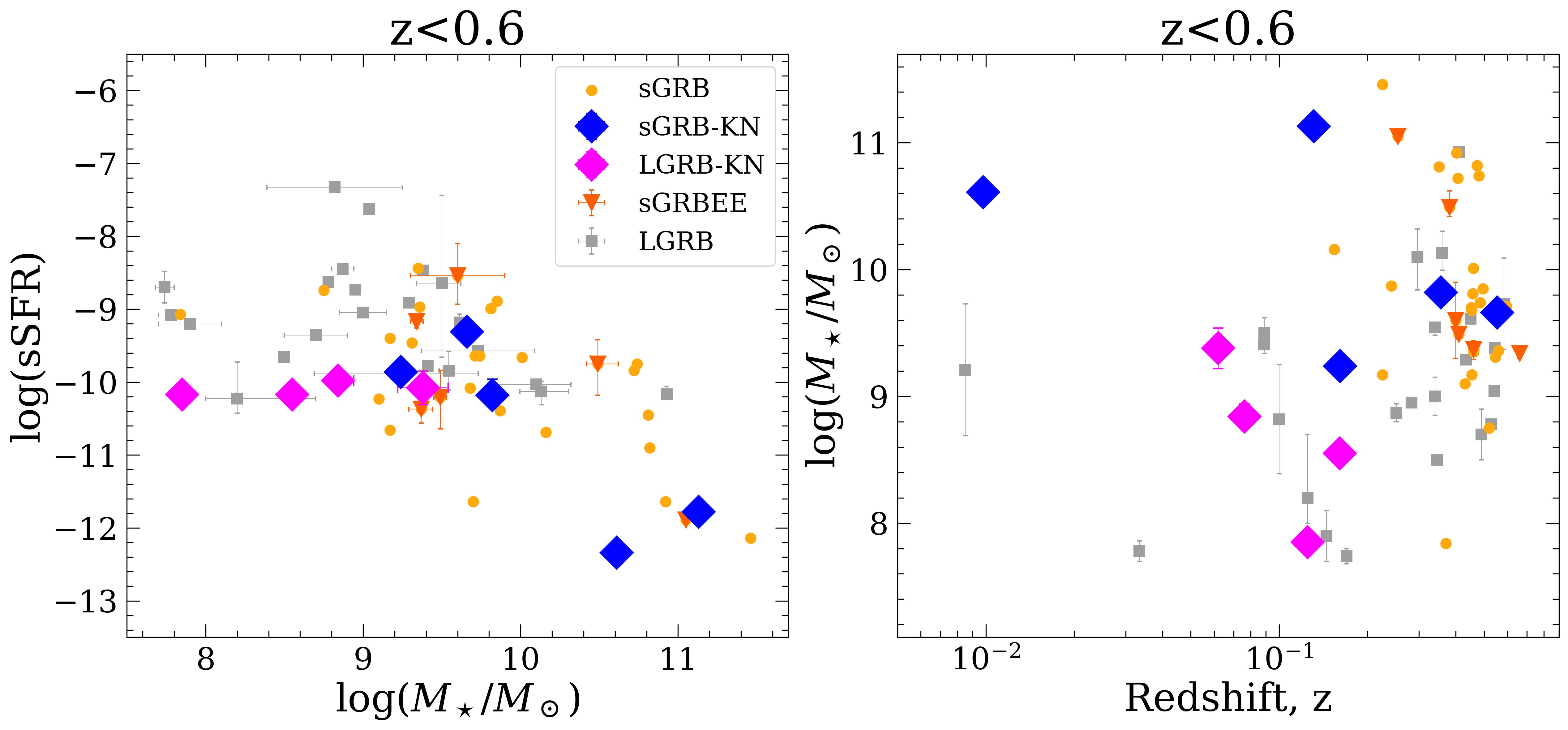}
    \caption{
    A comparison of the stellar masses against the sSFR (left panel) and the redshift, z, (right panel) for sGRBs (orange; \citealt{OConnor2020}, \citealt{Fong2022}, \citealt{Nugent2022}), LGRBs (grey; \citealt{Svensson2010}, \citealt{Perley2013}, \citealt{Vergani2015}, \citealt{Wang2014}, \citealt{Niino2017}), sGRBEEs (dark orange; \citealt{OConnor2020}, \citealt{Fong2022}, \citealt{Nugent2022}), sGRB-KNe (blue), and LGRB-KNe (magenta) with a redshift cut at $z < 0.6$.}
    \label{fig:logM_logsSFR}
\end{figure*}

We now discuss the host galaxy properties of our sample, such as star formation, redshift, stellar mass, and specific star formation rate (sSFR), and compare to other GRB host populations. Our sample consists of nearby galaxies of $z<0.6$ (see Table \ref{tab:offsets}), with GRBs 170817A, 211211A, and 230307A at $z<0.1$. We compile the properties of our galaxies, derived using optical spectroscopy and multi-band spectral energy distribution modeling, from \citet{Nugent2022}, supplemented by values derived by \citet{OConnor2021kn} and \citet{Yang2024} for GRBs 200522A and 211211A, respectively. Across all hosts in our sample, a Chabrier initial mass function \citep{Chabrier2003} was implemented in the fits (see \citealt{Nugent2022} for further details). Our sample of GRB-KNe have secure host associations and are included in the Gold sample of \citet{Fong2022,Nugent2022}, which require a probability of chance coincidence of $P_\textrm{cc}$\,$<$\,$0.02$ \citep{Bloom2002}. The sole exception is GRB 160821B in the Silver sample, which is defined as having $P_\textrm{cc}$\,$<$\,$0.10$. These low $P_\textrm{cc}$ values are to be expected as we are studying nearby GRBs which are found in brighter, low redshift galaxies that are known to have more robust associations as the probability of chance coincidence calculation is strongly dependent on galaxy brightness \citep{Bloom2002}. 

The star formation classification of our hosts is given in Table \ref{tab:properties}, based on results from \cite{Nugent2022}. The galaxies have been grouped by ``star-forming'' (SF), lying on the star-forming main sequence (SFMS), or ``quiescent'' (Q), lying off the SFMS, based on the calculation in \cite{Tacchella2022} using the sSFR and galaxy redshift. We find that $\sim 78\%$ of our hosts are star-forming and $\sim 22\%$ are quiescent. Our sample is consistent with previous studies that likewise find the majority of sGRB associated with star-forming hosts \citep{Fong2013,Fong2022,Nugent2022}. However, LGRBs are likewise predominately observed in star-forming galaxies \citep{Christensen2004, Levesque2014}. As such, we find no strong selection bias exists in the galaxy type of GRB-KNe; their host galaxies are generally consistent with the larger host population of sGRBs and LGRBs. We explore additional comparisons in stellar mass, sSFR, and redshift in Figures \ref{fig:CDF_mass_sSFR} and \ref{fig:logM_logsSFR}. 

Our host sample spans a considerable mass range (Figure \ref{fig:CDF_mass_sSFR}, left panel; Figure \ref{fig:logM_logsSFR}), from dwarf galaxies ($\rm{log}(M_*/M_\odot) < 9$) to high-mass systems ($\rm{log}(M_*/M_\odot) > 11$). The stellar mass CDF (Figure \ref{fig:CDF_mass_sSFR}, left panel) reveals a distinction between sGRB-KNe and LGRB-KNe ($P_{\rm{AD, mass}} \sim 0.02$), with LGRB-KN hosts less massive than sGRB-KN hosts. Both GRB-KN host samples are statistically consistent with their respective GRB host populations where we cannot reject the null hypothesis: sGRB-KN host galaxy masses match the sGRB host population ($P_{\rm{AD, mass}} \sim 0.3$) and LGRB-KNe masses match the LGRB host population ($P_{\rm{AD, mass}} \sim 0.3$). As discussed in \S \ref{sec:obs}, LGRB-KNe are thought to potentially  be related to sGRBEE as they show similar prompt emission properties and behavior \citep[see, e.g.,][]{Gehrels2006,Norris2006,Troja2022}. In Figure \ref{fig:CDF_mass_sSFR}, we compare our GRB-KNe sample, with a focus on the LGRB-KNe, to 6 sGRBEEs (050724A, 061006, 061210, 071227A, 180618A, 180805B) from \citealt{OConnor2020}, \citealt{Fong2022}, \citealt{Nugent2022}. Both visually and statistically, the LGRB-KNe are distinct from the sGRBEE population ($P_{\rm{AD, mass}} \sim 0.03$) and the sGRB population ($P_{\rm{AD, mass}} \sim 0.03$). Within the LGRB-KN hosts, 75\% of the galaxy masses are measured to be in the dwarf regime whereas 46\% of the LGRB host population is found with $\rm{log}(M_*/M_\odot) < 9$. Only GRB 230307A resides in a slightly more massive host at $\sim 10^{9.4} \ M_\odot$. On the other end of the mass range, the largest hosts in our sample are the two sGRB-KN 170817A and 150101B, both of which are quiescent galaxies. 

In the right panel of Figure \ref{fig:logM_logsSFR}, we compare stellar mass against redshift to see the effect of galaxy evolution and selection effects in our sample. Across the host populations, higher redshift galaxies are typically more massive than lower redshift galaxies with the exception of GRBs 150101B and 170817A. However, with our specific redshift cut at $z<0.6$, we miss that overall LGRB hosts are found at higher redshifts than sGRBs hosts.

The sSFR distributions (Figure \ref{fig:CDF_mass_sSFR}, right panel; Figure \ref{fig:logM_logsSFR}, left panel) show a distinction between the GRB progenitor populations. Both sGRB and sGRB-KN host populations exhibit a broad sSFR distribution spanning log(sSFR/$\rm{yr}^{-1}$) $\sim -9 \ \rm{to} -\!\!12$, reflecting a diversity of delay times in a variety of host galaxies (see Figure \ref{fig:CAS}). The sGRBEEs fall within the scope of the sGRB and sGRB-KN host populations. In contrast, the LGRB hosts are concentrated at higher sSFRs (log(sSFR/$\rm{yr}^{-1}$) $\sim -7 \ \rm{to} -10$), indicating ongoing star-forming to produce collapsar progenitors. Significantly, the LGRB-KN hosts have lower sSFRs than the LGRB host population, occupying the intermediate region at log(sSFR/$\rm{yr}^{-1}$) $\sim -10$. We performed an AD test and can reject the null hypothesis that the LGRB-KN hosts are from the same sSFR distribution as the LGRB host galaxies ($P_{\rm{AD, mass}} \sim 0.001$) and the sGRBEE sSFR distribution ($P_{\rm{AD, mass}} \sim 0.01$). However, the sSFR for LGRB-KNe and sGRBs are from the same underlying distribution ($P_{\rm{AD, mass}} \sim 0.3$). The lower sSFRs of the LGRB-KN hosts is indicative of an evolved stellar population, whereas the LGRB hosts are experiencing more recent star formation activity. The low sSFR of the LGRB-KN hosts suggests that they have less capacity to sustain the typically large star formation found in collapsar environments.

The galaxy mass and sSFR provide a clear boundary that the SFR and redshift could not. From the mass, we find that LGRB-KNe are more aligned with the LGRB host population towards the lower mass end. However, the sSFR shows that the distribution of LGRB-KN hosts at lower sSFR matches the sGRBs distribution. While these statements are conflicting, the analysis of these properties is needed alongside other observations to understanding how to classify GRBs

\begin{table*}
    \centering
    \caption{Stellar population properties of the GRB-KN host galaxy sample found in \cite{Nugent2022}, plus GRB 230307A.
    }
    \label{tab:properties}
    \begin{tabular}{lccccc}
        \hline
        \hline 
        \\[-2.5mm]
        \textbf{GRB} & \textbf{Type} & \textbf{log$(M_{\star}/M_{\odot})$} & \textbf{SFR} $(M_{\odot}\rm{yr^{-1}})$ & \textbf{log(sSFR/$\rm{yr}^{-1}$)}\\[1.5mm]
        \hline
        050709 & SF & $8.55_{-0.01}^{+0.01}$ & $0.024_{-0.001}^{+0.001}$ & $-10.17_{-0.01}^{+0.01}$ \\[1.5mm] 
        
        060614 & SF & $7.85_{-0.03}^{+0.04}$ & $0.005_{-0.001}^{+0.001}$ & $-10.17_{-0.06}^{+0.05}$ \\[1.5mm] 
        
        130603B & SF & $9.82_{-0.04}^{+0.05}$ & $0.44_{-0.09}^{+0.22}$ & $-10.18_{-0.13}^{+0.22}$ \\[1.5mm] 
        
        150101B & Q & $11.13_{-0.02}^{+0.02}$ & $0.22_{-0.02}^{+0.02}$ & $-11.78_{-0.05}^{+0.05}$ \\[1.5mm] 
        
        160821B & SF & $9.245_{-0.004}^{+0.003}$ & $0.24_{-0.01}^{+0.01}$ & $-9.86_{-0.02}^{+0.02}$ \\[1.5mm] 
        
        170817A & Q & $10.61_{-0.02}^{+0.01}$ & $0.019_{-0.005}^{+0.004}$ & $-12.34_{-0.12}^{+0.08}$ \\[1.5mm] 
        
        200522A & SF & $9.66_{-0.01}^{+0.01}$ & $2.23_{-0.05}^{+0.06}$ & $-9.31_{-0.01}^{+0.01}$ \\[1.5mm] 
        
        211211A & SF & $8.84_{-0.05}^{+0.10}$ & $0.07_{-0.01}^{+0.01}$ & $-9.98_{-0.11}^{+0.07}$ \\[1.5mm] 
        
        230307A\tablenotemark{b} & SF\tablenotemark{a} & $9.38\pm0.16$ & $0.20\pm0.03$ & $-10.08\pm0.23$ \\
        \hline
    \end{tabular}
    \tablenotetext{a}{Calculated following Eq. 2 in \cite{Tacchella2022}}
    \tablenotetext{b}{Properties taken and calculated from \cite{Yang2024}}
\end{table*}

\subsection{Comparison of Galactocentric Offsets} 
\label{subsec:offsets}

\begin{figure}
    \centering
    \includegraphics[width=\linewidth]{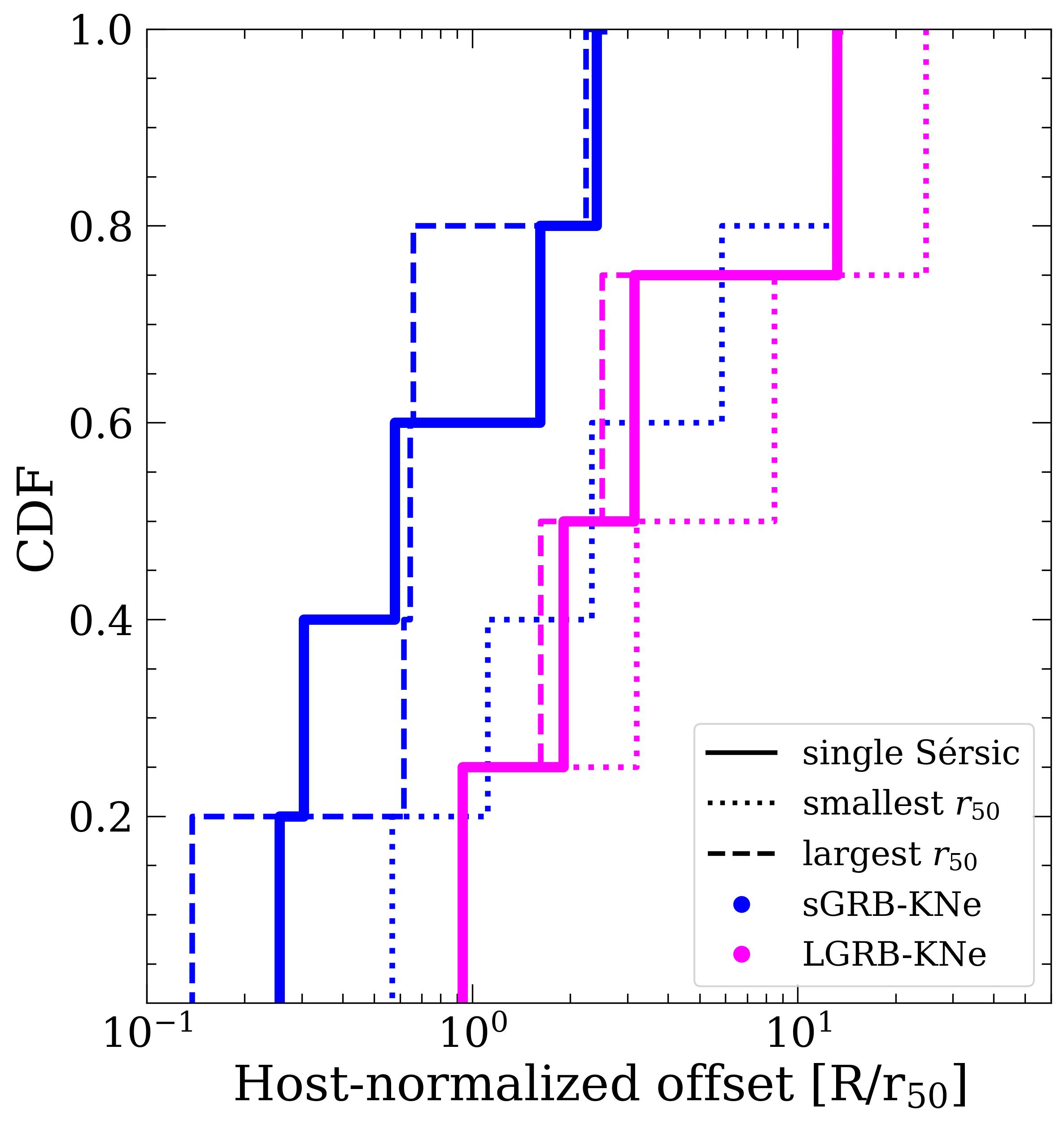}
    \caption{
    A comparison of the host-normalized offsets given a single (solid) or multi-Sérsic model. We show the smallest (dotted) and largest (dashed) $\rm{r_{50}}$ for the multi-Sérsic models. sGRBs are indicated by blue and LGRBs by magenta.
    }
    \label{fig:host-norm_compare}
\end{figure}

We measure the projected galactocentric offsets for all events in our host sample, considering both the physical and host-normalized offsets.  We use two complementary parameterizations of the host light within \texttt{GALFIT}. First, as discussed in \S \ref{subsubsec:par}, we modeled our GRB-KN hosts using multiple Sérsic profiles to reveal low surface brightness features of the galaxies in the residual image (see Figure \ref{fig:galfit} in Appendix \ref{app:galfit}). Second, we performed single Sérsic fits to provide an analogous comparison to literature samples, which typically define the $\rm{r_{50}}$ from a one-component profile \citep{Blanchard2016,OConnor2022,Fong2022}.

The multi-component models allow us to place each transient in the context of specific host components (e.g., bulge and disk) and any residual structures (e.g., tidal arms, streams). For the hosts exhibiting confirmed or likely merger signatures (GRBs 050709, 130603B, 160821B, 170817A, 200522A), the transient is localized to the portion of the galaxy disk where tidal features are visible (see Figure \ref{fig:grid}) and coincides with clear residuals visible in the model subtracted images (see Figure \ref{fig:galfit} in Appendix \ref{app:galfit}). The observed tidal features are commonly associated with recent interactions and (likely minor) mergers \citep[e.g.,][]{Lotz2008}. We discuss the possible association of these GRBs to merger-induced formation pathways in \S \ref{subsec:interpoff}. For what concerns the remaining systems without clear merger signatures, GRBs 060614 and 150101B lie along the periphery of the central host light, while GRBs 211211A and 230307A are located well outside the fitted host components, consistent with an extreme offset event.

We explore the effects multi-component models have on the calculated host-normalized offsets in Figure \ref{fig:host-norm_compare}. The blue highlights the sGRB-KNe offsets, and the magenta the LGRB-KNe. As discussed, the transient falls onto a galactic component or a residual structure from a galaxy merger. Given the component that best represents the environment of the GRB, the calculated host-normalized offset is dependent on the $\rm{r_{50}}$ of said component. This provides a more constrained host-normalized offset than using a single Sérsic profile to model the overall light of a galaxy. Figure \ref{fig:host-norm_compare} shows that the single Sérsic approach overestimates the largest galactic component, with the exception of GRBs 150101B and 170817A, but underestimate the smallest. Given the transients in our sample lie in the disk or at the edge or well beyond the host light, we use the largest $\rm{r}_{50}$ to compare against the single Sérsic measurements. The deconstruction of host morphology into its structural components is an important consideration when regarding host-normalized offsets. An analysis of GRB progenitor environments is influenced by this choice of $\rm{r_{50}}$, in particular when regarding the examination of kicks and their role in the offset of GRBs. A more detailed discussion on the relationship between offsets and formation pathways is in \S \ref{subsec:interpoff}.

Using the single Sérsic models, we compare the physical and host-normalized offsets against sGRB \citep{Gompertz2020, Nugent2022, OConnor2022} and LGRB host populations \citep{Blanchard2016}. As these past literature studies are based on single-component fits to galaxy light profiles to produce a $\rm{r_{50}}$, we utilize our single-component fits for an analogous comparison. From the physical offsets of our sample, we confirm a previously reported correlation between host galaxy type and offset, with quiescent galaxies exhibiting larger offsets than star-forming hosts \citep{OConnor2022,Nugent2022}. The median host-normalized offset for quiescent galaxies is smaller in our case too. This may plausibly reflect the larger characteristic size (and larger mass) of the quiescent galaxies in our sample, which reduce the contrast between populations when scaled by the $\rm{r_{50}}$. When considering the multi-Sérsic models, we also find that quiescent galaxies have a smaller median host-normalized offset than star-forming galaxies. However the difference in the medians of the two subsamples is smaller, potentially a result of the overestimation of the radii of star forming galaxies (which can contain a disky component in addition to a bulge) based on the single component fits. 

The CDFs (Figure \ref{fig:CDF_offsets}) show the relationship between the observed offsets and progenitors of these subpopulations. The sGRB-KNe follow this trend, spanning the entire range of the observed sGRB offset distribution. In contrast, the LGRB-KNe exhibit systematically larger offsets than the LGRB host population \citep[e.g.,][]{Blanchard2016,Lyman2017} and are more consistent with merger-driven events \citep[e.g.,][]{Rastinejad2022,Troja2022,Levan2023,Yang2024}. When comparing the LGRB-KN host-normalized offsets to those of the sGRB host population, the AD test confirms that the distributions are from the same underlying population ($P_{\rm{AD, host}} \sim 0.25$). However, when examining the right panel of Figure \ref{fig:CDF_offsets} by eye, the LGRB-KNe appear to be skewed towards higher host-normalized offsets. This skew is due to GRB 230307A, which has a host-normalized offset of $12.92\pm0.02 \ \rm{R/r}_{50}$ and a physical offset of $38.90\pm0.01  \ \rm{kpc}$. In any case, we suggest that these trends provide further support for the classification of these events as being produced by compact object mergers rather than originating from massive star collapse \citep{Rastinejad2022,Troja2022,Levan2023,Yang2024}.

\subsection{Offsets and Formation Pathways}
\label{subsec:interpoff}

The observed galactocentric offsets of a class of transients encodes both the birth sites of the progenitors as well as any subsequent migration prior to explosion (or merger) of the system. For example, LGRBs originate from the core-collapse of short-lived massive stars, resulting in smaller offsets from their birth sites \citep{Blanchard2016,Lyman2017} and a strong correlation with their host galaxy's ultraviolet light \citep{Fruchter2006,Blanchard2016}, while sGRBs are the result of compact object mergers that have diverse formation channels and a range of delay times allowing for larger physical offsets from their host galaxies \citep[e.g.,][]{Fryer1997,Bloom1999,Beniamini2016,Zevin2022}. 

There are numerous scenarios for the formation of compact object systems that eventually produce GRBs and kilonovae. The diverse environments of sGRBs, sGRB-KNe, and LGRB-KNe \citep[e.g.,][]{Berger2010a,Leibler2010,Fong2013,FongBerger2013,Tunnicliffe2014,OConnor2022,Fong2022,Nugent2022}, support the possibility that multiple pathways may potentially be invoked to explain all the available observations. The standard, and most straightforward, pathway is isolated formation of a binary neutron star system through standard stellar evolution \citep[e.g.,][]{Tauris2017}. There are alternative pathways such as accretion-induced collapse of white dwarfs (WD) \citep[e.g.,][]{Fryer1999AIC}, WD binaries \citep[e.g.,][]{Lloyd-Ronning2024}, and BH binaries \citep[e.g.,][]{Lee2007}. In the case of binary neutron stars, in order to reproduce the larger observed galactocentric offsets (Figure \ref{fig:CDF_offsets}), the system requires a significant natal kick at the birth of the second neutron star \citep[e.g.,][]{Fryer1997,Bloom1999}, as well as a usually long ($\sim$ Gyr or longer) delay time \citep[e.g,][]{Nakar2006,HaoYuan2013,Wanderman2015,Ghirlanda2016}. While these large kicks have been found to be capable of reproducing extremely large tens of kpc offsets from massive ($10^{9-11}M_\odot)$ host galaxies \citep[e.g.,][]{Zevin2020,Bom2025-kn,Chrimes2025}, recent analysis of the time delay distributions of neutron star binaries continue to point towards shorter delay times (steeper distributions) than previously expected \citep{Beniamini2016p2,Beniamini2016,BeniaminiPiran2019,Zevin2022,Beniamini2024,Pracchia2026}. This is due to both theoretical advances, an increased number of known Galactic binary neutron stars, and a larger number of high redshift ($z>1$) sGRBs identified in the last decade \citep{Selsing2018,OConnor2022,Fong2022,Nugent2022}. 

In an alternative scenario, the progenitor of the GRB is born in a dense globular cluster environment \citep{Salvaterra2010, Church2011}, which can show a range of offsets. However, simulations conducted by \cite{Ye2020} found that compact object merger rates are low in these environments and that it is therefore an unlikely channel. An additional scenario to consider is the birth of a progenitor in a low mass, ultra faint dwarf galaxy \citep[see, e.g.,][]{Beniamini2018} in the neighborhood of the more massive host (and likely gravitationally bound to it) that remains undetected \citep[see also][]{Dichiara2026}. The natal kicks in this case are limited to three paths: (\textit{i}) the kick is weak and the induced systemic velocity does not exceed the escape velocity of the dwarf galaxy, (\textit{ii}) the kick is strong enough to induce a large eccentricity that significantly shortens the delay time \citep{Beniamini2024}, producing a small offset despite the large systemic velocity, and (\textit{iii}) the system has a kick that easily escapes the dwarf galaxy, allowing it to travel to large distances from both the dwarf and the primary galaxy. While these scenarios are capable of producing a range of offsets, to date, no such ultra faint dwarf has been associated to a GRB. This may also be due to the extreme sensitivity required to detect such systems at the cosmological redshifts of GRBs. Approximately $20-30\%$ fall under this category of ``hostless'' GRBs \citep{FongBerger2013, Fong2022, OConnor2022}. Within a dwarf galaxy origin scenario, one could also speculate that some of the tidal features we observe in our host sample could be tidally disrupted low-mass galaxies.

Finally, in interacting hosts, dynamical perturbations can trigger bursts of star formation through gas inflows and turbulent mixing \citep[e.g.,][]{DiMatteo2007,DiMatteo2008,Lotz2008}. As multiple events (GRBs 050709, 130603B, 160821B, 170817A, 200522A) in our host sample originate within features indicative of a host interaction it is worth considering whether the formation and evolution of their progenitors could have been influenced by dynamical galaxy interactions or merger-induced star formation. If a GRB-KN site is physically associated with merger-triggered star formation, the implied delay time must be short enough that the progenitor remains near its birth environment, and the required systemic velocity is potentially modest. It is also possible that the progenitor is an older (long delay time) system originating from within a galaxy shredded by the dynamical interactions. This appears to be the case for GRB 170817A, since no recent star formation appears to have been triggered by the merger and the stellar population is rather old (median age $\sim 11$ Gyr; \citealt{Palmese2017}). 

Alternatively, the apparent alignment of the GRB localization with such tidal features could simply be a chance alignment \citep[e.g.,][]{Bloom2002} rather than in-situ formation within the debris, though the exact probability of chance coincidence is not trivial to calculate. We carry out a simple estimation of the probability by taking the ratio of the area of the tidal feature to the area of the galaxy light. In our host sample, GRBs 050709, 130603B, and 200522A fall on or next to a tidal feature in their host galaxy (see Figure \ref{fig:galfit} in Appendix \ref{app:galfit}). From our estimates, the probability of chance alignment for each GRB is $P = 0.07, \ P = 0.06, \ P = 0.10$ respectively.

Following the same line of thinking, the presence of recent or ongoing star formation somewhere in a host does not by itself establish a causal link to the GRB site. Standard ``star-forming'' classifications reflect activity over the past $\sim$30--100~Myr \citep{Nugent2022}, and a transient offset from the central stellar light (or localized to faint tidal structures) may instead originate from an older formation episode with a longer delay time. The same can be said for the observation of sGRBs in currently quiescent galaxies, which does not immediately require that the system has a large merger delay time. While the median offset for quiescent galaxies is indeed larger than in star forming hosts \citep{Paterson2020,OConnor2022,Nugent2022}, this is not necessarily an impact of long delay times. Indeed, population synthesis simulations have shown the opposite trend, with higher redshift systems that necessarily require a shorter delay time at larger offsets \citep{Perna2021}. The large observed offsets in quiescent galaxies could also be attributed to the multiple scenarios discussed above (i.e. globular clusters, undetected dwarf galaxy), or, the additional possibility of formation within the extended halo, which would not require kicks \citep{PeretsBeniamini2021}. Another argument is that the movement of the progenitors within a star-forming versus quiescent galaxy relative to the galaxy radius is not necessarily as significant for the physical offsets due to the typically larger sizes and masses of quiescent galaxies \citep{Nugent2022}, hence a larger difference in the physical offsets specifically. 

Additionally, the observed offset distributions provide an independent consistency check on the physical interpretation of the LGRB-KNe. Their larger offsets relative to classical LGRBs disfavor massive-star progenitors (which instead tracks star-forming regions and yields small offsets; \citealt{Fruchter2006}) and instead support merger-driven origins with a potentially broader range of delay times and migration histories \citep[e.g.,][]{Troja2022,Yang2024,Rastinejad2022,Levan2023}. The apparent skew toward large $\rm{R/r}_{50}$ is driven by a very small number of extreme offset events (especially GRB 230307A), emphasizing that multiple formation channels (or multiple dynamical pathways within a single channel) likely contribute to the merger-driven GRB population.

\subsection{Correlation Between High-energy Properties and Physical Offsets}
\label{subsec:hep}

\begin{figure}
    \centering
    \includegraphics[width=\linewidth]{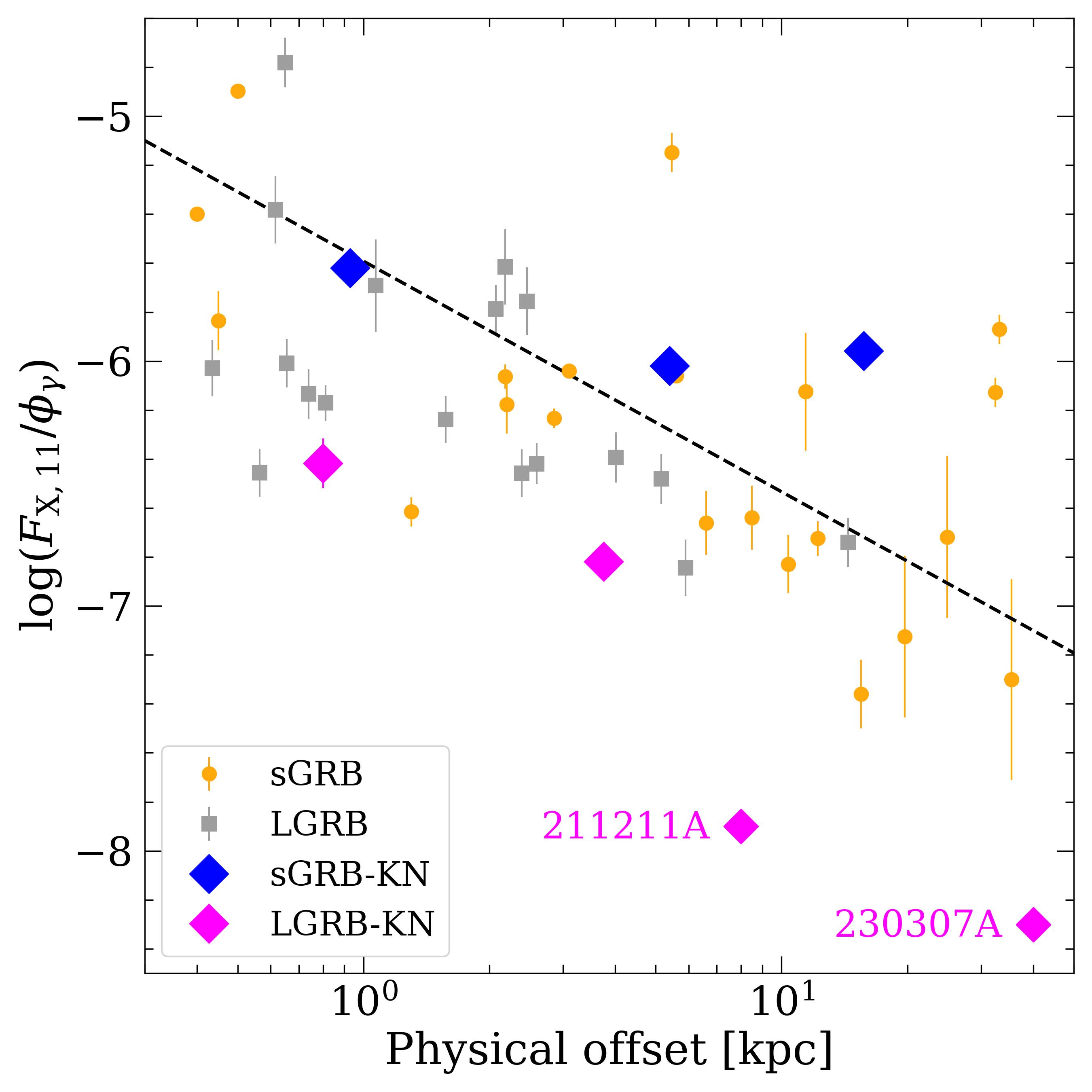}
    \caption{Ratio of $0.3$\,$-$\,$10$ keV X-ray flux at 11 hours, $F_{X,11}$, to the $15$\,$-$\,$150$ keV gamma-ray fluence, $\phi_\gamma$, versus the projected physical offset from the GRB host galaxy. We highlight the events in our sample, divided by their gamma-ray duration into sGRB-KNe (blue) and LGRB-KNe (magenta), and compare them to a population of short and long GRBs (orange and gray, respectively). We note that some events in our sample (GRBs 150101B and 170817A) are not shown as they lack X-ray data at 11 hours. The black line is shown to guide the eye. The figure is reproduced from \citet{OConnor2022,Yang2024}.
    }
    \label{fig:fluxfluence}
\end{figure}

\begin{figure*}
    \centering
    \includegraphics[width=\linewidth]{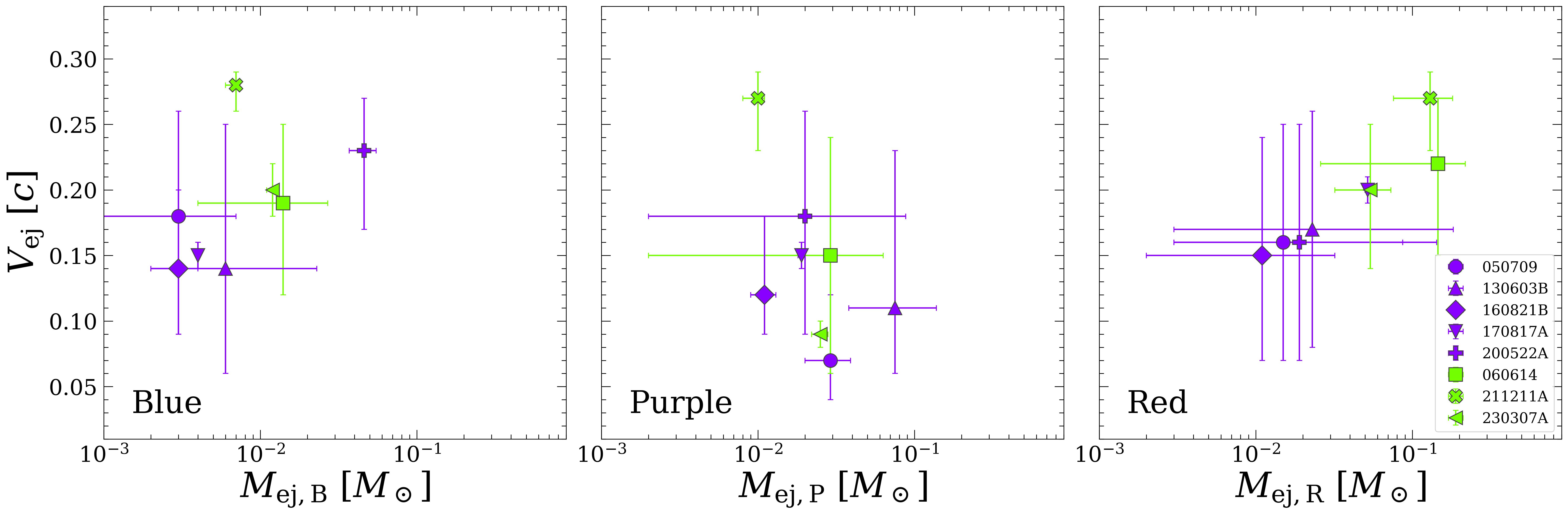}
    \caption{A comparison of the median $M_{ej}$ and $V_{ej}$ for the blue (left), purple (middle), and red (right) components of the GRB associated kilonova. Likely galaxy mergers or galaxies with visible tidal features are colored purple and non-merging galaxies are in green. The ejecta masses and velocities are gathered from \cite{Rastinejad2025}.
    }
    \label{fig:kn}
\end{figure*}

The density of the surrounding local environment (the circumburst environment) influences the brightness of the GRB afterglow \citep{Sari1998,Wijers1999,Granot2002}, see, e.g., \citet[][]{OConnor2020} for further discussion. As such, the afterglow brightness can potentially trace the circumburst environments of GRBs, which in the case of sGRBs and merger-driven events that experience natal kicks and long delay times is likely the interstellar medium (ISM). Therefore, a correlation may exist between the afterglow brightness and the physical offset of the GRB from the center of its host galaxy, which is a good tracer of ISM density \citep[for further discussion, see][]{OConnor2020}. 

A possible correlation between the X-ray flux at 11 hours $F_{X,11}$ (the afterglow brightness) and the prompt gamma-ray fluence $\phi_\gamma$ (the gamma-ray energy release) and the physical offset of short GRBs was presented by \citet{OConnor2022}. This ratio is commonly used to relate to the surrounding circumburst density \citep{Nysewander2009,Berger2014,OConnor2020}. While there is considerable scatter around the possible correlation \citep[see Figure \ref{fig:fluxfluence};][]{OConnor2022}, the recent LGRB-KN GRBs 211211A and 230307A clearly show a significant deviation from the rest of the population \citep[e.g.,][]{Troja2022,Levan2023,Yang2024}, which is likely related to their extremely large offsets from their host galaxies (8 and 40 kpc, respectively; \citealt{Rastinejad2022,Troja2022,Levan2023,Yang2024}) that could indicate lower circumburst environments \citep{OConnor2020}. We find that other LGRB-KNe (GRBs 050709 and 060614) also lie towards the bottom of the larger population of events, though consistent with the trend (Figure \ref{fig:fluxfluence}). On the other hand, sGRB-KNe are found to lie towards the top of the population. While this is clearly a small sample of events, further investigation of this trend is warranted as additional kilonova candidates are discovered and remains a useful pathway to identifying outliers in the long GRB population that require further investigation. 

\subsection{Comparison of Kilonova Properties}
\label{subsec:KN}

We evaluate the kilonova properties to determine if there are any trends among the host galaxies that are likely mergers and/or have visible tidal features or among the non-merging host galaxies. We veer away from comparing sGRB-KNe and LGRB-KNe to determine if within these subgroups, merging and non-merging galaxies, there is evidence for a difference in GRB progenitors. From \citealt{Rastinejad2025}, we gather the kilonova properties, particularly the ejecta masses and velocities, of our GRB-KN host sample. \citealt{Rastinejad2025} modeled the afterglows of eight GRBs with associated kilonova (excluding GRB 150101B) using a three-component model in \texttt{MOSFiT} (see \cite{Rastinejad2025} for further details). The three components are associated with ejecta: ``red" maps to tidal dynamical ejecta, ``purple" to disk wind ejecta, and ``blue" to dynamical ejecta shocked at the NS contact interface and ejected near the poles or ejected in a magnetized wind from the neutron star remnant prior to black hole formation as stated in \citealt{Rastinejad2025}. We present these properties in Figure \ref{fig:kn} separating into the blue (left panel), purple (middle panel), and red (right panel) components. Within the blue and purple components, there is a clear homogeneous spread among the hosts. For the red component, there is a clumping of all merging galaxies towards lower ejecta mass and velocities. Another study by \citealt{Singh2025} modeled the kilonova ejecta parameters and compared against \citealt{Rastinejad2025}. While GRB 170817A results were consistent, the total ejecta mass was lower for GRBs 160821B and 230307A, and even smaller for GRB 211211A, suggesting a similar behavior of lower ejecta masses for the interacting galaxies. With the extent of the errors, however, we cannot make a definitive claim that there is a distinction of ejecta mass and velocity between the two subgroups. 

\section{Conclusions} 
\label{sec:conclusions}

In this work, we performed a morphological analysis of nine host galaxies of GRBs with robust kilonova associations. We analyzed the available \textit{HST} and \textit{JWST} images of these GRB-KN hosts. We divided our GRB-KN host sample into subclasses based on their prompt gamma-ray durations with five events designated as sGRB-KN and four events as LGRB-KN. We additionally compiled and compared properties such as galactocentric offsets, stellar mass, sSFR, and redshift between these subclasses and against the overall classical sGRB and LGRB host populations. Our major findings are as follows:

\begin{itemize}
    \item Non-parametric measurements of galaxy structure such as the CAS, namely C and A, and the Gini-$\rm{M_{20}}$ statistics reveal that our GRB-KN host sample is diverse in morphology, spanning a range of galaxy types from ellipticals, spirals, and mergers. There is no morphological distinction between the two subclasses sGRB-KN and LGRB-KN hosts and against the sGRB and LGRB host populations. The diversity in morphology and merger environments suggests diverse progenitor formation pathways with varying delay times.
    \item The non-parametric morphological parameters are not mutually exclusive. The CAS statistics classify any major galaxy mergers given a high enough asymmetry measurement, while Gini-$\rm{M}_{20}$ is sensitive to both major and minor mergers. Given choice of filter, galaxy mass ratio, and point in the merger relaxation time, these statistics can fail to detect mergers. We find that for the host of GRB 170817A, it is classified as an early-type (elliptical) and a merger, in the CA and Gini-$\rm{M}_{20}$ parameter space, respectively. GRB 160821B likewise is classified as a major merger when evaluated in the $F606W$ band, and GRB 050709 is a merger based on its Gini-$\rm{M}_{20}$ parameters. Thus, three out of nine hosts are classified as mergers based on these statistics. We find that the population is also skewed towards more disturbed morphologies when compared with a generic PanSTARRS population, although the stellar mass range available from the latter sample limits the extent of the comparison.
    \item Five GRB-KN hosts possess visible tidal features and have likely undergone recent mergers. Similar to GRB 170817A, the recent merger history offers additional progenitor formation pathways driven by the dynamic nature of these merger environments. Interestingly, the kilonova location falls on top of such features. The probability of chance alignment for these GRBs with the tidal feature is $P \lesssim 0.1$.
    \item The stellar mass of our sample shows a distinction in the underlying mass distributions between sGRB-KN and LGRB-KN hosts. We find that LGRB-KN host galaxies are less massive than sGRB-KN hosts, with 75\% of our LGRB-KN hosts measured in the dwarf galaxy mass range of log($\rm{M}_\star / \rm{M_\odot} < 9$). The sGRB-KN hosts are from the same underlying galaxy distribution as the sGRB host population, as are the LGRB-KN hosts with the LGRB host population.
    \item The sSFR distributions show that the LGRB-KN subclass is consistent with being sampled from the same distribution as both the sGRB-KN and sGRB host populations. These exhibit a broader sSFR distribution than the LGRB host population, suggesting a diversity of delay times. The LGRB-KN hosts have lower sSFR than LGRB hosts.
    \item The physical and host-normalized offsets were the most notable distinction among our GRB-KN sample and the GRB populations. The LGRB-KNe have the largest median physical and host-normalized offsets, mainly due to GRB 230307A. Both the sGRB-KNe and sGRBs display a wider range of physical and host-normalized offsets, likely due to the diversity of delay times, as opposed to LGRBs which explode on shorter timescales, remaining near their birthplace. The LGRB-KN offsets are consistent with being drawn from the same underlying distribution of both the sGRB-KN and sGRB populations.
    \item Analysis of multi-Sérsic models reveals both over- and underestimation of host-normalized offsets when employing a single Sérsic approach. Considering which galactic component that is likely the home environment of the GRB is necessary to calculate a more precise host-normalized offset. This ultimately impacts the interpretation of the progenitor formation.
\end{itemize}

We conclude that both subclasses of our sample, sGRB-KNe and LGRB-KNe, are sampled from the same underlying distribution of galaxies and progenitors as the sGRB population. Though morphology cannot alone distinguish GRB-KN and GRB host populations, evaluating this alongside various galaxy properties can help elucidate these distinctions. Our findings support that LGRB-KNe are likely compact object merger-driven and share similar progenitors as sGRB-KNe and sGRBs. Moreover, our work strongly suggests that the majority of sGRBs are likely produced by compact object mergers and have associated kilonova but are not detected due to observational limitations, e.g., a limited sample observed by space-based facilities and with \textit{HST} limited to $z<0.5$. Now with state-of-the-art facilities such as the Rubin Observatory, \textit{JWST}, and the expected release of the Nancy Grace Roman Telescope, we can capture both GRB afterglows and kilonovae out to $z\sim1$ \citep{Chase2022,Kunnumkai:2024tuq,2026Kaur}, improving our capabilities, and hopefully allowing for an expanded sample of GRB-KNe in the future. 


\begin{acknowledgments}

We acknowledge useful discussions with Vicente Rodriguez-Gomez and thank them for sharing the \texttt{Statmorph} population data. We additionally want to thank the anonymous referee who carefully read our manuscript and provided insightful comments that improved this paper.

B.O. gratefully acknowledges support from the McWilliams Postdoctoral Fellowship in the McWilliams Center for Cosmology and Astrophysics at Carnegie Mellon University. A.P. is supported by NSF Grant No. 2308193.

This research is based on observations made with the NASA/ESA Hubble Space Telescope obtained from the Space Telescope Science Institute, which is operated by the Association of Universities for Research in Astronomy, Inc., under NASA contract NAS 5–26555. The HST data used in this work was obtained from the Mikulski Archive for Space Telescopes (MAST). STScI is operated by the Association of Universities for Research in Astronomy, Inc., under NASA contract NAS5-26555. These observations are associated with programs 10624, 10917, 12307, 13497, 13941, 14237, 14607, 15965, 16846, 16923, and 17298.

This work is based in part on observations made with the NASA/ESA/CSA James Webb Space Telescope. The data were obtained from the Mikulski Archive for Space Telescopes at the Space Telescope Science Institute, which is operated by the Association of Universities for Research in Astronomy, Inc., under NASA contract NAS 5-03127 for JWST. These observations are associated with program 4434.

The specific observations analyzed can be accessed via \dataset[doi:10.17909/y0py-4005]{https://doi.org/10.17909/y0py-4005}.

\end{acknowledgments}

%

\vspace{5mm}
\facilities{\textit{HST}, \textit{JWST}}


\software{\texttt{APLpy} \citep{RobitailleBressert2012}, \texttt{Astrodrizzle} \citep{Gonzaga2012}, \texttt{astropy} \citep{astropy:2013,astropy:2018,astropy:2022}, \texttt{GALFIT} \citep{Peng2002}, \texttt{Jupyter} \citep{2007CSE.....9c..21P,kluyver2016jupyter}, \texttt{matplotlib} \citep{Hunter:2007}, \texttt{morfometryka} \citep{Ferrari2015}, \texttt{numpy} \citep{numpy}, \texttt{pandas} \citep{mckinney-proc-scipy-2010,pandas_19340003}, \texttt{python} \citep{python}, \texttt{scipy} \citep{2020SciPy-NMeth,scipy_18736568}, \texttt{scikit-image} \citep{scikit-image}, \texttt{seaborn} \citep{Waskom2021}, \texttt{Source Extractor} \citep{Bertin1996}, \texttt{spike} \citep{Polzin2025}, and \texttt{statmorph} \citep{Rodriguez_Gomez_2018}. This research made use of Photutils, an Astropy package for detection and photometry of astronomical sources \citep{Photutils_17129028}.}

Software citation information aggregated using \texttt{\href{https://www.tomwagg.com/software-citation-station/}{The Software Citation Station}} \citep{software-citation-station-paper,software-citation-station-zenodo}.



\appendix

\section{Log of Observations} \label{app:obs}

We present the log of \textit{HST} and \textit{JWST} imaging for our sample of events (see \S \ref{sec:obs}) in Table \ref{tab:obslog}.

\begin{table*}
    \centering
    \caption{Log of \textit{HST} and \textit{JWST} imaging of the GRB-KN host galaxies in our sample. 
    }
\label{tab:obslog}
    \begin{tabular}{lcccccccc}
        \hline
        \hline \\[-2.5mm]
        \textbf{GRB} &\textbf{Obs. Date}  &   \textbf{Telescope} & \textbf{Filter} & \textbf{Exp.} & \textbf{ObsID} & \textbf{Prog.} &\textbf{PI}  \\
        & & \textbf{(UT)} & & & \textbf{(s)} & & & \\ [2mm]
        \hline
        
        050709 & 2005-08-13 16:31:05 & \textit{HST}/ACS/UVIS & $F814W$ & 2088 & J9H524040 & 10624 & Fox \\
        
        \hline
         
        060614 & 2010-10-08 21:36:37 & \textit{HST}/WFC3/IR & $F160W$ & 906 & IBJV71010 & 12307 & Levan \\	
        & 2006-09-08 3:36:45 & \textit{HST}/ACS/UVIS & $F606W$ & 1110 & J9R931060 & 10917 & Fox \\
        & 2006-11-01 0:15:00 & \textit{HST}/ACS/UVIS & $F814W$ & 1240 & J9H568040 & 10624 & Fox \\
        
        \hline
         
        130603B & 2013-07-03 5:33:03 & \textit{HST}/WFC3/IR & $F160W$ &	2612 & IC81W2010 & 13497 & Tanvir \\	
        & 2013-07-03 7:09:12 & \textit{HST}/ACS/UVIS & $F606W$ &	2216 & JC81A2010 & 13497 & Tanvir \\	
        
        \hline
        
        150101B & 2015-12-16 1:55:05 & \textit{HST}/WFC3/IR & $F160W$ & 2398 & ICS201010 & 13941 & Troja \\	
        & 2015-12-16 3:23:56 & \textit{HST}/WFC3/UVIS & $F606W$ & 2520 & ICS201020 & 13941 & Troja \\	
        
        \hline
        
        160821B & 2016-11-29 3:41:53 & \textit{HST}/WFC3/IR & $F110W$ & 5395 & ICY3X9010 & 14237 & Tanvir \\		
        & 2016-12-03 7:47:10 & \textit{HST}/WFC3/UVIS & $F606W$ & 2484 & ICY3XA010 & 14237 & Tanvir \\		
        & 2018-08-03 15:45:59 & \textit{HST}/WFC3/IR & $F160W$ & 2797 & ID7D04010 & 14607 & Troja \\	
        
        \hline
        
        170817A & 2017-12-08 20:33:09 & \textit{HST}/WFC3/IR & $F110W$ & 2412 & IDPM06010 & 15329 & Berger \\	
        & 2017-12-08 22:03:58 & \textit{HST}/WFC3/IR & $F110W$ & 2612 & IDPM06020 & 15329 & Berger \\	
        & 2017-12-08 23:39:19 & \textit{HST}/WFC3/IR & $F110W$ & 2612 & IDPM06030 & 15329 & Berger \\	
        & 2017-08-27 7:06:58 & \textit{HST}/WFC3/IR & $F160W$ & 1012 & IDPM01020 & 15329 & Berger \\
        & 2017-08-28 3:25:31 & \textit{HST}/WFC3/IR & $F160W$ & 298 & ID8CA2010 & 14771 & Tanvir \\
        & 2017-08-22 10:45:01 & \textit{HST}/WFC3/IR & $F160W$ & 298 & IDP7G2010 & 14804 & Levan \\
        & 2017-08-26 22:49:12 & \textit{HST}/WFC3/IR & $F160W$ & 298 & IDP7G4010 & 14804 & Levan \\
        & 2017-12-06 1:45:52 & \textit{HST}/WFC3/IR & $F160W$ & 2397 & IDFF04010 & 14270 & Levan \\
        & 2017-12-08 17:23:18 & \textit{HST}/WFC3/IR & $F160W$ & 2412 & IDP903010 & 15346 & Kasliwal \\
        & 2021-01-06 1:48:08 & \textit{HST}/WFC3/IR & $F160W$ & 5212 & IE5503010 & 15886 & Fong \\
        & 2021-01-06 4:58:58 & \textit{HST}/WFC3/IR & $F160W$ & 2409 & IE5503020 &  15886 & Fong \\
        & 2021-01-06 6:34:19 & \textit{HST}/WFC3/IR & $F160W$ & 203 & IE5503UIQ & 15886 & Fong \\
        & 2018-01-01 13:24:14 & \textit{HST}/ACS/UVIS & $F606W$ & 2120 & JDPM07010 & 15329 & Berger \\
        & 2018-03-23 21:07:38 & \textit{HST}/ACS/UVIS & $F606W$ & 2120 & JDPM08010 & 15329 & Berger \\	
        & 2018-07-20 8:12:51 & \textit{HST}/ACS/UVIS & $F606W$ & 2120 & JDPM09010 & 15329 & Berger \\	
        & 2019-03-21 17:38:22 & \textit{HST}/ACS/UVIS & $F606W$ & 6728 & JDWV01010 & 15606 & Margutti \tablenotemark{a}\\	
        & 2019-03-27 10:18:10 & \textit{HST}/ACS/UVIS & $F606W$ & 6728 & JDWV02010 & 15606 & Margutti \tablenotemark{b}{}\\
        
        \hline
        200522A & 2020-07-16 19:16:34 & \textit{HST}/WFC3/IR & $F125W$ & 2812 & IE0152020 & 15964 & Berger \\
        & 2020-07-16 19:16:34 & \textit{HST}/WFC3/IR & $F160W$ & 2812 & IE0103040 & 15964 & Berger \\	
        
        \hline
         
        211211A & 2022-04-21 17:08:42 & \textit{HST} /WFC3/IR & $F160W$ & 2412 & IERB06010 & 16846 & Troja \\		
        & 2022-04-02 15:48:05 & \textit{HST}/WFC3/UVIS & $F814W$ & 2160 & IERB04010 & 16846 & Troja \\
        & 2022-04-12 17:11:04 & \textit{HST}/WFC3/IR & $F140W$ & 2411 & IETK02010 & 16923 & Rastinejad \\	
        & 2022-04-14 1:20:01 & \textit{HST}/ACS/UVIS & $F606W$ & 2000 & JETK01010 & 16923 & Rastinejad \\		
        \hline
        
        230307A & 2023-05-02 10:56:08 & \textit{HST}/WFC3/IR & $F105W$ & 1509 & IF3T05020 & 17298 & Troja \\
        & 2023-05-02 10:32:47 & \textit{HST}/WFC3/IR & $F140W$ & 1209 & IF3T05010 & 17298 & Troja \\
        & 2023-04-05 13:03:08 & \textit{JWST}/NIRCAM & $F070W$ & 1868 & 128207458 & 4434 & Levan \\
        & 2023-04-05 13:03:08 & \textit{JWST}/NIRCAM & $F115W$ & 1868 & 128207458 & 4434 & Levan \\
        & 2023-04-05 13:03:08 & \textit{JWST}/NIRCAM & $F150W$ & 1868 & 128207458 & 4434 & Levan \\
        & 2023-04-05 13:03:08 & \textit{JWST}/NIRCAM & $F277W$ & 1868 & 128207458 & 4434 & Levan \\
        & 2023-04-05 13:03:08 & \textit{JWST}/NIRCAM & $F356W$ & 1868 & 128207458 & 4434 & Levan \\
        & 2023-04-05 13:03:08 & \textit{JWST}/NIRCAM & $F444W$ & 1868 & 128207458 & 4434 & Levan \\
        
        \hline
  \end{tabular}
  \tablenotetext{a}{Only used one exposure in observation program.}
  \tablenotetext{b}{Used 12 out of 24 exposures in observation program.}
\end{table*}

\section{\texttt{GALFIT} RESULTS} \label{app:galfit}

\begin{figure*}
\centering
\figsetstart
\figsetnum{10}
\figsettitle{GALFIT results}

\figsetgrpstart
\figsetgrpnum{10.1}
\figsetgrptitle{GRB 050709, F814W}
\figsetplot{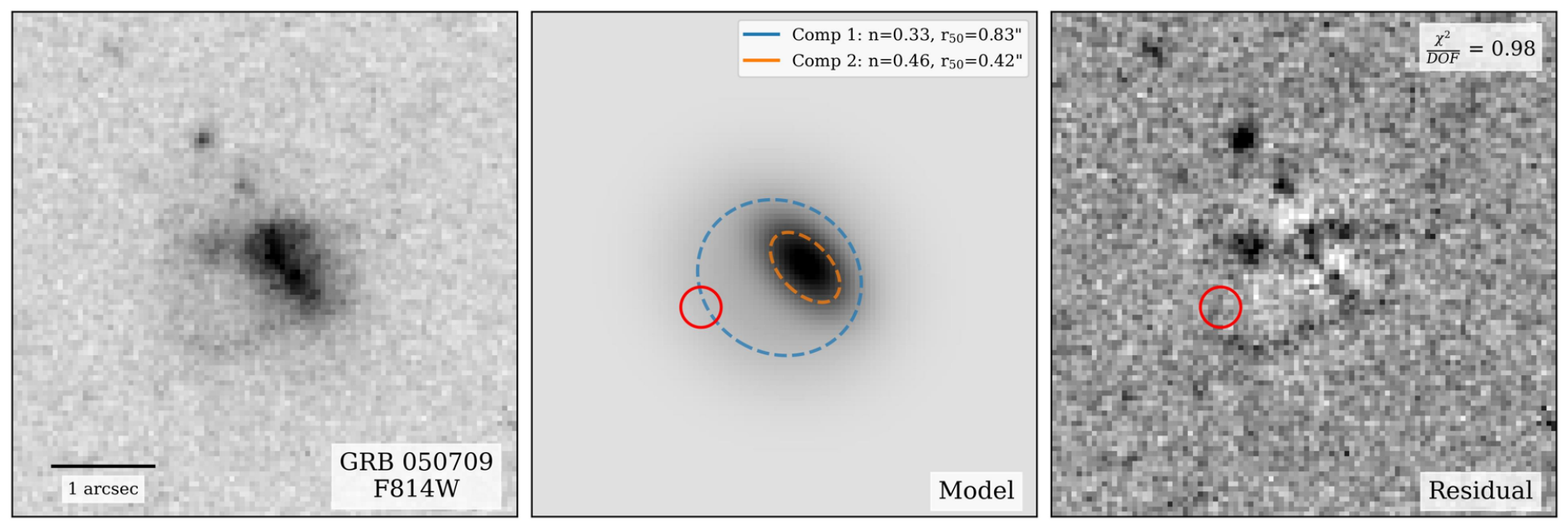}
\figsetgrpnote{\texttt{GALFIT} results. \textit{Left panel:} original data; \textit{Middle panel:} \texttt{GALFIT} model of best-fitting S\'ersic light profiles; \textit{Right panel:} the residual image of \texttt{GALFIT} after image subtraction. The S\'ersic profile components are marked with dotted lines with the approximate location of transient indicated by the red circle. GRB 230307A falls outside the cutout and is located in the direction of the red arrow. The complete figure set (28 images) is available in the online journal.}
\figsetgrpend

\figsetgrpstart
\figsetgrpnum{10.2}
\figsetgrptitle{GRB 060614, F160W}
\figsetplot{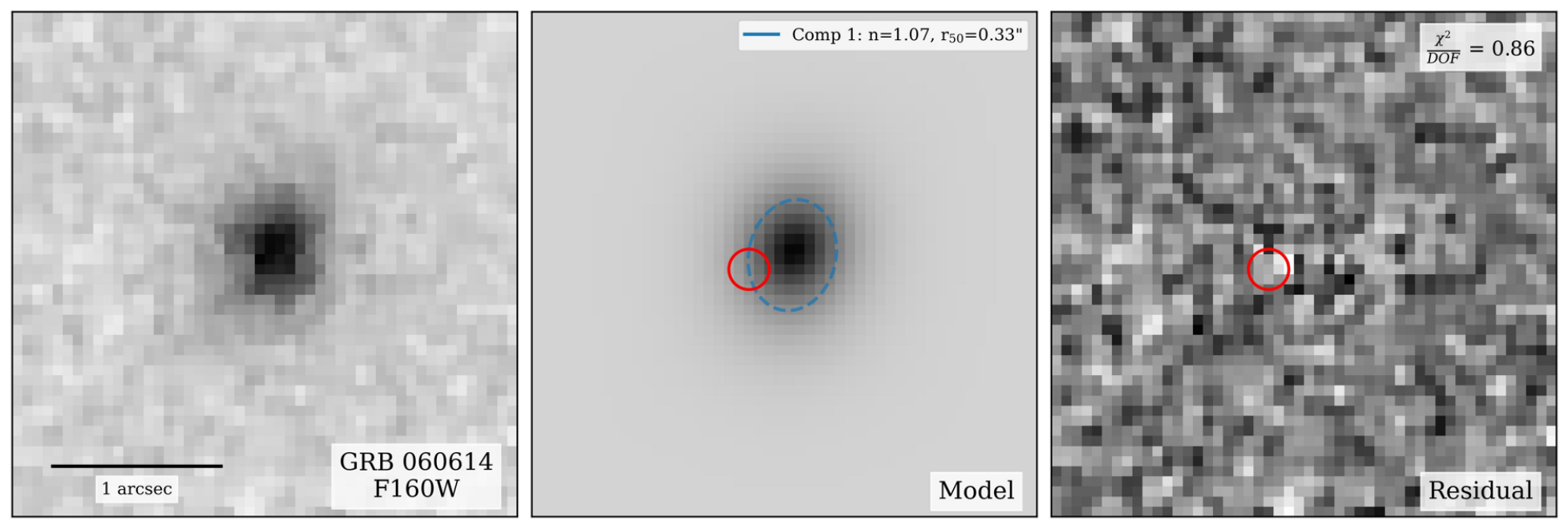}
\figsetgrpnote{\texttt{GALFIT} results. \textit{Left panel:} original data; \textit{Middle panel:} \texttt{GALFIT} model of best-fitting S\'ersic light profiles; \textit{Right panel:} the residual image of \texttt{GALFIT} after image subtraction. The S\'ersic profile components are marked with dotted lines with the approximate location of transient indicated by the red circle. GRB 230307A falls outside the cutout and is located in the direction of the red arrow.}
\figsetgrpend

\figsetgrpstart
\figsetgrpnum{10.3}
\figsetgrptitle{GRB 060614, F606W}
\figsetplot{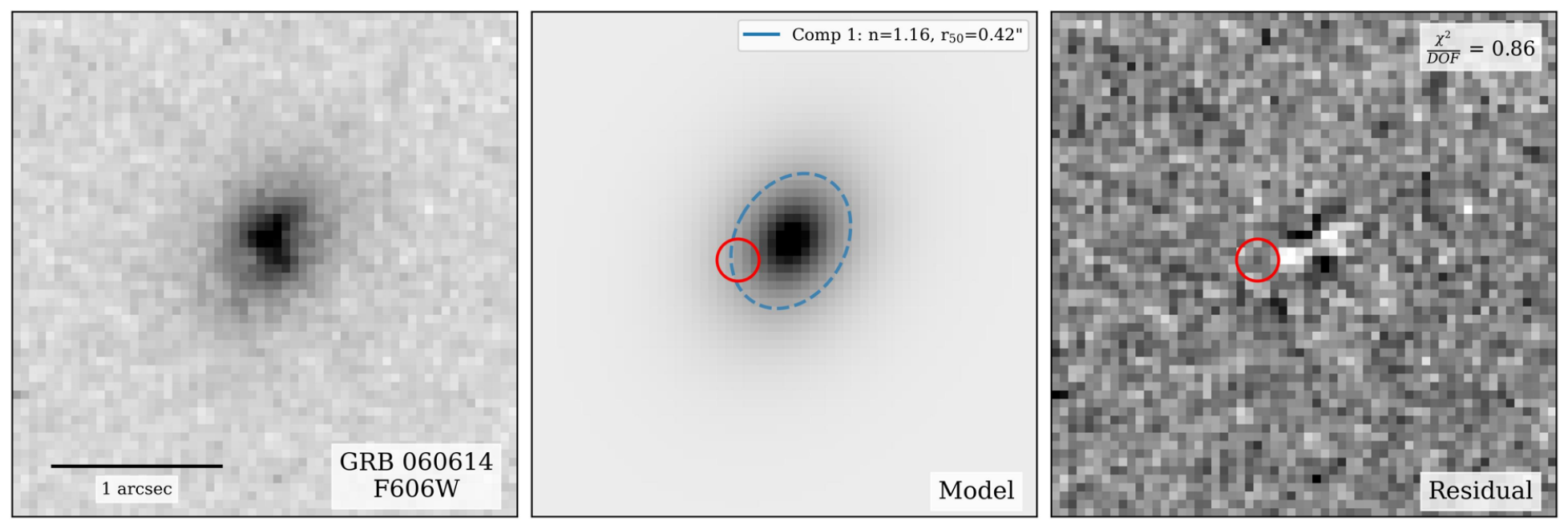}
\figsetgrpnote{\texttt{GALFIT} results. \textit{Left panel:} original data; \textit{Middle panel:} \texttt{GALFIT} model of best-fitting S\'ersic light profiles; \textit{Right panel:} the residual image of \texttt{GALFIT} after image subtraction. The S\'ersic profile components are marked with dotted lines with the approximate location of transient indicated by the red circle. GRB 230307A falls outside the cutout and is located in the direction of the red arrow.}
\figsetgrpend

\figsetgrpstart
\figsetgrpnum{10.4}
\figsetgrptitle{GRB 060614, F814W}
\figsetplot{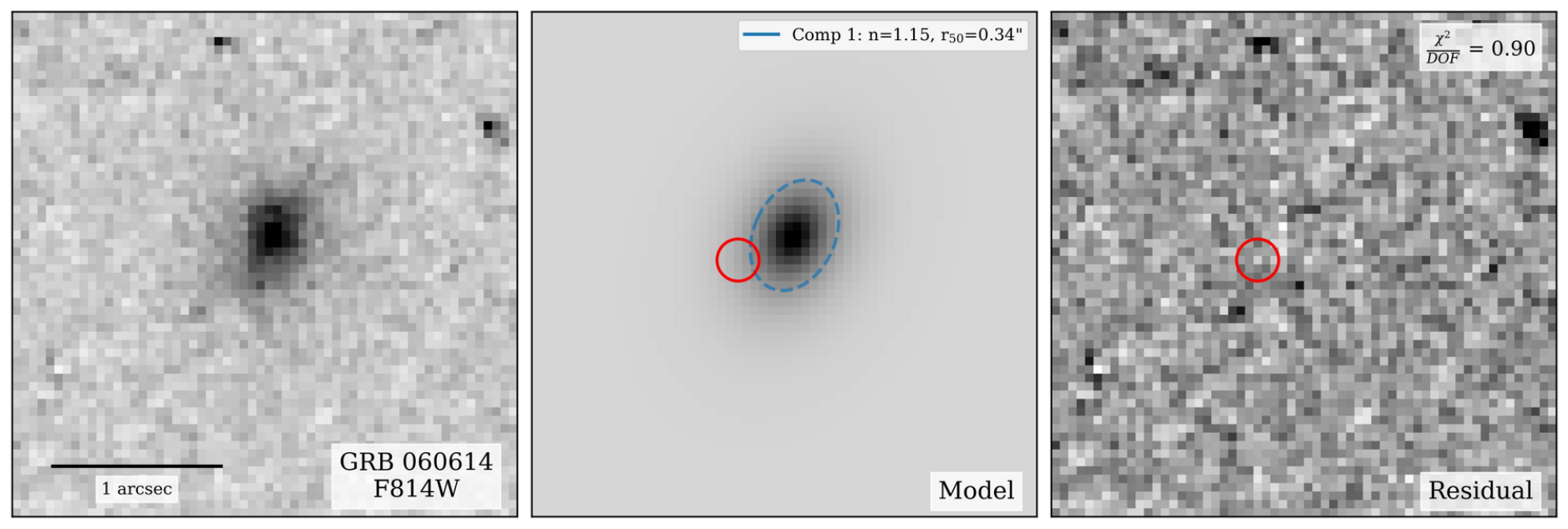}
\figsetgrpnote{\texttt{GALFIT} results. \textit{Left panel:} original data; \textit{Middle panel:} \texttt{GALFIT} model of best-fitting S\'ersic light profiles; \textit{Right panel:} the residual image of \texttt{GALFIT} after image subtraction. The S\'ersic profile components are marked with dotted lines with the approximate location of transient indicated by the red circle. GRB 230307A falls outside the cutout and is located in the direction of the red arrow.}
\figsetgrpend

\figsetgrpstart
\figsetgrpnum{10.5}
\figsetgrptitle{GRB 130603B, F160W}
\figsetplot{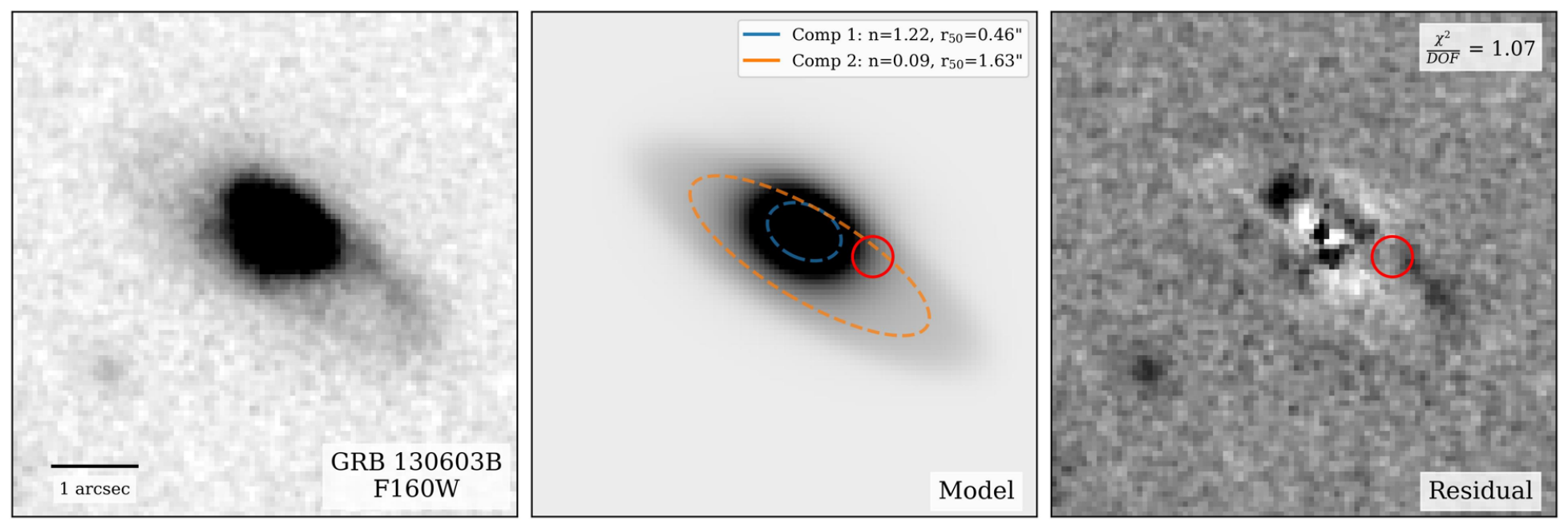}
\figsetgrpnote{\texttt{GALFIT} results. \textit{Left panel:} original data; \textit{Middle panel:} \texttt{GALFIT} model of best-fitting S\'ersic light profiles; \textit{Right panel:} the residual image of \texttt{GALFIT} after image subtraction. The S\'ersic profile components are marked with dotted lines with the approximate location of transient indicated by the red circle. GRB 230307A falls outside the cutout and is located in the direction of the red arrow.}
\figsetgrpend

\figsetgrpstart
\figsetgrpnum{10.6}
\figsetgrptitle{GRB 130603B, F606W}
\figsetplot{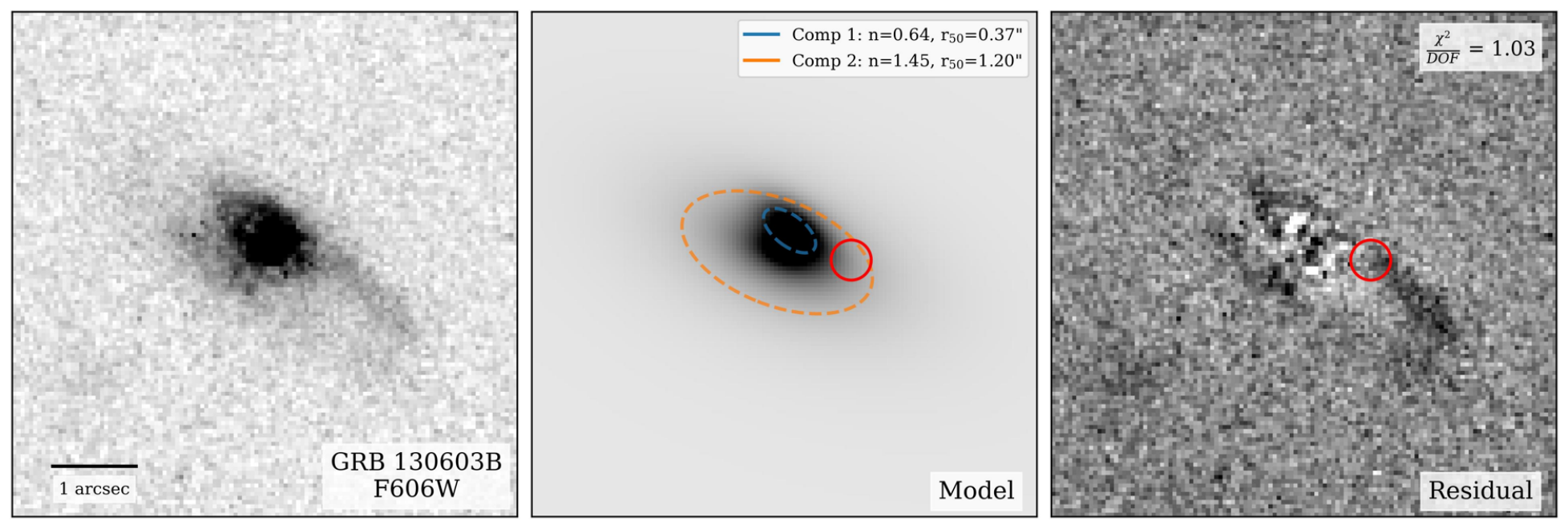}
\figsetgrpnote{\texttt{GALFIT} results. \textit{Left panel:} original data; \textit{Middle panel:} \texttt{GALFIT} model of best-fitting S\'ersic light profiles; \textit{Right panel:} the residual image of \texttt{GALFIT} after image subtraction. The S\'ersic profile components are marked with dotted lines with the approximate location of transient indicated by the red circle. GRB 230307A falls outside the cutout and is located in the direction of the red arrow.}
\figsetgrpend

\figsetgrpstart
\figsetgrpnum{10.7}
\figsetgrptitle{GRB 150101B, F160W}
\figsetplot{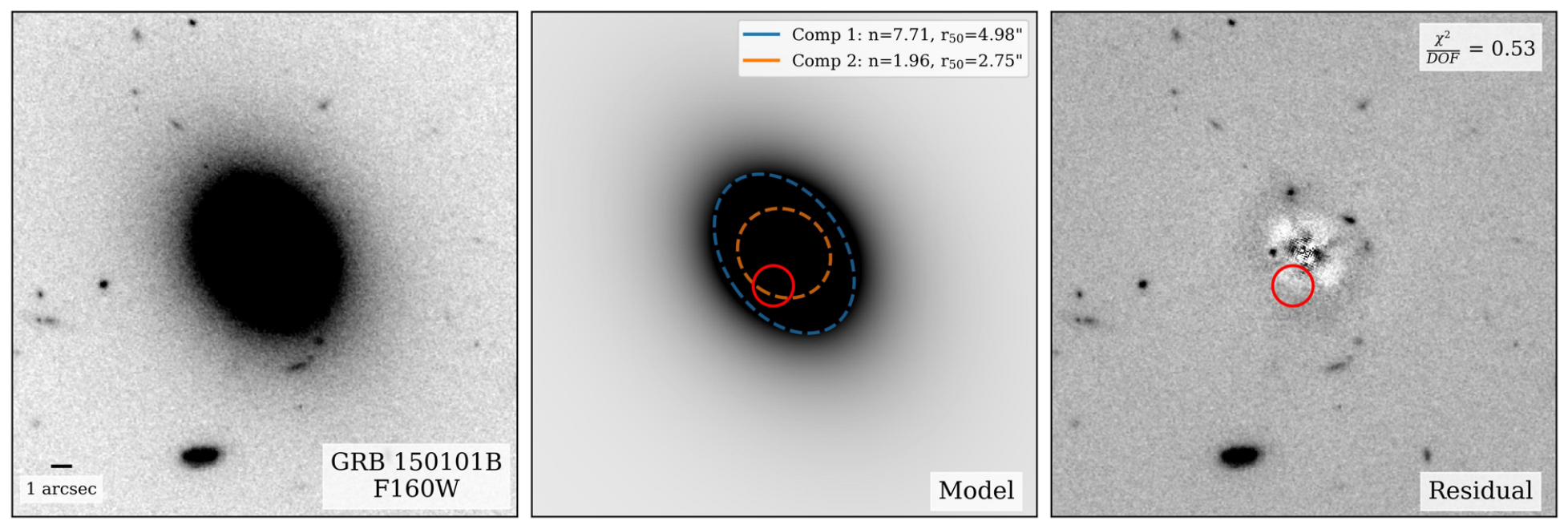}
\figsetgrpnote{\texttt{GALFIT} results. \textit{Left panel:} original data; \textit{Middle panel:} \texttt{GALFIT} model of best-fitting S\'ersic light profiles; \textit{Right panel:} the residual image of \texttt{GALFIT} after image subtraction. The S\'ersic profile components are marked with dotted lines with the approximate location of transient indicated by the red circle. GRB 230307A falls outside the cutout and is located in the direction of the red arrow.}
\figsetgrpend

\figsetgrpstart
\figsetgrpnum{10.8}
\figsetgrptitle{GRB 150101B, F606W}
\figsetplot{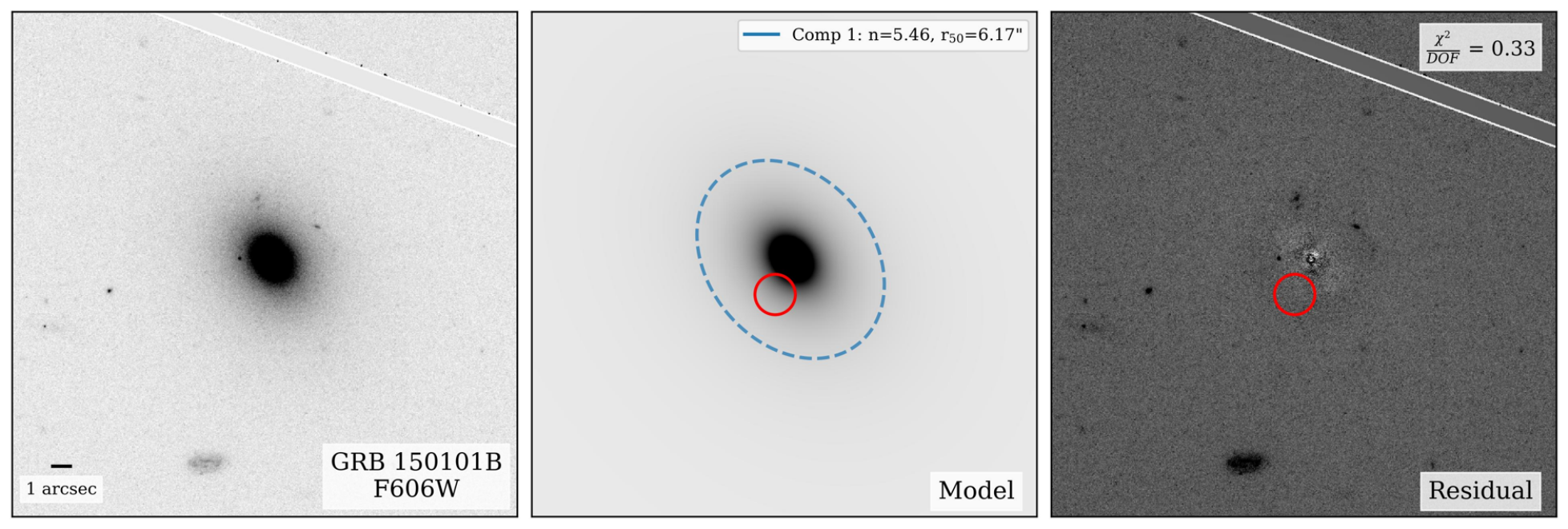}
\figsetgrpnote{\texttt{GALFIT} results. \textit{Left panel:} original data; \textit{Middle panel:} \texttt{GALFIT} model of best-fitting S\'ersic light profiles; \textit{Right panel:} the residual image of \texttt{GALFIT} after image subtraction. The S\'ersic profile components are marked with dotted lines with the approximate location of transient indicated by the red circle. GRB 230307A falls outside the cutout and is located in the direction of the red arrow.}
\figsetgrpend

\figsetgrpstart
\figsetgrpnum{10.9}
\figsetgrptitle{GRB 160821B, F110W}
\figsetplot{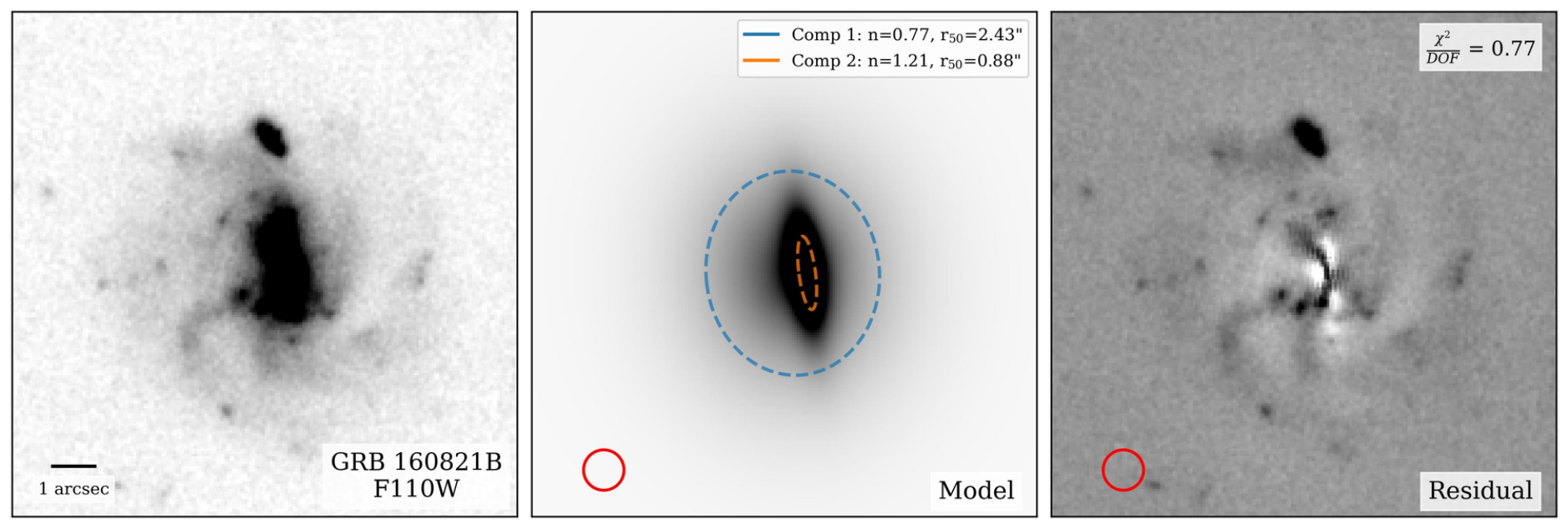}
\figsetgrpnote{\texttt{GALFIT} results. \textit{Left panel:} original data; \textit{Middle panel:} \texttt{GALFIT} model of best-fitting S\'ersic light profiles; \textit{Right panel:} the residual image of \texttt{GALFIT} after image subtraction. The S\'ersic profile components are marked with dotted lines with the approximate location of transient indicated by the red circle. GRB 230307A falls outside the cutout and is located in the direction of the red arrow.}
\figsetgrpend

\figsetgrpstart
\figsetgrpnum{10.10}
\figsetgrptitle{GRB 160821B, F160W}
\figsetplot{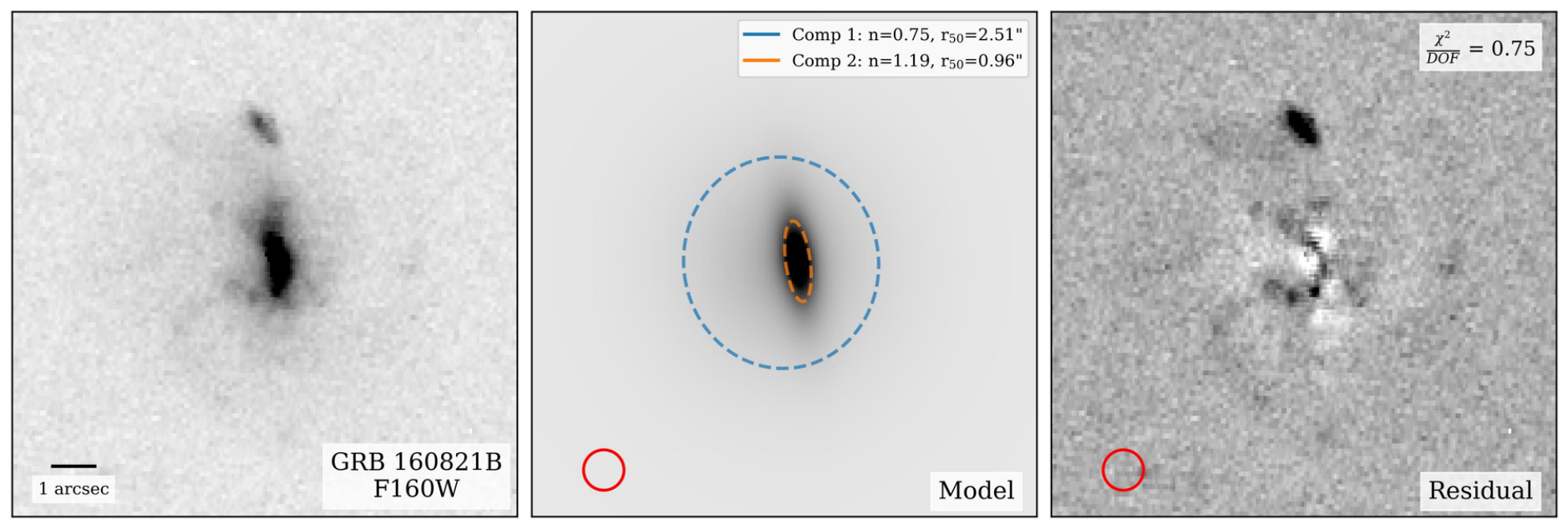}
\figsetgrpnote{\texttt{GALFIT} results. \textit{Left panel:} original data; \textit{Middle panel:} \texttt{GALFIT} model of best-fitting S\'ersic light profiles; \textit{Right panel:} the residual image of \texttt{GALFIT} after image subtraction. The S\'ersic profile components are marked with dotted lines with the approximate location of transient indicated by the red circle. GRB 230307A falls outside the cutout and is located in the direction of the red arrow.}
\figsetgrpend

\figsetgrpstart
\figsetgrpnum{10.11}
\figsetgrptitle{GRB 160821B, F606W}
\figsetplot{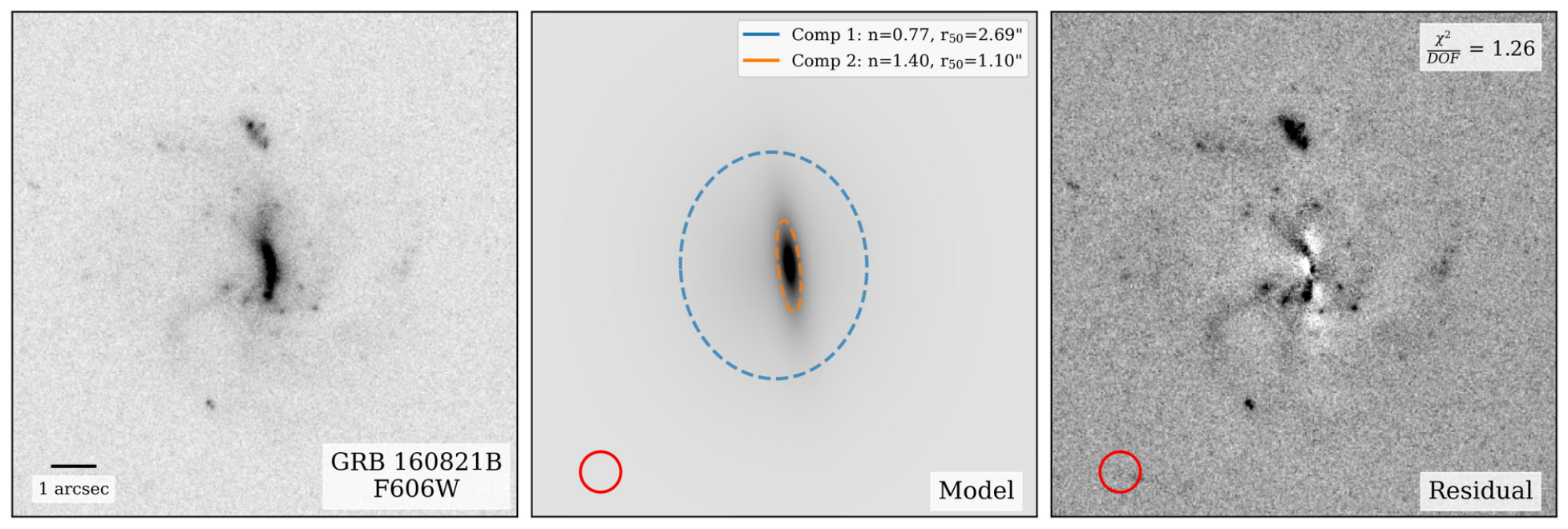}
\figsetgrpnote{\texttt{GALFIT} results. \textit{Left panel:} original data; \textit{Middle panel:} \texttt{GALFIT} model of best-fitting S\'ersic light profiles; \textit{Right panel:} the residual image of \texttt{GALFIT} after image subtraction. The S\'ersic profile components are marked with dotted lines with the approximate location of transient indicated by the red circle. GRB 230307A falls outside the cutout and is located in the direction of the red arrow.}
\figsetgrpend

\figsetgrpstart
\figsetgrpnum{10.12}
\figsetgrptitle{GRB 170817A, F110W}
\figsetplot{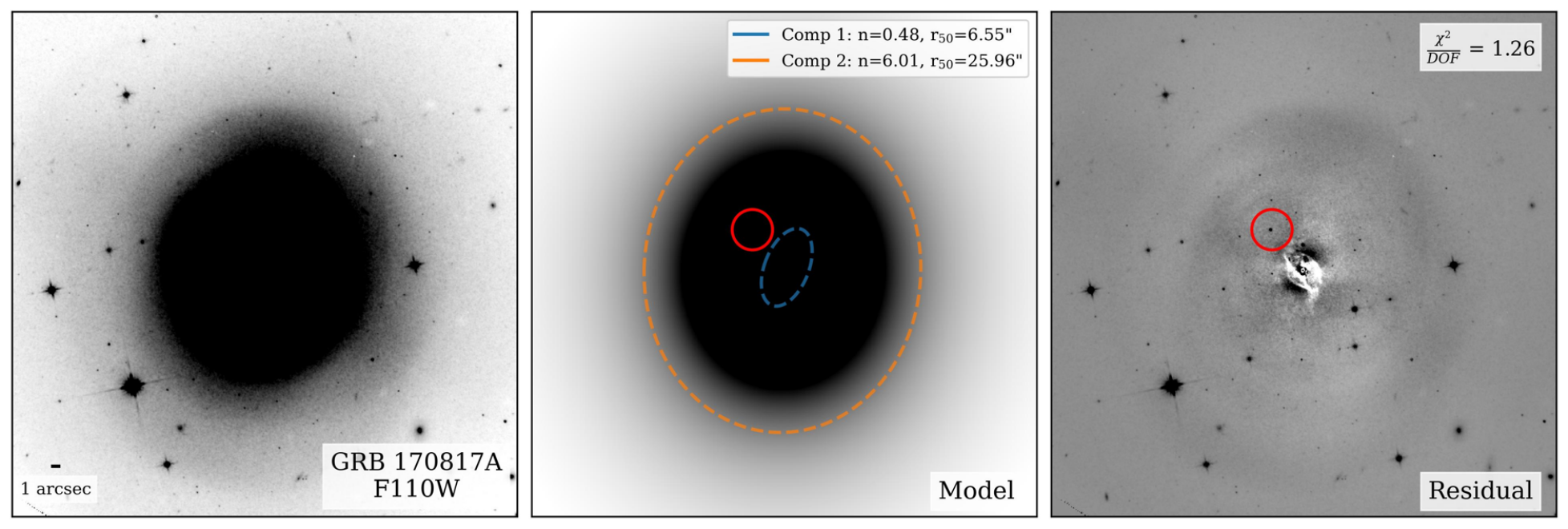}
\figsetgrpnote{\texttt{GALFIT} results. \textit{Left panel:} original data; \textit{Middle panel:} \texttt{GALFIT} model of best-fitting S\'ersic light profiles; \textit{Right panel:} the residual image of \texttt{GALFIT} after image subtraction. The S\'ersic profile components are marked with dotted lines with the approximate location of transient indicated by the red circle. GRB 230307A falls outside the cutout and is located in the direction of the red arrow.}
\figsetgrpend

\figsetgrpstart
\figsetgrpnum{10.13}
\figsetgrptitle{GRB 170817A, F160W}
\figsetplot{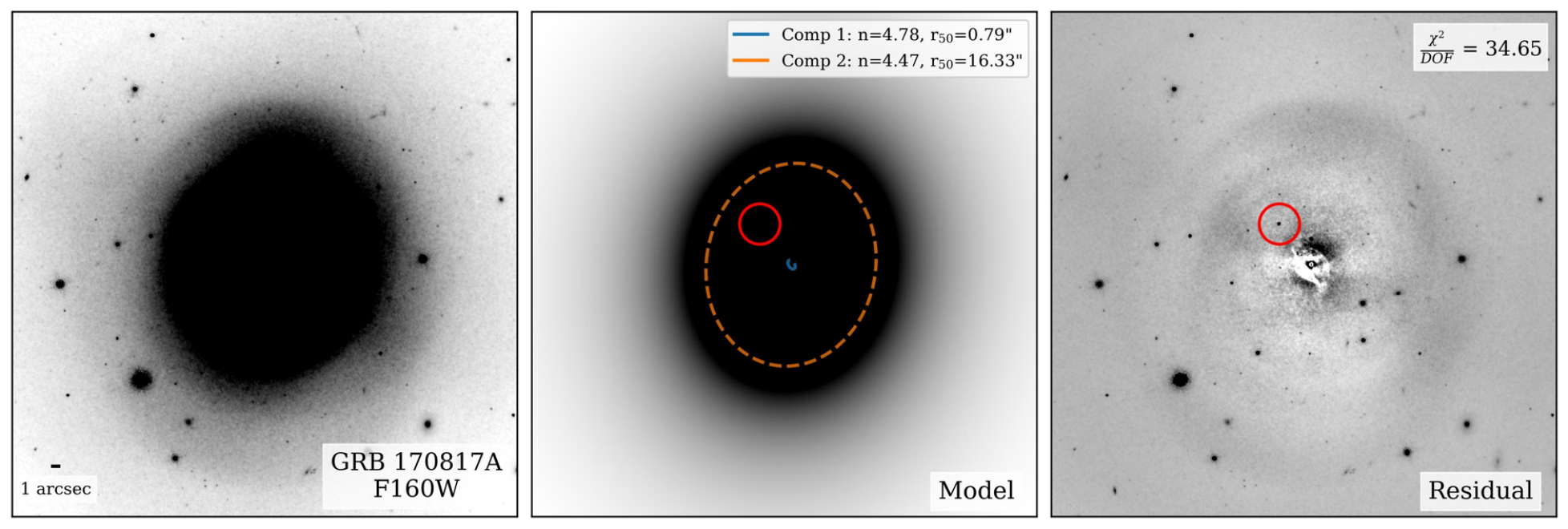}
\figsetgrpnote{\texttt{GALFIT} results. \textit{Left panel:} original data; \textit{Middle panel:} \texttt{GALFIT} model of best-fitting S\'ersic light profiles; \textit{Right panel:} the residual image of \texttt{GALFIT} after image subtraction. The S\'ersic profile components are marked with dotted lines with the approximate location of transient indicated by the red circle. GRB 230307A falls outside the cutout and is located in the direction of the red arrow.}
\figsetgrpend

\figsetgrpstart
\figsetgrpnum{10.14}
\figsetgrptitle{GRB 170817A, F606W}
\figsetplot{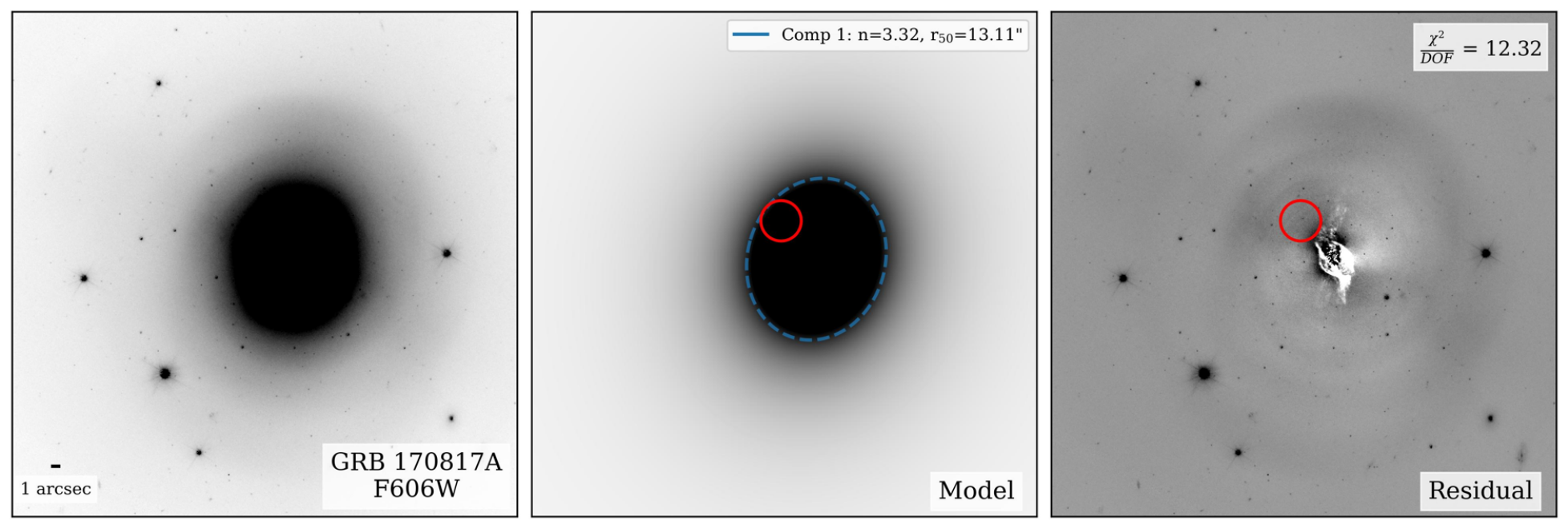}
\figsetgrpnote{\texttt{GALFIT} results. \textit{Left panel:} original data; \textit{Middle panel:} \texttt{GALFIT} model of best-fitting S\'ersic light profiles; \textit{Right panel:} the residual image of \texttt{GALFIT} after image subtraction. The S\'ersic profile components are marked with dotted lines with the approximate location of transient indicated by the red circle. GRB 230307A falls outside the cutout and is located in the direction of the red arrow.}
\figsetgrpend

\figsetgrpstart
\figsetgrpnum{10.15}
\figsetgrptitle{GRB 200522A, F125W}
\figsetplot{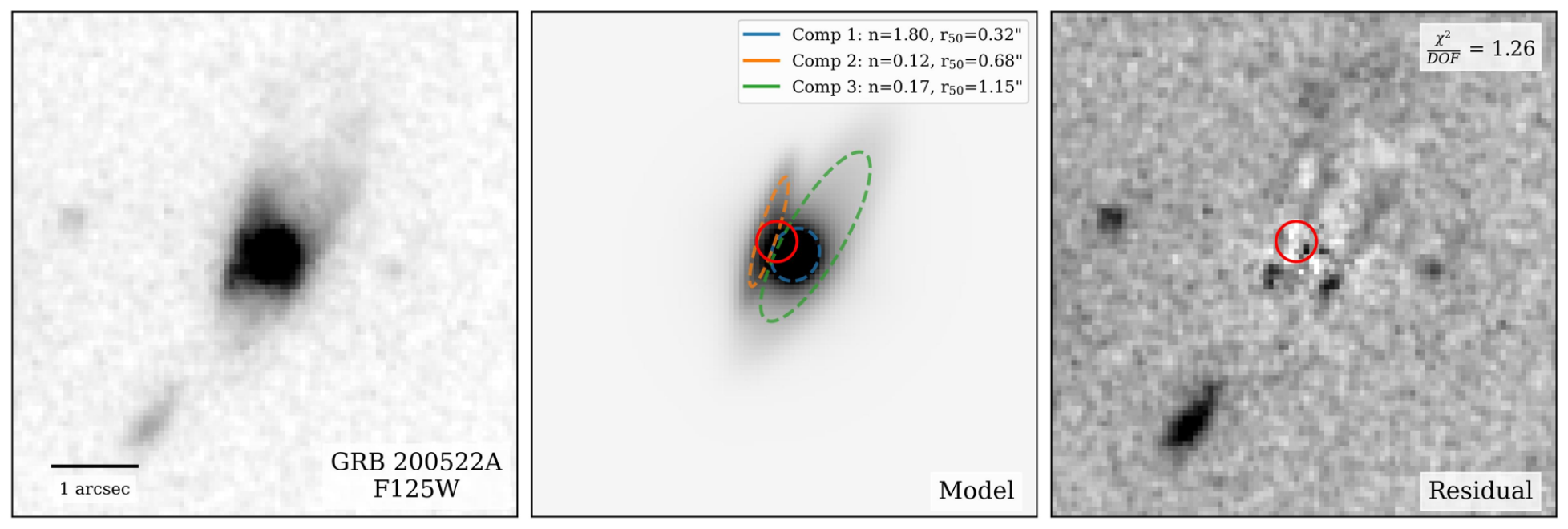}
\figsetgrpnote{\texttt{GALFIT} results. \textit{Left panel:} original data; \textit{Middle panel:} \texttt{GALFIT} model of best-fitting S\'ersic light profiles; \textit{Right panel:} the residual image of \texttt{GALFIT} after image subtraction. The S\'ersic profile components are marked with dotted lines with the approximate location of transient indicated by the red circle. GRB 230307A falls outside the cutout and is located in the direction of the red arrow.}
\figsetgrpend

\figsetgrpstart
\figsetgrpnum{10.16}
\figsetgrptitle{GRB 200522A, F160W}
\figsetplot{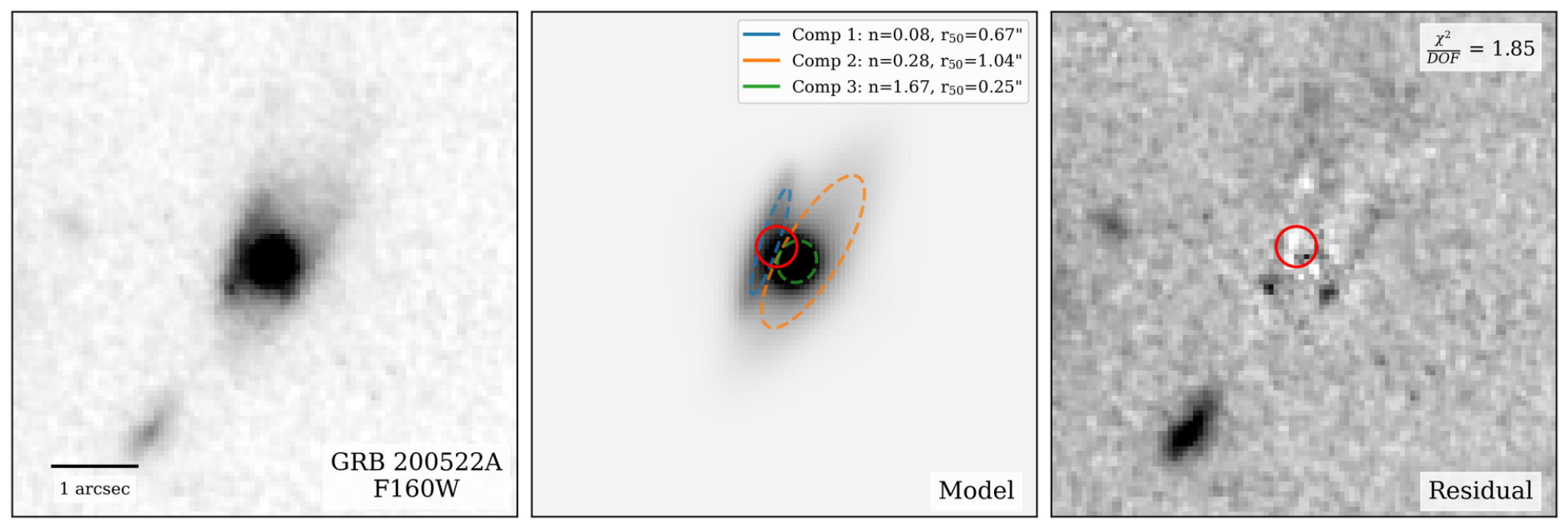}
\figsetgrpnote{\texttt{GALFIT} results. \textit{Left panel:} original data; \textit{Middle panel:} \texttt{GALFIT} model of best-fitting S\'ersic light profiles; \textit{Right panel:} the residual image of \texttt{GALFIT} after image subtraction. The S\'ersic profile components are marked with dotted lines with the approximate location of transient indicated by the red circle. GRB 230307A falls outside the cutout and is located in the direction of the red arrow.}
\figsetgrpend

\figsetgrpstart
\figsetgrpnum{10.17}
\figsetgrptitle{GRB 211211A, F140W}
\figsetplot{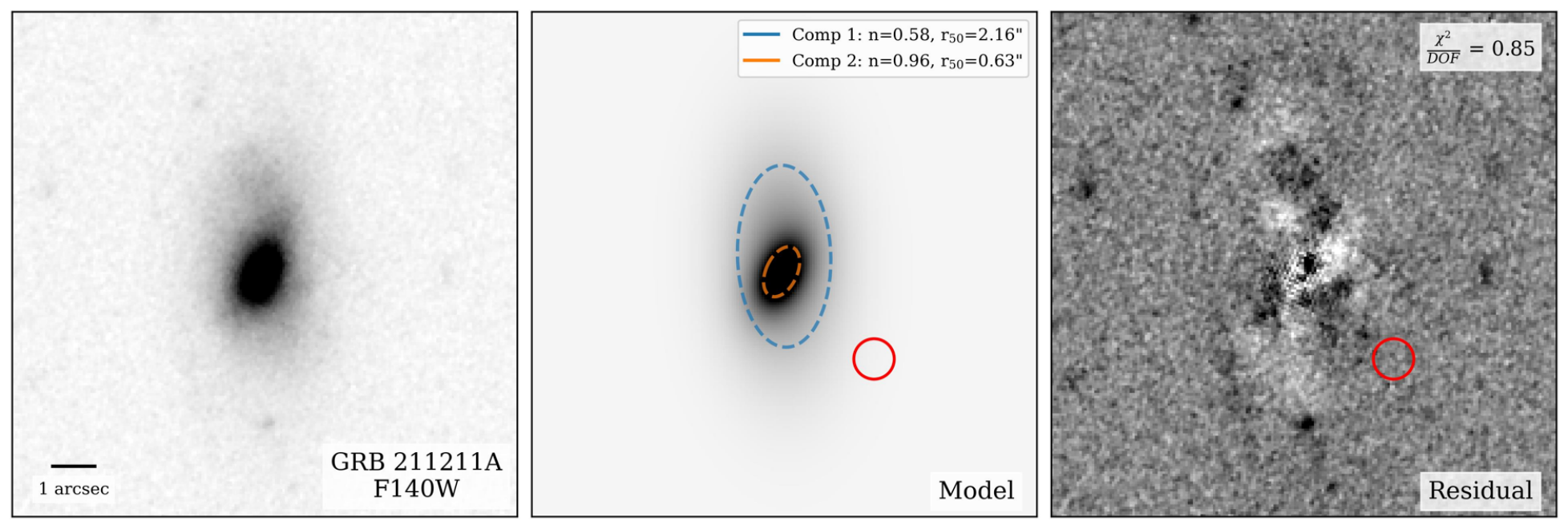}
\figsetgrpnote{\texttt{GALFIT} results. \textit{Left panel:} original data; \textit{Middle panel:} \texttt{GALFIT} model of best-fitting S\'ersic light profiles; \textit{Right panel:} the residual image of \texttt{GALFIT} after image subtraction. The S\'ersic profile components are marked with dotted lines with the approximate location of transient indicated by the red circle. GRB 230307A falls outside the cutout and is located in the direction of the red arrow.}
\figsetgrpend

\figsetgrpstart
\figsetgrpnum{10.18}
\figsetgrptitle{GRB 211211A, F160W}
\figsetplot{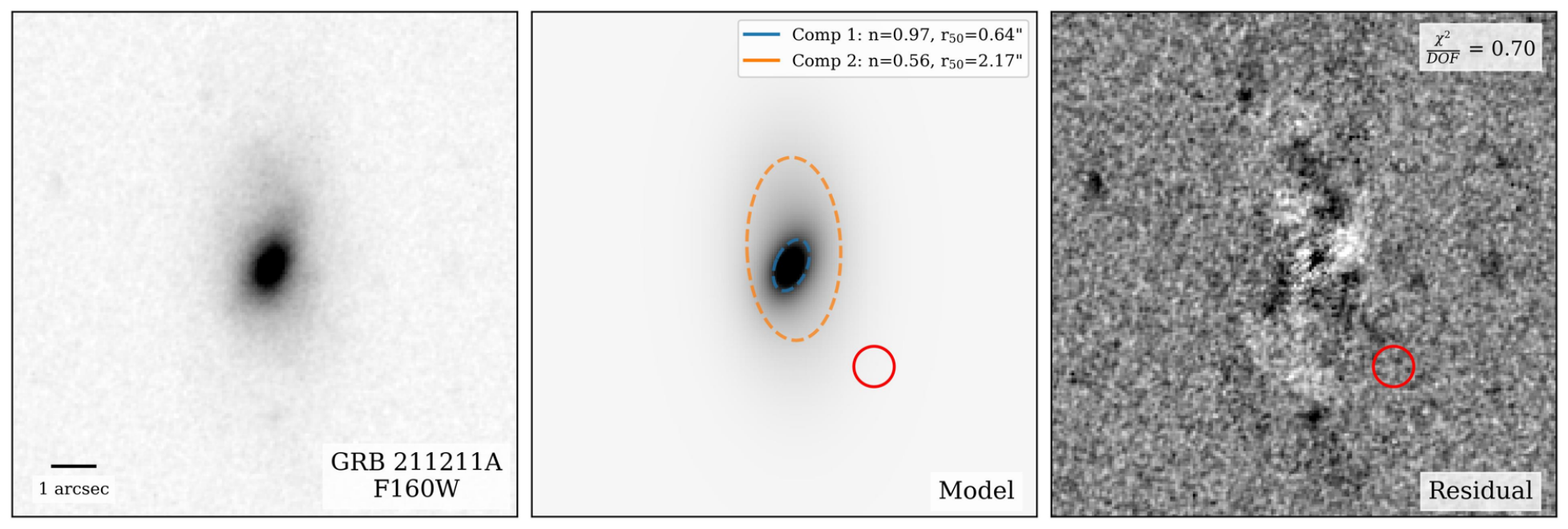}
\figsetgrpnote{\texttt{GALFIT} results. \textit{Left panel:} original data; \textit{Middle panel:} \texttt{GALFIT} model of best-fitting S\'ersic light profiles; \textit{Right panel:} the residual image of \texttt{GALFIT} after image subtraction. The S\'ersic profile components are marked with dotted lines with the approximate location of transient indicated by the red circle. GRB 230307A falls outside the cutout and is located in the direction of the red arrow.}
\figsetgrpend

\figsetgrpstart
\figsetgrpnum{10.19}
\figsetgrptitle{GRB 211211A, F606W}
\figsetplot{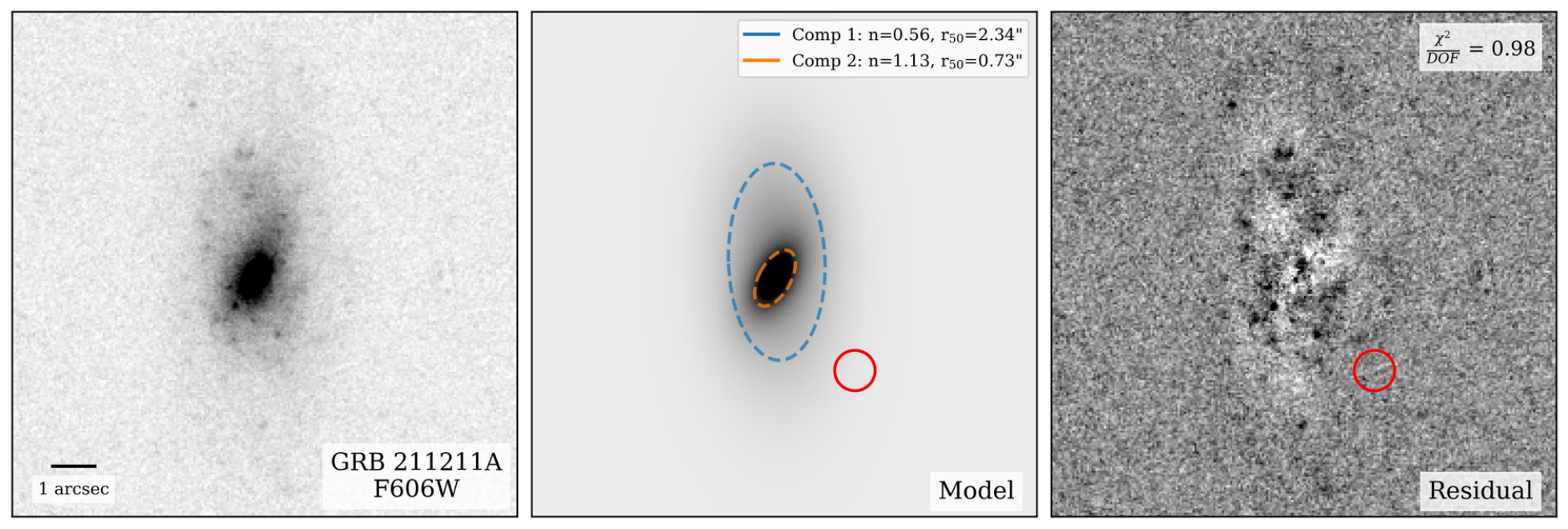}
\figsetgrpnote{\texttt{GALFIT} results. \textit{Left panel:} original data; \textit{Middle panel:} \texttt{GALFIT} model of best-fitting S\'ersic light profiles; \textit{Right panel:} the residual image of \texttt{GALFIT} after image subtraction. The S\'ersic profile components are marked with dotted lines with the approximate location of transient indicated by the red circle. GRB 230307A falls outside the cutout and is located in the direction of the red arrow.}
\figsetgrpend

\figsetgrpstart
\figsetgrpnum{10.20}
\figsetgrptitle{GRB 211211A, F814W}
\figsetplot{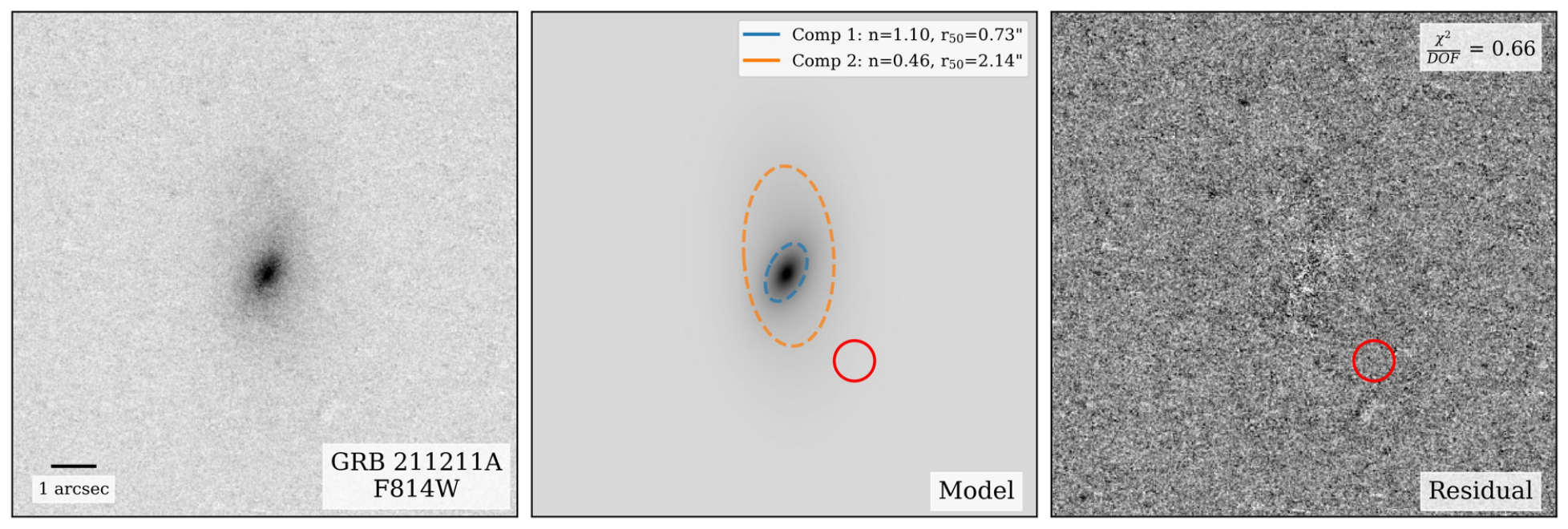}
\figsetgrpnote{\texttt{GALFIT} results. \textit{Left panel:} original data; \textit{Middle panel:} \texttt{GALFIT} model of best-fitting S\'ersic light profiles; \textit{Right panel:} the residual image of \texttt{GALFIT} after image subtraction. The S\'ersic profile components are marked with dotted lines with the approximate location of transient indicated by the red circle. GRB 230307A falls outside the cutout and is located in the direction of the red arrow.}
\figsetgrpend

\figsetgrpstart
\figsetgrpnum{10.21}
\figsetgrptitle{GRB 230307A, F105W}
\figsetplot{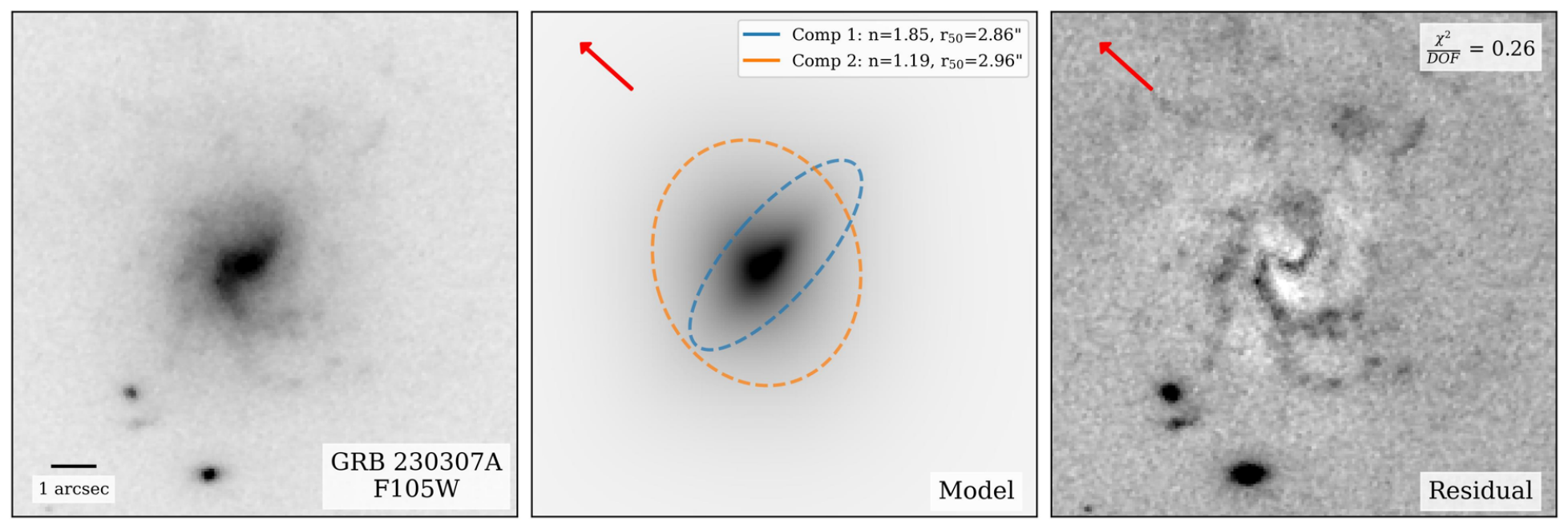}
\figsetgrpnote{\texttt{GALFIT} results. \textit{Left panel:} original data; \textit{Middle panel:} \texttt{GALFIT} model of best-fitting S\'ersic light profiles; \textit{Right panel:} the residual image of \texttt{GALFIT} after image subtraction. The S\'ersic profile components are marked with dotted lines with the approximate location of transient indicated by the red circle. GRB 230307A falls outside the cutout and is located in the direction of the red arrow.}
\figsetgrpend

\figsetgrpstart
\figsetgrpnum{10.22}
\figsetgrptitle{GRB 230307A, F140W}
\figsetplot{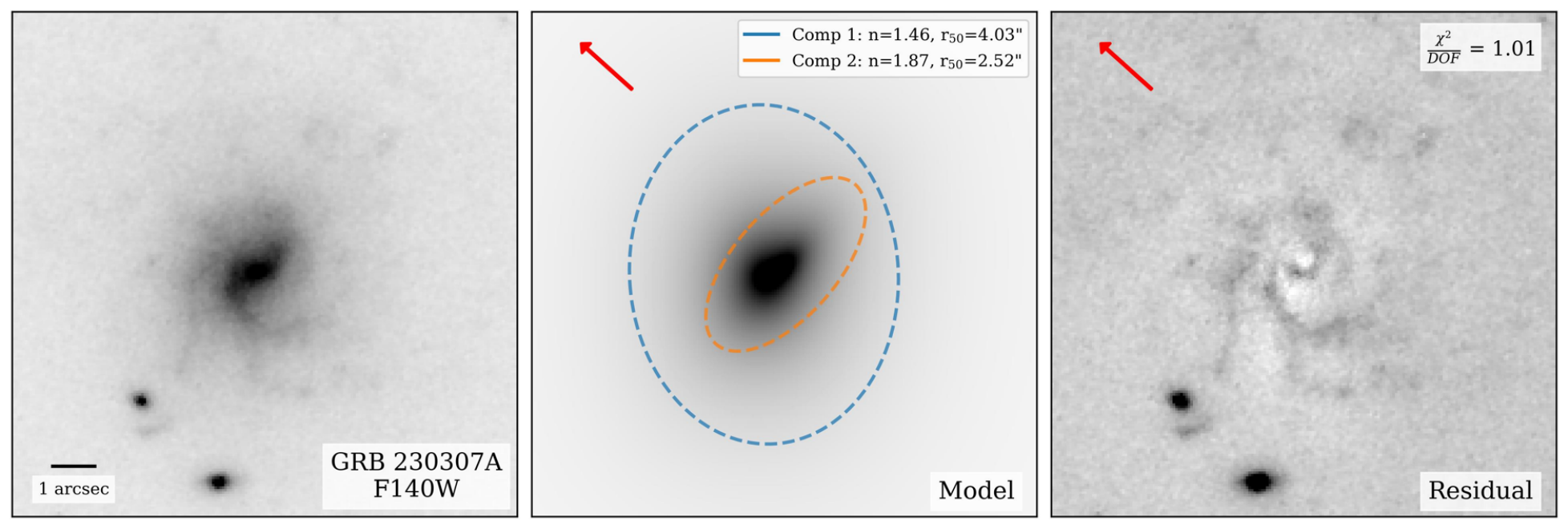}
\figsetgrpnote{\texttt{GALFIT} results. \textit{Left panel:} original data; \textit{Middle panel:} \texttt{GALFIT} model of best-fitting S\'ersic light profiles; \textit{Right panel:} the residual image of \texttt{GALFIT} after image subtraction. The S\'ersic profile components are marked with dotted lines with the approximate location of transient indicated by the red circle. GRB 230307A falls outside the cutout and is located in the direction of the red arrow.}
\figsetgrpend

\figsetgrpstart
\figsetgrpnum{10.23}
\figsetgrptitle{GRB 230307A, F070W}
\figsetplot{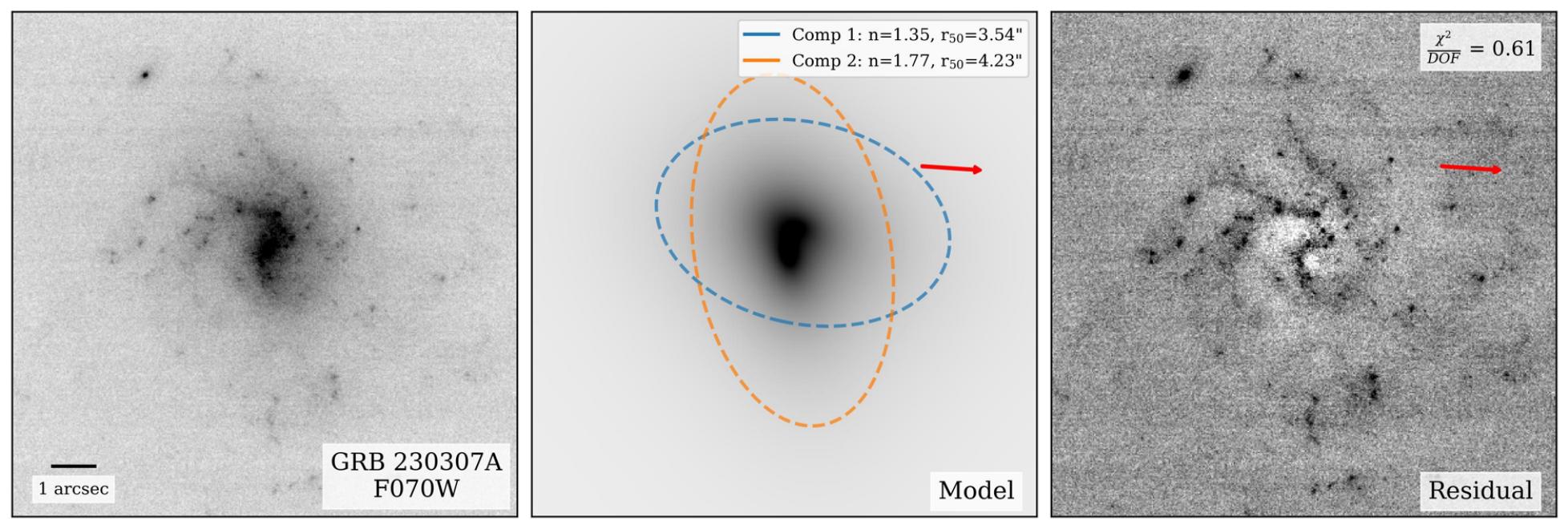}
\figsetgrpnote{\texttt{GALFIT} results. \textit{Left panel:} original data; \textit{Middle panel:} \texttt{GALFIT} model of best-fitting S\'ersic light profiles; \textit{Right panel:} the residual image of \texttt{GALFIT} after image subtraction. The S\'ersic profile components are marked with dotted lines with the approximate location of transient indicated by the red circle. GRB 230307A falls outside the cutout and is located in the direction of the red arrow.}
\figsetgrpend

\figsetgrpstart
\figsetgrpnum{10.24}
\figsetgrptitle{GRB 230307A, F115W}
\figsetplot{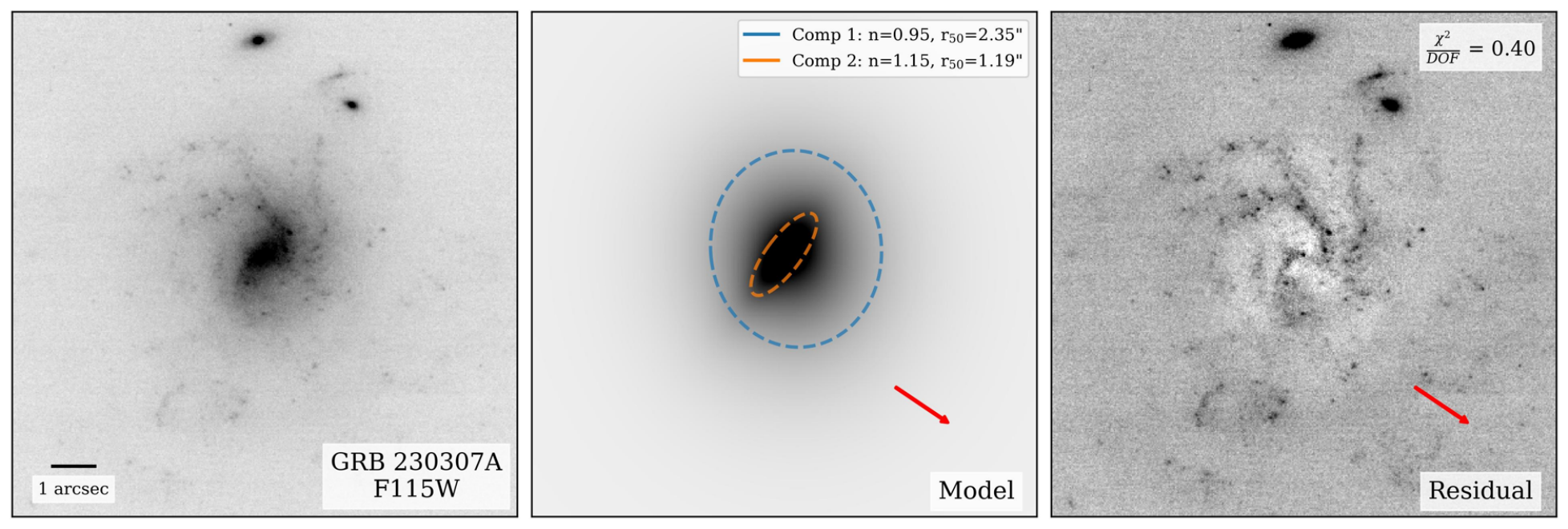}
\figsetgrpnote{\texttt{GALFIT} results. \textit{Left panel:} original data; \textit{Middle panel:} \texttt{GALFIT} model of best-fitting S\'ersic light profiles; \textit{Right panel:} the residual image of \texttt{GALFIT} after image subtraction. The S\'ersic profile components are marked with dotted lines with the approximate location of transient indicated by the red circle. GRB 230307A falls outside the cutout and is located in the direction of the red arrow.}
\figsetgrpend

\figsetgrpstart
\figsetgrpnum{10.25}
\figsetgrptitle{GRB 230307A, F150W}
\figsetplot{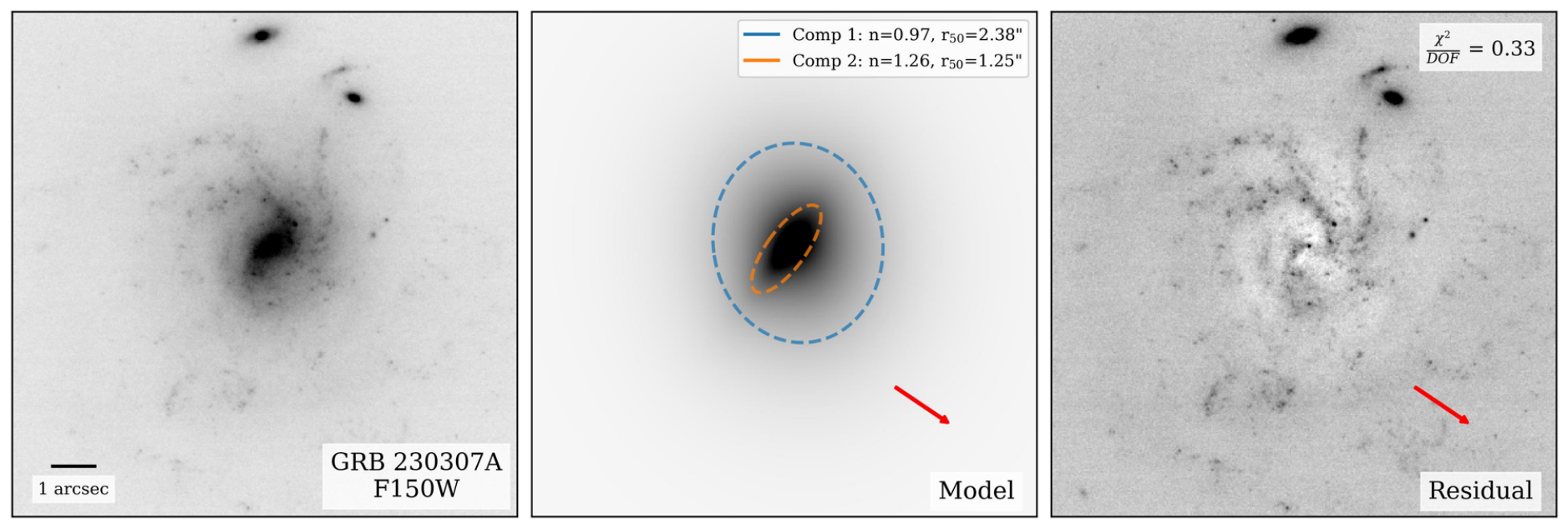}
\figsetgrpnote{\texttt{GALFIT} results. \textit{Left panel:} original data; \textit{Middle panel:} \texttt{GALFIT} model of best-fitting S\'ersic light profiles; \textit{Right panel:} the residual image of \texttt{GALFIT} after image subtraction. The S\'ersic profile components are marked with dotted lines with the approximate location of transient indicated by the red circle. GRB 230307A falls outside the cutout and is located in the direction of the red arrow.}
\figsetgrpend

\figsetgrpstart
\figsetgrpnum{10.26}
\figsetgrptitle{GRB 230307A, F277W}
\figsetplot{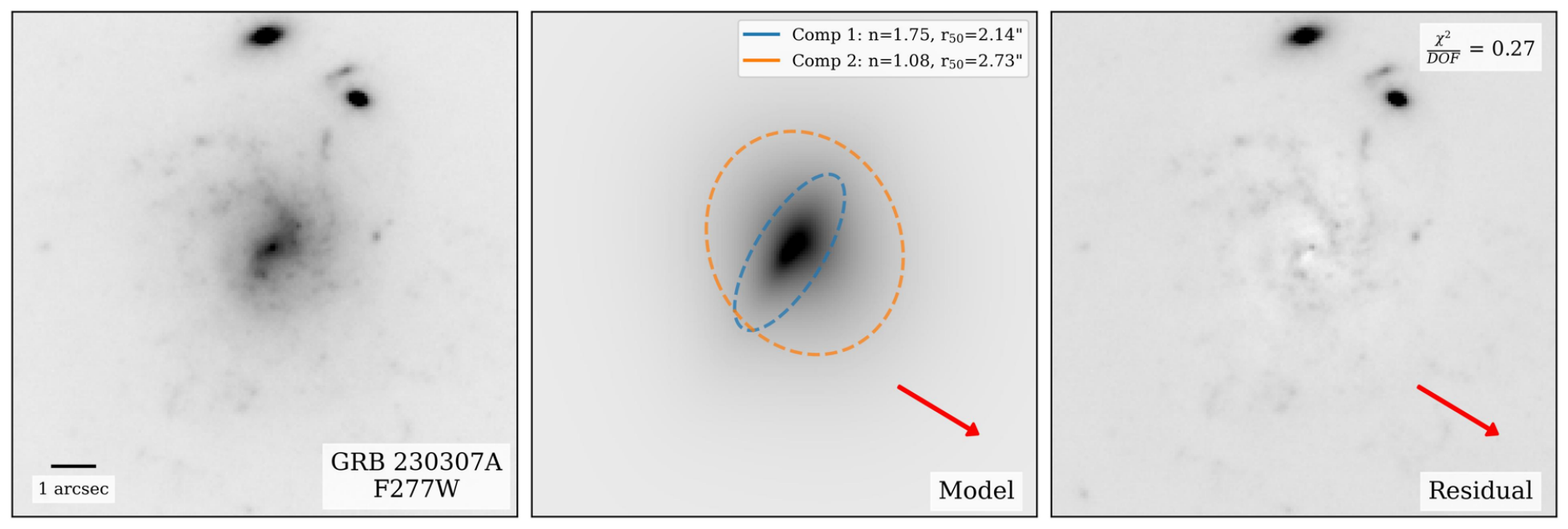}
\figsetgrpnote{\texttt{GALFIT} results. \textit{Left panel:} original data; \textit{Middle panel:} \texttt{GALFIT} model of best-fitting S\'ersic light profiles; \textit{Right panel:} the residual image of \texttt{GALFIT} after image subtraction. The S\'ersic profile components are marked with dotted lines with the approximate location of transient indicated by the red circle. GRB 230307A falls outside the cutout and is located in the direction of the red arrow.}
\figsetgrpend

\figsetgrpstart
\figsetgrpnum{10.27}
\figsetgrptitle{GRB 230307A, F356W}
\figsetplot{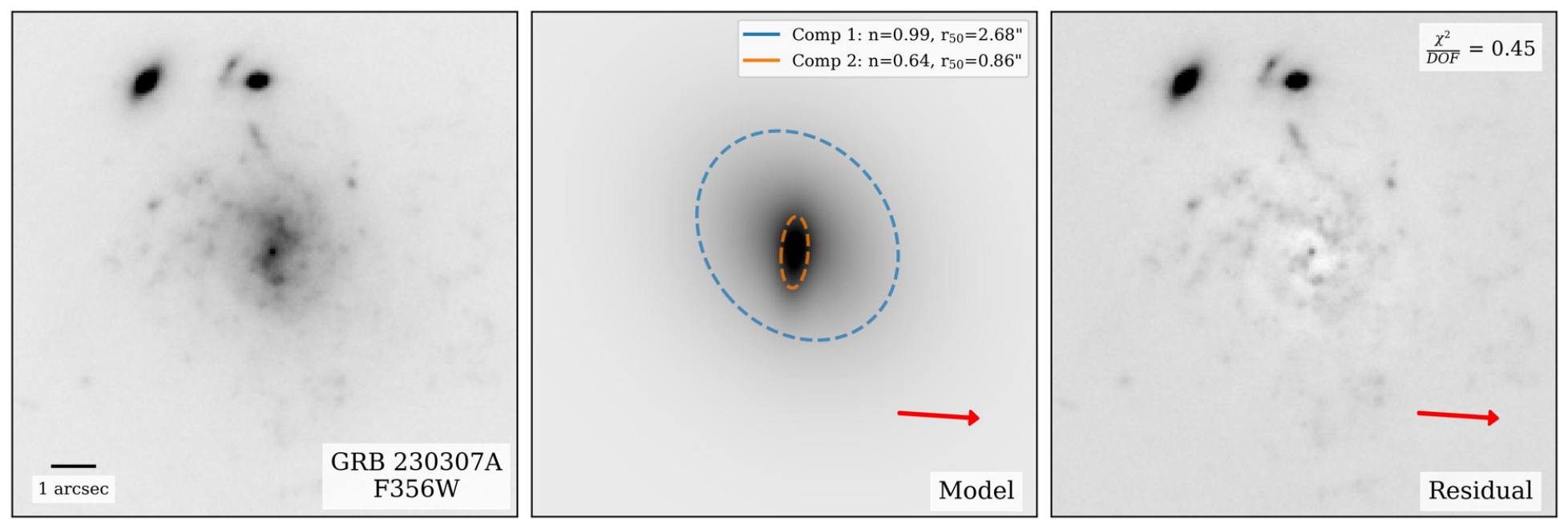}
\figsetgrpnote{\texttt{GALFIT} results. \textit{Left panel:} original data; \textit{Middle panel:} \texttt{GALFIT} model of best-fitting S\'ersic light profiles; \textit{Right panel:} the residual image of \texttt{GALFIT} after image subtraction. The S\'ersic profile components are marked with dotted lines with the approximate location of transient indicated by the red circle. GRB 230307A falls outside the cutout and is located in the direction of the red arrow.}
\figsetgrpend

\figsetgrpstart
\figsetgrpnum{10.28}
\figsetgrptitle{GRB 230307A, F444W}
\figsetplot{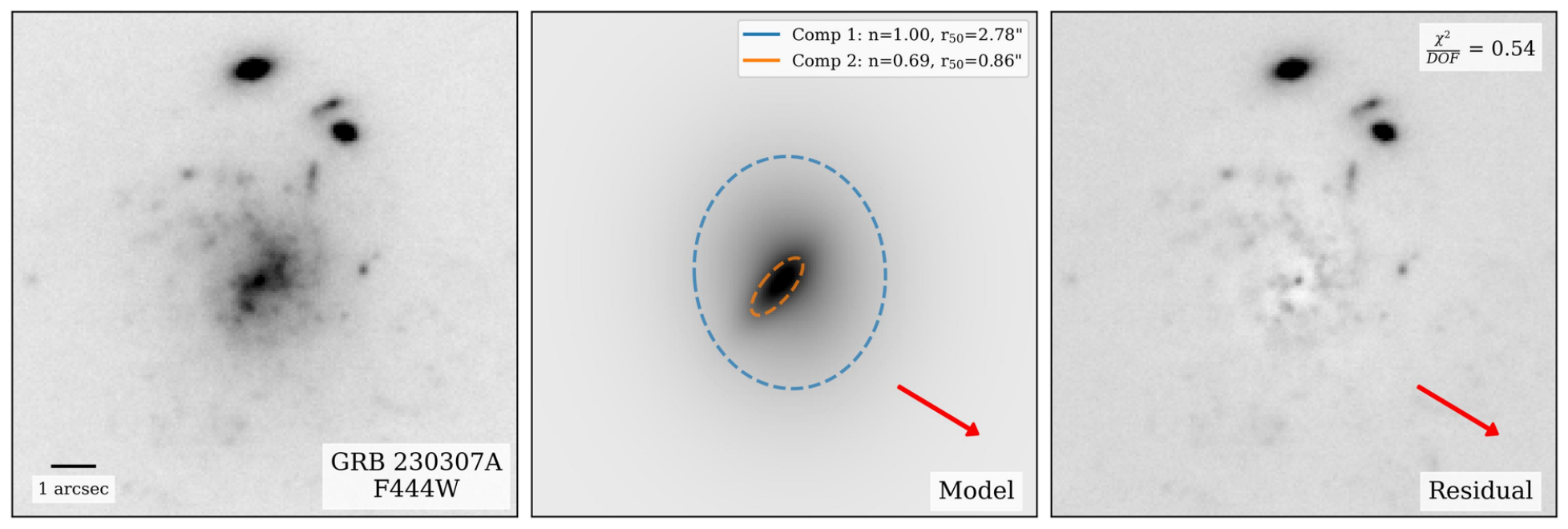}
\figsetgrpnote{\texttt{GALFIT} results. \textit{Left panel:} original data; \textit{Middle panel:} \texttt{GALFIT} model of best-fitting S\'ersic light profiles; \textit{Right panel:} the residual image of \texttt{GALFIT} after image subtraction. The S\'ersic profile components are marked with dotted lines with the approximate location of transient indicated by the red circle. GRB 230307A falls outside the cutout and is located in the direction of the red arrow.}
\figsetgrpend

\figsetend

\includegraphics[width=0.92\linewidth]{050709_F814W_galfit_ellipse_multi.pdf}
\caption{\texttt{GALFIT} results. \textit{Left panel:} original data; \textit{Middle panel:} \texttt{GALFIT} model of best-fitting S\'ersic light profiles; \textit{Right panel:} the residual image of \texttt{GALFIT} after image subtraction. The S\'ersic profile components are marked with dotted lines with the approximate location of transient indicated by the red circle. GRB 230307A falls outside the cutout and is located in the direction of the red arrow. The complete figure set (28 images) is available in the online journal.}
\label{fig:galfit}
\end{figure*}

In Figure \ref{fig:galfit} we show the results of our \texttt{GALFIT} modeling for each event in our sample. 

\subsection{GRB 050709} 

The host galaxy of GRB 050709 visibly shows an irregular morphology. We model the light with two Sérsic profiles that highlights the disturbed nature of the galaxy.

\subsection{GRB 060614}

The host galaxy of GRB 060614 appears to be elliptical. We model the light with a single Sérisc profile.

\subsection{GRB 130603B}

The host galaxy of GRB 130603B shows an irregular morphology. We model the light with two Sérsic profiles, highlighting disturbed structures in the galaxy.

\subsection{GRB 150101B}

The host galaxy of GRB 150101B has no distinct features observed by eye. We model the light with two Sérsic profiles to reveal an AGN.

\subsection{GRB 160821B}

The host galaxy of GRB 160821B visibly shows spiral morphology. We model the light with two Sérsic profiles that show pockets of star formation in the arms.

\subsection{GRB 170817A}

The host galaxy of GRB 170817A appears to have no features. We model the light with two Sérsic profiles to reveal concentric shell structures and numerous globular clusters \citep{Blanchard2017, Levan2017, Pan2017, Lee2018, Fong2019, Kilpatrick2022}. 

\subsection{GRB 200522A}

The host galaxy of GRB 200522A visibly shows irregular morphology. We model the light with two Sérsic profiles that show the disturbed features of the galaxy.

\subsection{GRB 211211A}

The host galaxy of GRB 211211A appears elliptical. We mode the light with two Sérsic profiles that reveals spiral arms branching out from the center.

\subsection{GRB 230307A}

The host galaxy of GRB 230307A shows a spiral morphology. We model the light with two Sérsic profiles, highlighting the spiral structure and pockets of star formation.

\bibliography{knbib}{}
\bibliographystyle{aasjournal}



\end{document}